\documentclass[12pt]{report}
\usepackage[dvips]{graphicx}
\usepackage{latexsym}
\usepackage{amssymb}

\newcommand{\C}{\mathbb{C}}

\newcommand{\R}{\mathbb{R}}

\newcommand{\Z}{\mathbb{Z}}

\let\a=\alpha

\def\pr{{\bf pr}}

\def\t{\tilde}
\def\ov{\overline}
\def\co{\mbox{const}}
\def\p{\partial}

\def\g{\mathfrak{g}}
\def\pr{{\bf pr}}
\def\opsi{\overline{\psi}}
\def\ophi{\overline{\phi}}

\newcommand{\bea}{\begin{eqnarray}}
\newcommand{\eea}{\end{eqnarray}}

\newcommand{\koniec}{\begin{flushright}  $\Box $ \end{flushright}}
\def\be{\begin{equation}}
\def\ee{\end{equation}}

\def\theequation{\thesection.\arabic{equation}}

\def\t{\tilde}

\def\ve{\varepsilon}

\def\th{\theta}
\def\const{\mbox{const}}

\def\ov{\overline}
\def\g{\mathfrak{g}}

\def\p{\partial}

\def\ov{\overline}

\def\a{\alpha}

\newtheorem{theo}{Theorem}[section] 
\newtheorem{prop}[theo]{Proposition}  

\newtheorem{defi}[theo]{Definition}

\usepackage{graphicx}
\begin{document}
\pagestyle{plain}
\title{\vskip -70pt
\begin{flushright}
\end{flushright}
\vskip 80pt
{\bf Integrable systems} \vskip 20pt}
\author{Maciej Dunajski \\[10pt]
{\sl Department of Applied Mathematics and Theoretical Physics} \\[5pt]
{\sl University of Cambridge} \\[5pt]
{\sl Wilberforce Road, Cambridge CB3 0WA, UK} \\[15pt]} 
\maketitle
\tableofcontents
\newpage


\chapter*{Preface}
These notes are based on lecture courses I gave to 
third year mathematics students at Cambridge in years 2003–-2008 and
2025--2026.
They could form a basis of an elementary one--term lecture course on 
integrable systems covering the Arnold--Liouville theorem,
inverse scattering transform, Hamiltonian methods in soliton theory 
and Lie point symmetries.  No knowledge beyond basic calculus and ordinary 
differential equation is assumed.
These notes are likely to contain fewer errors
than the  first four 
chapters of \cite{Dunajski}, but possibly more errors than \cite{Dunajski1}, where some of this material has appeared.

There are some  excellent text books which treat the material presented here
in great depth. 
Readers should consult \cite{arnold} for action-angle variables, \cite{NMPZ}
for inverse scattering transform,  and \cite{olver}  for the symmetry methods.
Several other books, as well as some landmark papers on integrability are 
listed in the Bibliography.

While preparing the notes I have been influenced by work and lectures of
many colleagues. In particular I would like to thank
Boris Dubrovin, Jenya Ferapontov,  Piotr Kosinski, Nick Manton,  Paolo Santini,
David Stuart and Paul Tod, as well as many  Cambridge  students whose names I do not remember, but who pointed out mistakes
in earlier drafts.

\chapter*{Introduction}
Integrable systems are nonlinear differential equations which
`in principle' can be solved analytically. This means that the solution can 
be reduced to a finite number of algebraic operations and integrations.
Such systems are very rare - most nonlinear differential 
equations admit chaotic behaviour and no explicit solutions can be written 
down. Integrable systems nevertheless lead to a very interesting mathematics
ranging from differential geometry and complex analysis to quantum field
theory and fluid dynamics. 
The main reference for the course is \cite{Dunajski}. There are other books
which cover particular topics treated in the course:
\begin{itemize}
\item {\bf Integrability of ODEs} \cite{arnold}
(Hamiltonian formalism, Arnold--Liouville theorem,  action--angle variables). 
The integrability of ordinary differential equations is a fairly clear 
concept (i.e. it can be defined)  based on existence of sufficiently 
many well behaved first integrals, or (as a physicist would put it) 
constant of motions. 

\item  {\bf Integrability of PDEs} \cite{NMPZ}, \cite{DJ}(Solitons, Inverse Scattering Transform).
The universally accepted definition of integrability does not exist in this 
case. The phase space is infinite dimensional but having `infinitely many'
first integrals may not be enough - we could have missed every second one.
Here one focuses on properties of solutions and solutions generation 
techniques. We shall study solitons - 
solitary non-linear waves which preserve their shape (and other 
characteristics) in the evolution. These soliton solutions will be constructed
by means of an inverse problem: recovering 
a potential from the scattering data.

\item {\bf Lie symmetries} \cite{Hydon}, \cite{olver} (Group invariant solutions, vector fields,
symmetry reduction,  Painlev\'e equations). The powerful symmetry methods
can be applied to ODEs and PDEs alike. 
In case of ODEs a knowledge of 
sufficiently large symmetry group  
allows a construction of the most general solution.
For PDEs the knowledge of symmetries is not sufficient to construct the most
general solution, but it can be used to find new solutions from given ones
and to reduce PDEs to more tractable ODEs. The PDEs integrable by inverse
problems reduce to equations with Painlev\'e property.
\end{itemize}

\chapter{Integrability in classical mechanics}
\label{chapter_int_clas}
In this Chapter we shall introduce the integrability of ordinary 
differential equations. It is a fairly clear 
concept  based on existence of sufficiently 
many well behaved first integrals. 
\section{Hamiltonian formalism}
\begin{figure}
\caption{Level surface}
\label{IS_levelsurface}
\begin{center}
\includegraphics[width=10cm,height=5cm,angle=0]{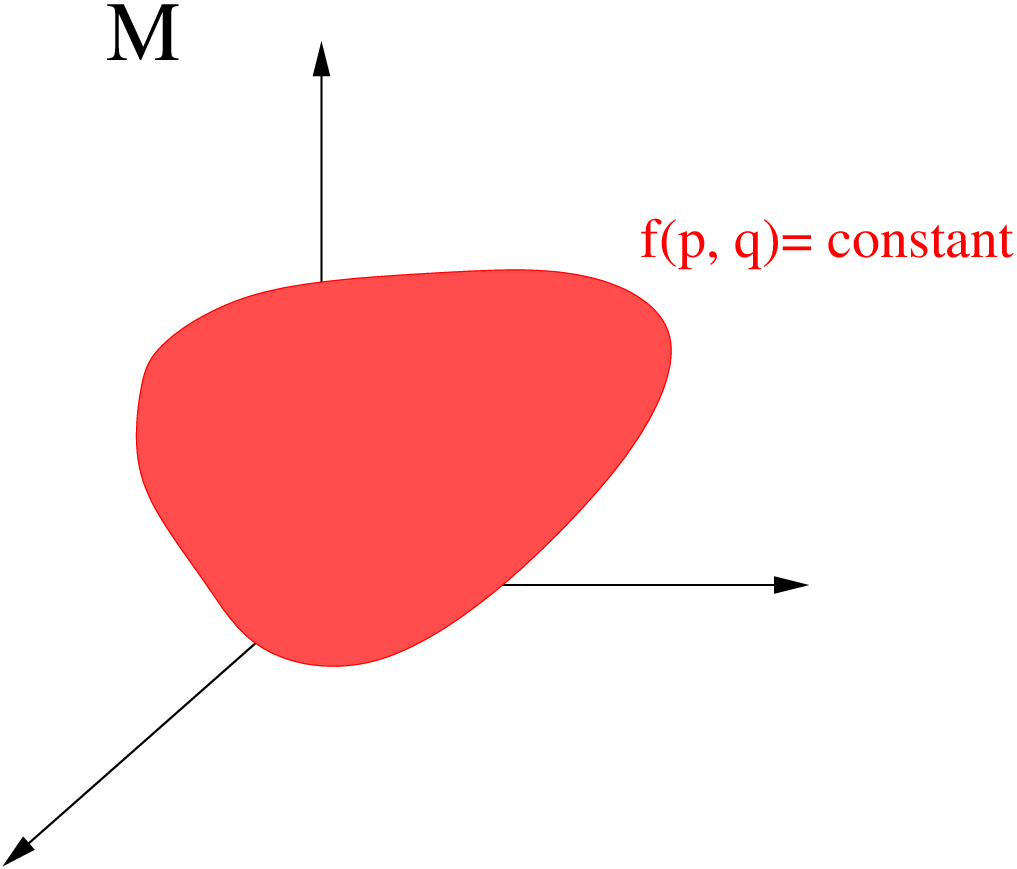}
\end{center}
\end{figure}

Motion of a system with $n$ degrees of freedom  is described by a trajectory in a $2n$ dimensional phase space $M$ 
(locally think of an open set in $\R^{2n}$ but globally 
it can be topologically 
non-trivial manifold  - e.g. a sphere or a torus. 
See Appendix \ref{appendix_mdf}) with local coordinates
\[
(p_j, q_j), \quad j=1, 2, \ldots, n.
\]
The dynamical variables are 
functions $f:M\times \R\longrightarrow \R$, so that
$f=f(p, q, t)$ where $t$ is called `time'.
Let $f, g:M\times\R\longrightarrow \R$. Define a Poisson 
bracket of $f, g$ to be a function 
\be
\label{poisson}
\{f, g\}:=\sum_{k=1}^n\frac{\p f}{\p q_k}\frac{\p g}{\p p_k}-
\frac{\p f}{\p p_k}\frac{\p g}{\p q_k}.
\ee
It satisfies
\[
\{f, g\}=-\{g, f\},\qquad
\{f,\{g, h\}\}+\{g,\{h, f\}\}+\{h,\{f, g\}\}=0.
\]
The second property is called the Jacobi identity.
The coordinate functions $(p_j, q_j)$ satisfy the canonical commutation
relations
\[
\{p_j, p_k\}=0, \quad \{q_j, q_k\}=0, \quad
\{q_j, p_k\}=\delta_{jk}.
\]
Given a Hamiltonian $H=H(p, q, t)$ (usually
$H(p, q)$) the dynamics is determined by
\[
\frac{d f}{d t}=\frac{\p f}{\p t}+\{ f, H\},\qquad\mbox{for any}\qquad
f=f(p, q, t).
\]
Setting $f=p_j$ or $f=q_j$ yields Hamilton's equations of motion
\be
\label{caneq}
\dot{p}_j=-\frac{ \p H}{\p q_j},\qquad \dot{q}_j=\frac{\p H}{\p p_j}. 
\ee
The system (\ref{caneq}) of $2n$ ODEs is {\em deterministic} in a sense
that $(p_j(t), q_j(t))$ are uniquely determined  by $2n$ initial conditions
$(p_j(0), q_j(0))$. Equations (\ref{caneq}) also imply that
volume elements in phase space are conserved. This system is essentially 
equivalent to Newton's equations of motion. The Hamiltonian formulation
allows a more geometric insight to classical mechanics. It is also
the starting point to quantisation.
\begin{defi}
A function  $f=f(p_j, q_j, t)$ which satisfies $\dot f=0$ 
when equations {\em(\ref{caneq})} hold
is called
a first integral or a constant of motion. Equivalently,
\[
f(p(t), q(t), t)=\co
\]
if $p(t), q(t)$ are solutions of {\em(\ref{caneq})}.
\end{defi}
In general the system 
(\ref{caneq}) will be solvable if it admits  `sufficiently many' first 
integrals and the reduction of order can be applied. This is because
any first integral eliminates one equation.
\begin{itemize}
\item {\bf Example.} Consider a system with one degree of freedom with
$M=\R^2$ and the Hamiltonian
\[
\label{1Dfreedom}
H(p, q)=\frac{1}{2}p^2+V(q).
\]
Hamilton's equations (\ref{caneq}) give 
\[
\dot{q}=p, \quad \dot{p}=-\frac{dV}{dq}.
\]
The Hamiltonian itself is a first integral as $\{H, H\}=0$. Thus
\[
\frac{1}{2}p^2+V(q)=E
\]
where $E$ is a constant called energy. Now
\[
\dot{q}=p, \quad p=\pm\sqrt{2(E-V(q)}
\]
and one integration gives a solution in the implicit form
\[
t=\pm\int\frac{dq}{\sqrt{2(E-V(q))}}.
\]
The explicit solution could be found if we can perform the integral
on the RHS and invert the relation $t=t(q)$ to find $q(t)$. These two steps
are not always possible to take but nevertheless we would certainly regard
this system as integrable.
\end{itemize}
It is useful to adopt a more geometric approach. Assume that a first
integral $f$ does not explicitly depend on time, and that it defines a 
hypersurface $f(p, q)=\co$ in $M$ (Figure.\ref{IS_levelsurface}).
Two hypersurfaces corresponding to two independent first
integrals generically intersect in a surface of co--dimension 2 in $M$.
In general the trajectory lies on a surface of dimension $2n-L$ where
$L$ is the number of independent first integrals. If $L=2n-1$ this surface
is a curve - a solution to (\ref{caneq}).

\vskip10pt
How to find first integrals? Given two first integrals
which do not explicitly depend on time their Poisson bracket will
also be a first integral if it is not zero. This follows from the Jacobi 
identity and the fact all first  integrals Poisson commute with the 
Hamiltonian.
More generally, the  Noether theorem gives some first integrals  (they
correspond to symmetries Hamilton's equation 
(\ref{caneq}) may possess e.g. time translation, rotations) but not enough.  
The difficulty with finding the first integrals has deep significance.
For assume  we use some existence theorem for ODEs and apply it
to (\ref{caneq}). Now solve the algebraic equations
\[
q_k=q_k(p^0, q^0, t), \quad p_k=p_k(p^0, q^0, t),
\]
for the initial conditions $(p^0, q^0)$ thus giving
\[
{q^0}_k={q^0}_k(p, q, t), \quad {p^0}_k={p^0}_k(p, q, t).
\]
This gives $2n$ first integrals as obviously $(p^0, q^0)$ are
constants which we can freely specify. One of these integrals
determines the time parametrisations and others could perhaps be used
to construct the trajectory in the phase space. However
for some of the integrals the equations
\[
f(p, q)=\co
\]
may not define a `nice' surface in the phase space. Instead
it defines a pathological (at least from the applied 
mathematics point of view) 
set which densely covers the phase space.
Such integrals do not separate points in $M$.
\vskip10pt
One first integral - energy - always
exist for Hamiltonian systems giving the energy surface
$H(p, q)=E$, but often it is the only first integral.
Sufficiently complicated, deterministic, systems may behave according
to the laws of thermodynamics:  probability that the system 
is contained in some element of the energy surface is proportional to the 
normalised volume  of this element. This means that the time evolution  
covers uniformly 
the entire region of the constant energy surface in the phase space.
It is not known wether this ergodic postulate can be derived from 
Hamilton's equations.
\vskip10pt
Early computer simulations in the 1960s revealed that some nonlinear
systems (with infinitely many degrees of freedom!) are not ergodic.
Soliton equations
\[
u_t=6uu_x-u_{xxx}, \quad u=u(x, t), \qquad KdV
\]
or 
\[
\phi_{xx}-\phi_{tt}=\sin{\phi}, \quad \phi=\phi(x, t), \qquad Sine-Gordon
\]
are examples of such systems. Both posses infinitely many first
integrals. We shall study them in Chapter \ref{chapter_sol}.
\section{Integrability and action--angle variables}
\label{act_ang_sec}
Given a  system of Hamilton's equations 
(\ref{caneq}) it is often sufficient to know $n$ (rather than $2n-1$) first
integrals as each of them reduces the order of the system by two.
This underlies the following  definition of an {\em integrable system}.
\begin{defi}
\label{definition_al}
An integrable system consists of a $2n$-dimensional phase-space ${M}$ together
with $n$  independent functions (in a sense that the gradients
$\nabla f_j$ are linearly independent vectors on a tangent space
to any point in ${M}$)
$f_1, \dots, f_n:{M}\rightarrow \R$ such 
that 
\be
\label{involution}
\{f_j, f_k\}=0, \qquad j, k=1, \ldots, n.
\ee
\end{defi}
The vanishing of Poisson brackets (\ref{involution}) means that the first 
integrals are in involution.  We shall  
show that integrable
systems lead to completely solvable Hamilton's equations of motion.
Let us first
explore the freedom in (\ref{caneq})
given by a coordinate transformation of a phase-space
\[
Q_k=Q_k(p, q), \qquad P_k=P_k(p,q).
\]
This transformation  is called
{\em canonical} if it preserves the Poisson bracket
\[
\sum_{k=1}^n\frac{\p f}{\p q_k}\frac{\p g}{\p p_k}-
\frac{\p f}{\p p_k}\frac{\p g}{\p q_k}=
\sum_{k=1}^n\frac{\p f}{\p Q_k}\frac{\p g}{\p P_k}-
\frac{\p f}{\p P_k}\frac{\p g}{\p Q_k}
\]
for all $f, g:{M}\longrightarrow \R$. Canonical transformations
preserve Hamilton's equation (\ref{caneq}).

Given a function 
$
S(q, P, t)
$
such that \[
\mbox{det}\;\Big(\frac{\p^2 S}{\p q_j\p P_k}\Big)\neq 0
\] 
we can construct a canonical transformation by setting
\[
p_k=\frac{\p S}{\p q_k}, \quad Q_k=\frac{\p S}{\p P_k},\quad
\widetilde{H}=H+\frac{\p S}{\p t}.
\]
The function $S$ is an example of a generating function 
\cite{arnold,landau,woodhouse}. 
The idea behind the following Theorem  is to 
seek a canonical transformation such that in the new variables
$H=H(P_1, \dots, P_n)$ so that
\[
P_k(t)=P_k(0)=\mbox{const}, \qquad Q_k(t)=Q_k(0)+t\frac{\p H}{\p P_k}.
\] 
Finding a generating function for such canonical transformation 
is in practise very difficult, and 
deciding whether a given Hamiltonian system is integrable (without a priori knowledge of $n$ Poisson commuting integrals) is still an open problem.
\begin{theo}[Arnold, Liouville]
\label{al}
Let  \[(M, f_1, \dots, f_n)\] be an integrable system with a Hamiltonian
$H=f_1$, and let 
\[
{M}_f:=\{ (p, q)\in {M}; f_k(p, q)=c_k\}, 
\qquad c_k=\co, \qquad k=1,\dots,n
\]
be an $n$-dimensional level surface of first integrals $f_k$.
Then
\begin{itemize}
\item
If ${M}_f$ is compact and connected then it is diffeomorphic
to a torus 
\[
T^n:=S^1\times S^1\times \dots\times S^1,
\]
 
and (in a neighbourhood of this torus in $M$)
one can introduce the `action-angle' coordinates
\[
I_1, \dots, I_n, \phi_1, \dots, \phi_n, \qquad 0\leq\phi_k\leq 2\pi,
\]
such that {\em angles} $\phi_k$ are coordinates on $M_f$ and {\em actions} 
$I_k=I_k(f_1, \dots, f_n)$ are first integrals.
\item 
The canonical equations of motion {\em(\ref{caneq})} become
\be
\label{periodic}
\dot{I}_k=0,\qquad \dot{\phi}_k=\omega_k(I_1, \dots, I_n), \qquad k=1,\dots, n
\ee
and so the integrable systems are solvable by 
quadratures (a finite number of algebraic operations, and integrations
of known functions).
\end{itemize}
\end{theo}
{\bf Proof.}
We shall follow the proof   given  in \cite{arnold}, but try to make it more
accessible by
avoiding  the language of
differential forms
\begin{itemize}
\item The motion takes place on the surface
\[
f_1(p, q)=c_1, f_2(p, q)=c_2, \dots, f_n(p, q)=c_n
\]
of dimension $2n-n=n$. The first part of the Theorem says that
this surface is a torus\footnote{This part of the proof requires some knowledge
of Lie groups and Lie algebras. It is given in 
Appendix \ref{appendix_mdf}.}.
For each point in $M$ there exists precisely one
torus $T^n$ passing through that point. This means that $M$
admits a foliation by $n$--dimensional leaves.  Each leaf is a torus
and different tori correspond to different choices of the constants
$c_1, \dots, c_n$.

Assume 
\[
\mbox{det}\Big(\frac{\p f_j}{\p p_k}\Big)\neq 0
\]
so that the system $f_k(p, q)=c_k$ can be solved for the momenta
$p_i$ 
\[
p_i=p_i(q, c)
\]
and the relations $f_i(q, p(q, c))=c_i$ hold identically. 
Differentiate these identities with respect to  $q_j$ 
\[
\frac{\p f_i}{\p q_j}+\sum_k\frac{\p f_i}{\p p_k}\frac{\p p_k}{\p q_j}=0
\]
and multiply the resulting
equations by $\p f_m/\p p_j$ 
\[
\sum_j\frac{\p f_m}{\p p_j}\frac{\p f_i}{\p q_j}
+\sum_{j,k}\frac{\p f_m}{\p p_j}\frac{\p f_i}{\p p_k}
\frac{\p p_k}{\p q_j}=0.
\]
Now swap the indices and subtract $(mi)-(im)$. This yields
\[
\{ f_i, f_m\}+\sum_{j,k}\Big(\frac{\p f_m}{\p p_j}
\frac{\p f_i}{\p p_k} \frac{\p p_k}{\p q_j}-
\frac{\p f_i}{\p p_j}
\frac{\p f_m}{\p p_k} \frac{\p p_k}{\p q_j}
\Big)
=0.
\]
The first term vanishes as the first integrals are in involution. Rearranging
the indices in the second term gives
\[
\sum_{j,k}\frac{\p f_i}{\p p_k}\frac{\p f_m}{\p p_j}
\Big( \frac{\p p_k}{\p q_j}- \frac{\p p_j}{\p q_k}\Big)=0
\]
and, as the matrices ${\p f_i}/{\p p_k}$ are invertible,
\be
\label{curl}
\frac{\p p_k}{\p q_j}- \frac{\p p_j}{\p q_k}=0.
\ee
This condition implies that 
\[
\oint\sum_{j}p_j d q_j=0
\]
for any closed contractible curve on the torus $T^n$. This is
a consequence of the Stokes theorem. To see it recall that in $n=3$ 
\[
\oint_{\delta D} {\bf p}\cdot d{\bf q} =\int_D(\nabla\times{\bf p})\cdot d {\bf q}
\]
where $\delta D$ is a boundary of a surface $D$ and 
\[
(\nabla\times{\bf p})_m= 
\frac{1}{2}\epsilon_{jkm}\Big(\frac{\p p_k}{\p q_j}
- \frac{\p p_j}{\p q_k}\Big).
\]
\item There are $n$ closed curves which can not be contracted down to a point, 
so that the corresponding integrals do not vanish.
\begin{center}
\includegraphics[width=7cm,height=5cm,angle=0]{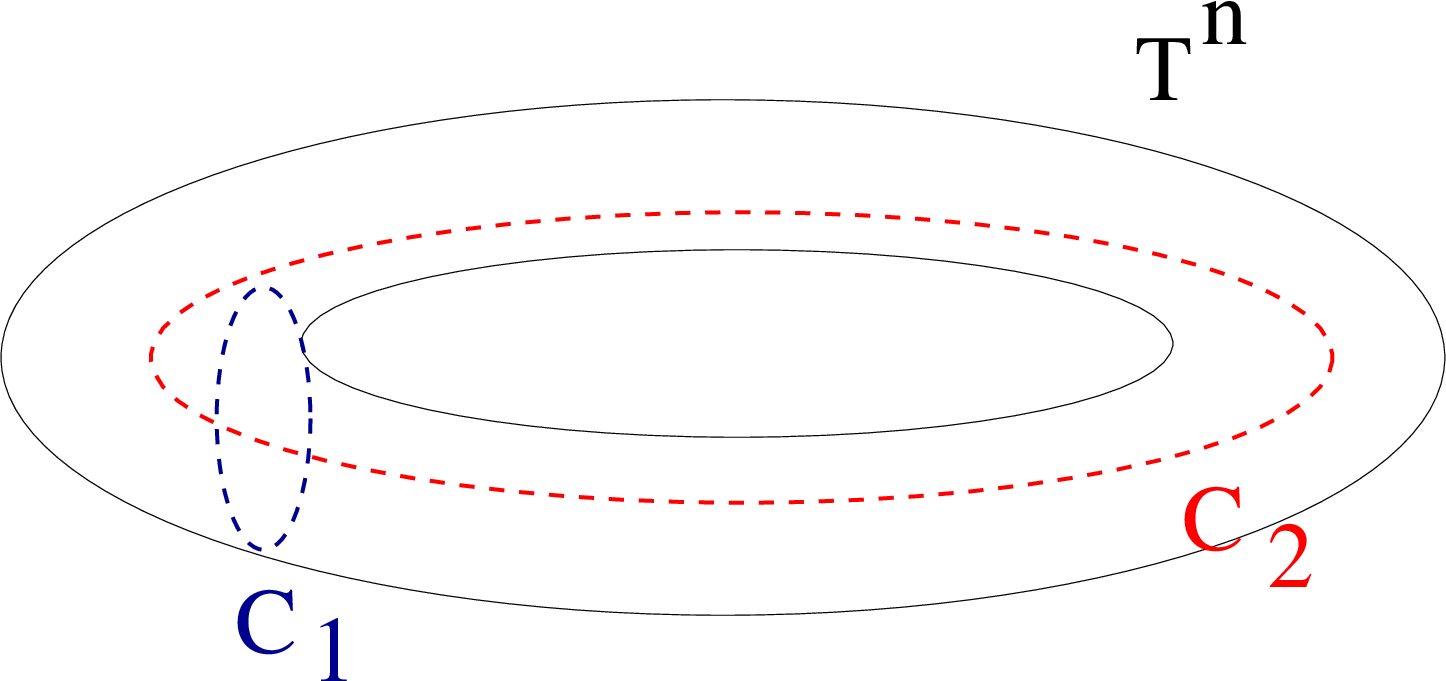}
{\bf  Cycles on a Torus}
\end{center}
Therefore we can define the action 
coordinates  
\be
\label{action}
I_k:=\frac{1}{2\pi}\oint_{\Gamma_k}\sum_{j}p_jd q_j,
\ee
where the closed curve $\Gamma_k$ is the $k$-th basic cycle 
(the term `cycle' in general means `submanifold without boundary') 
of the torus $T^n$
\[
\Gamma_k=\{(\tilde{\phi}_1, \dots, \tilde{\phi}_n)\in T^n; 
0\leq\tilde{\phi}_k\leq2\pi, \tilde{\phi}_j=\co
\;\;\mbox{for}\; j\neq k\},
\]
where $\tilde{\phi}$ are some coordinates\footnote{This is a non-trivial step. 
In practice it is unclear how to explicitly 
describe the $n$--dimensional torus and the curves $\Gamma_k$ in
$2n$ dimensional phase space. Thus, to some extend the Arnold--Liouville
theorem has a character of the existence theorem.}
 on $T^n$.

The Stokes theorem implies that the actions (\ref{action}) are 
independent on the
choice of $\Gamma_k$.
\begin{center}
\includegraphics[width=7cm,height=5cm,angle=0]{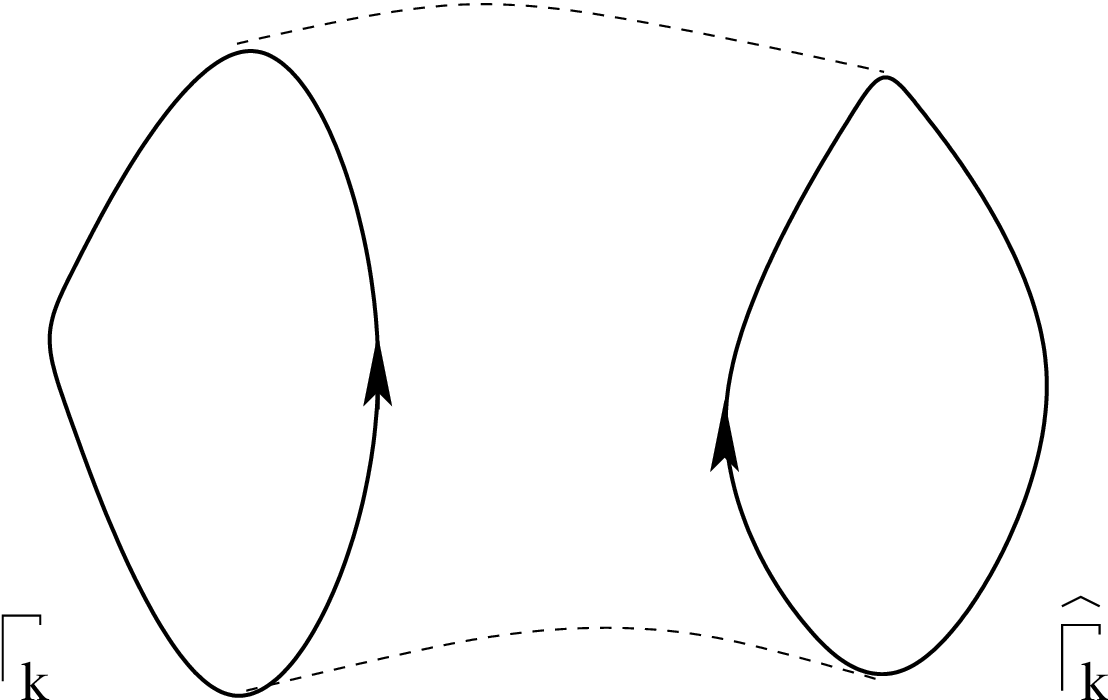}
\vskip5pt
{\bf Stokes Theorem}
\end{center}
This is because
\[
\oint_{\Gamma_k}\sum_{j}p_jd q_j+\oint_{\hat{\Gamma}_k}\sum_{j}p_jd q_j
=\int \Big(\frac{\p p_i}{\p q_j}- \frac{\p p_j}{\p q_i}\Big)
dq_j\wedge dq_i=0
\]
where we have chosen $\Gamma$ and $\hat{\Gamma}$ to have opposite orientations.
\item The actions (\ref{action}) are also first integrals
as $\oint p(q, c)dq$ only depends on $c_k=f_k$
and $f_k$s are first integrals.
The actions are Poisson commuting
\[
\{I_i, I_j\}=\sum_{r, s, k}\frac{\p I_i}{\p f_r}\frac{\p f_r}{\p q_k}
\frac{\p I_j}{\p f_s}\frac{\p f_s}{\p p_k}
-\frac{\p I_i}{\p f_r}\frac{\p f_r}{\p p_k}
\frac{\p I_j}{\p f_s}\frac{\p f_s}{\p q_k}
=\sum_{r, s}\frac{\p I_i}{\p f_r}\frac{\p I_j}{\p f_s}\{f_r, f_s\}=0
\]
and in particular $\{I_k, H\}=0$.

The torus $M_f$ can be equivalently represented
by
\[
I_1= \tilde{c}_1,\qquad \ldots,\qquad I_1= \tilde{c}_n.
\]
for some constants $\tilde{c}_1, \dots, \tilde{c}_n$
(We might have been 
tempted just to define $I_k=f_k$ but then the transformation
$(p, q)\rightarrow (I, \phi)$ would not be canonical in general.)

\item We shall construct the angle coordinates $\phi_k$ canonically
conjugate to the actions using  a generating function
\[
S(q, I)=\int_{q_0}^q \sum_j p_j dq_j,
\]
where $q_0$ is some chosen point on the torus.
This definition does not depend on a path joining $q_0$ and $q$ as
a consequence of (\ref{curl}) and Stokes's theorem. Choosing a different
$q_0$ just adds a constant to $S$ thus leaving
the {\it angles}
\[
\phi_i=\frac{\p S}{\p I_i}
\]
invariant.
\item 
The angles are periodic coordinates with a period $2\pi$. To see it
consider two paths $C$ and $C\cup C_k$ 
(where $C_k$ represents the $k$th cycle)
between $q_0$ and $q$
and calculate
\[
S(q, I)=\int_{C\cup C_k}\sum_j p_j dq_j =
\int_{C}\sum_j p_j dq_j+ \int_{ C_k}\sum_j p_j dq_j
=S(q, I)+2\pi I_k
\]
so 
\[
\phi_k=\frac{\p S}{\p I_k}=\phi_k+2\pi.
\]
\begin{center}
\includegraphics[width=7cm,height=5cm,angle=0]{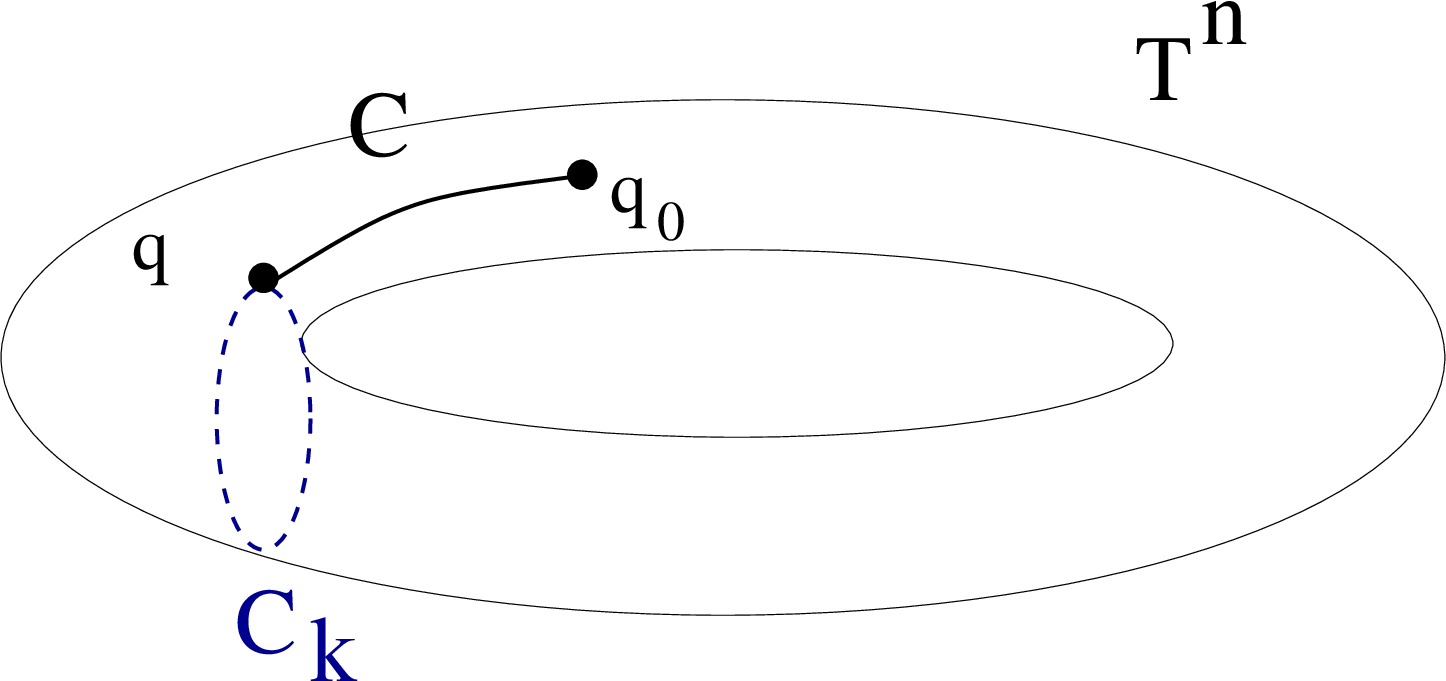}
{\bf Generating Function}
\end{center}
\item
The transformations 
\[
q=q(\phi, I), \quad p=p(\phi, I), \qquad\mbox{and}\qquad
\phi=\phi(q, p), \quad I=I(q, p)
\]
are canonical (as they are defined by a generating function) 
and invertible. Thus 
\[
\{ I_j, I_k\}=0, \quad \{ \phi_j, \phi_k\}=0,\quad \{ \phi_j, I_k\}=\delta_{jk}
\]
and the dynamics is given by 
\[
\dot{\phi}_k=\{ \phi_k, \widetilde{H}\},\quad
\dot{I}_k=\{ I_k, \widetilde{H}\},
\]
where
\[
\widetilde{H}(\phi, I)=H(q(\phi, I),p(\phi, I)).
\] 
The  $I_k$s are first integrals, therefore \[
\dot{I}_k=
-\frac{\p \widetilde{H}}{\p \phi_k}=0
\]
so $\widetilde{H}=\widetilde{H}(I)$ and
\[
 \dot{\phi}_k=\frac{\p \widetilde{H}}{\p I_k}=\omega_k(I)
\]
where the $\omega_k$s are also first integrals. This proves (\ref{periodic}).
Integrating these canonical equations of motion
yields
\be
\label{periodic_traj_al}
{\phi}_k(t)=\omega_k(I)t+\phi_k(0) ,\quad I_k(t)=I_k(0).
\ee
These are $n$ circular motions with constant angular velocities.
\koniec
\end{itemize}
The trajectory (\ref{periodic_traj_al}) may be closed on 
the torus or it may cover it densely.
That depends on the values of the angular velocities. If $n=2$
the trajectory will be closed if $\omega_1/\omega_2$ is rational and
dense otherwise.

 Interesting things happen  to the tori under a small perturbation
of the integrable Hamiltonian
\[
H(I)\longrightarrow H(I)+\epsilon K(I, \phi).
\]
In some circumstances the motion is still periodic and most tori do not vanish
but become deformed. This is governed by the  Kolmogorov--Arnold--Moser (KAM)
theorem -  not covered in this course.  
Consult  the popular book  by Schuster \cite{schuster}, 
or  read the complete account  given by Arnold \cite{arnold}.
\begin{itemize}
\item  {\bf Example.}
All time-independent Hamiltonian system with two-dimensional phase
spaces are integrable. 
Consider the harmonic oscillator with the Hamiltonian
\[
H(p, q)=\frac{1}{2}(p^2+\omega^2 q^2).
\]
Different choices of  the energy  $E$ give a foliation of $M_f$
by ellipses
 \[
\frac{1}{2}(p^2+\omega^2q^2)=E.
\]

For a fixed value of $E$ we can take $\Gamma={M}_f$.
Therefore
\[
I=\frac{1}{2\pi}\oint_{M_f} p d q=\frac{1}{2\pi}\int\int_S dp dq=\frac{E}{\omega}
\]
where we used the Stokes's theorem to  express
the line integral in terms of the  area enclosed by $M_f$.

The Hamiltonian expressed in the new variables
is $\widetilde{H}=\omega I$ and
\[
\dot{\phi}=\frac{\p \widetilde{H}}{\p I}=\omega, \qquad
\phi=\omega t+\phi_0.
\]
To complete the picture we need to express $(I, \phi)$ 
in terms of $(p, q)$. We already know
\[
I=\frac{1}{2}\Big(\frac{1}{\omega}p^2+\omega q^2\Big).
\]
Thus the generating function is
\[
S(q, I)=\int pdq=\pm\int\sqrt{2I\omega-\omega^2 q^2}dq
\]
and (choosing a sign)
\[
\phi=\frac{\p S}{\p I}=\int\frac{\omega dq}{\sqrt{2I\omega-\omega^2 q^2}}
=  \arcsin{\Big(q\sqrt\frac{\omega}{2I}\Big)}-\phi_0.
\]
This gives
\[
q=\sqrt{\frac{2I}{\omega}}\sin{(\phi+\phi_0)}
\]
and finally we recover the familiar solution
\[
p=\sqrt{2E}\cos{(\omega t+\phi_0)},
\qquad q=\sqrt{2E/\omega^2}\sin{(\omega t+\phi_0)}.
\]
\item {\bf Example.} The Kepler problem is another doable example. Here the four--dimensional
phase space is coordinatised by $(q_1=\phi, q_2=r, p_1=p_\phi, p_2=p_r)$
and the Hamiltonian is
\[
H=\frac{{p_\phi}^2}{2r^2}+\frac{{p_r}^2}{2}-\frac{\alpha}{r}
\]
where $\alpha>0$ is a constant. One readily verifies that
\[
\{H, p_\phi\}=0
\]
so the system is integrable in the sense of Definition 
\ref{definition_al}. The level
set $M_f$ of first integrals is given by
\[
H=E,\quad p_\phi=\mu
\]
which gives
\[
p_\phi=\mu, \quad p_r=\pm\sqrt{2E-\frac{\mu^2}{r^2}+\frac{2\alpha}{r}}.
\]
This leaves $\phi$ arbitrary and gives one constraint on $(r, p_r)$. Thus
$\phi$ and one function of $(r, p_r)$ parametrise $M_f$. Varying $\phi$
and fixing the other coordinate gives one cycle $\Gamma_\phi\subset M_f$
and
\[
I_\phi=\frac{1}{2\pi}\oint_{\Gamma_\phi} p_\phi d\phi +p_r dr=
\frac{1}{2\pi}\int_0^{2\pi}p_\phi d\phi=p_\phi.
\]
To find the second action coordinate fix $\phi$ (on top of $H$ and $p_\phi$).
This gives another cycle $\Gamma_r$ and
\begin{eqnarray*}
I_r&=&\frac{1}{2\pi}\oint_{\Gamma_r}p_r dr\\
&=&2\frac{1}{2\pi}
\int_{r_{-}}^{r_{+}}\sqrt{2E-\frac{\mu^2}{r^2}+\frac{2\alpha}{r}}dr\\
&=&\frac{\sqrt{-2E}}{\pi}\int_{r_{-}}^{r_{+}}
\frac{\sqrt{(r-r_{-})(r_{+}-r)}}{r} dr
\end{eqnarray*}
where the periodic orbits have $r_{-}\leq r\leq r_{+}$ and
\[
r_{\pm}=\frac{-\alpha\pm\sqrt{\alpha^2+2\mu^2E}}{2E}.
\]
The integral can be performed using the residue calculus and 
choosing a contour with a branch cut from $r_-$ to $r_+$ on the real axis
\footnote{The following method is taken from Max Born's {\em The Atom}
published in 1927. I thank Gary Gibbons for pointing out this 
reference to me.}.
\begin{figure}
\caption{Branch cut for the Kepler integral.}
\label{Kepler_integral_fig}
\begin{center}
\includegraphics[width=10cm,height=8cm,angle=0]{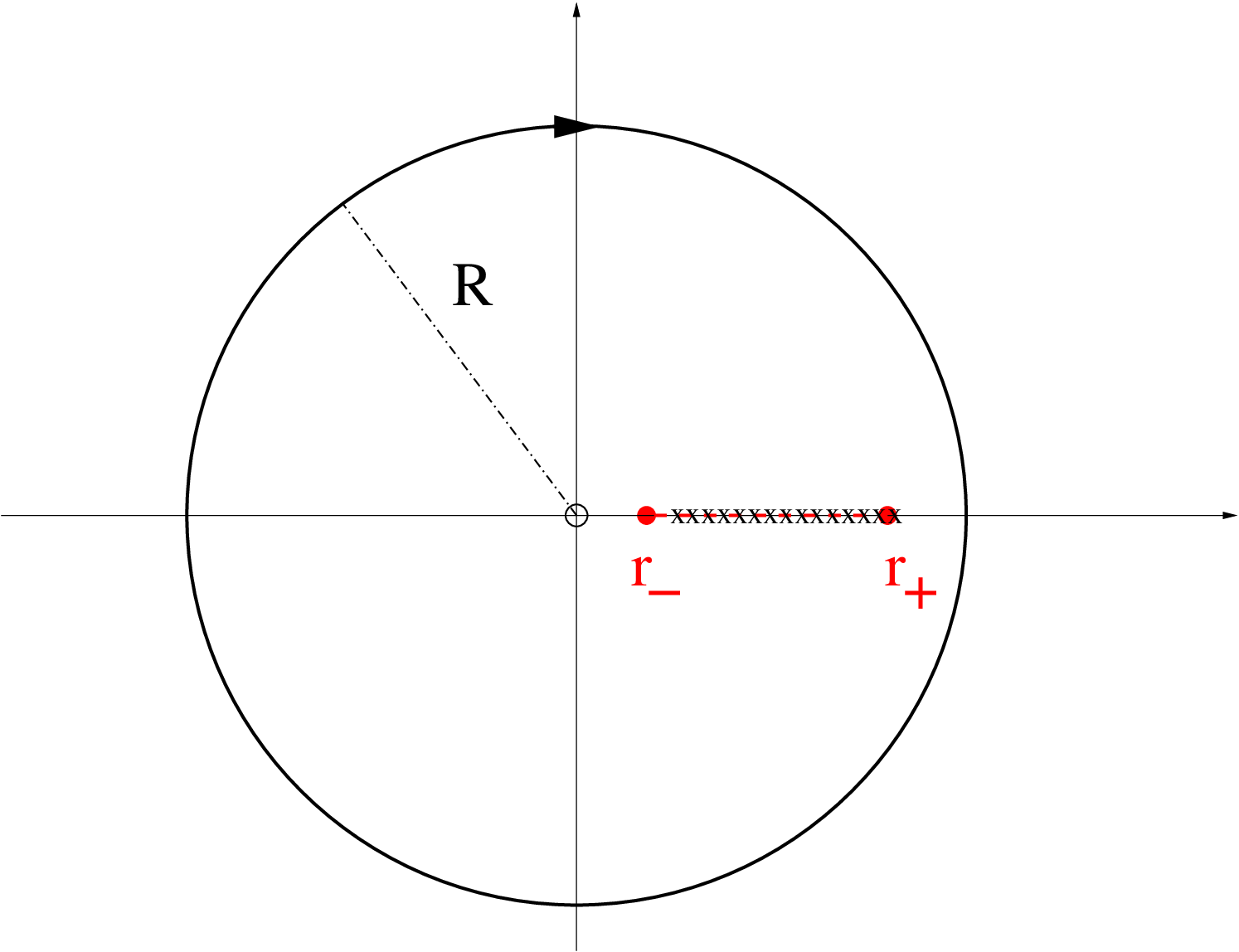}
\end{center}
\end{figure}
Consider a branch of \[f(z)=\sqrt{(z-r_-)(r_+-z)}\] defined by a branch cut
from $r_-$ to $r_+$ with $f(0)=i\sqrt(r_+r_-)$ on the top side of the cut.
We evaluate the integral over a large circular contour $|z|=R$ integrating the 
Laurent expansion
\begin{eqnarray*}
\int_{|z|=R}z^{-1}f(z)dz&=&\int_{0}^{2\pi}\sqrt{-1}\Big(1-\frac{r_-}{R}e^{-i\theta}\Big)^{1/2}
\Big(1-\frac{r_+}{R}e^{-i\theta}\Big)^{1/2}iRe^{i\theta}d \theta\\
&=&\pi(r_+ + r_-) \qquad \mbox{when}\quad R\rightarrow \infty,
\end{eqnarray*}
since all terms containing powers of $\exp{(i\theta)}$ are periodic and 
do not contribute to the integral.
The same value must arise from a residue at $0$ and collapsing the contour
onto the branch cut (when calculating the residue remember that $z=0$ is on
the left hand side of the cut and thus $\sqrt{-1}=-i$. Integration along
the big circle is equivalent to taking a residue at $\infty$ which
is on the right side of the cut where $\sqrt{-1}=i$). Thus
\[
\pi(r_+ +r_-)=2\pi\sqrt{r_+r_-}+\int_{r_{-}}^{r_{+}}
\frac{\sqrt{(r-r_{-})(r_{+}-r)}}{r} dr
-\int_{r_{+}}^{r_{-}}
\frac{\sqrt{(r-r_{-})(r_{+}-r)}}{r} dr
\]
a rational number (here it is equal to 1). The orbits are therefore 
closed - a remarkable result known
lue must arise from a residue at $0$ and collapsing the contour
onto the branch cut (when calculating the residue remember that $z=0$ is on
the left hand side of the cut and thus $\sqrt{-1}=-i$. Integration along
the big circle is equivalent to taking a residue at $\infty$ which
is on the right side of the cut where $\sqrt{-1}=i$). Thus
\[
\pi(r_+ +r_-)=2\pi\sqrt{r_+r_-}+\int_{r_{-}}^{r_{+}}
\frac{\sqrt{(r-r_{-})(r_{+}-r)}}{r} dr
-\int_{r_{+}}^{r_{-}}
\frac{\sqrt{(r-r_{-})(r_{+}-r)}}{r} dr
\]
and
\begin{eqnarray*}
I_r&=&\frac{\sqrt{-2E}}{\pi}\; \frac{\pi}{2}(r_+ +r_- -2\sqrt{r_+r_-})\\
&=&\alpha\sqrt{\frac{1}{2|E|}}-\mu.
\end{eqnarray*}
The Hamiltonian becomes
\[
\widetilde{H}=-\frac{\alpha^2}{2(I_r+I_\phi)^2}
\]
and we conclude that the absolute values of frequencies are equal
and given by
\[
\frac{\p\widetilde{H}}{\p I_r}=\frac{\p\widetilde{H}}{\p I_\phi}
=\frac{\alpha^2}{(I_r+I_\phi)^3}=\Big(\frac{r_++r_-}{2}\Big)^{-3/2}
\sqrt{\alpha}.
\]
This is a particular case when the ratio of two frequencies is
a rational number (here it is equal to 1). The orbits are therefore 
closed - a remarkable result known to Kepler. 
\end{itemize}
\section{Poisson structures}
\label{sec_poiss}
There is a natural way to extend  the Hamiltonian formalism by generalising
the notion of Poisson bracket (\ref{poisson}). 
A geometric approach is given by symplectic
geometry \cite{arnold}. 
We shall take a lower level (but a slightly more general ) point of view
and introduce the Poisson structures. The phase space $M$ is 
$m$ dimensional with local coordinates $(\xi^1, \ldots, \xi^{m})$. In particular
we do not distinguish between the positions and momenta.

\begin{defi}
A skew--symmetric matrix $\omega^{ab}=\omega^{ab}(\xi)$ is called a Poisson
structure if the Poisson bracket defined by
\be
\label{poisson_bracket_poisson}
\{f, g\}=\sum_{a,b=1}^{m}\omega^{ab}(\xi)\frac{\p f}{\p \xi^a}
\frac{\p g}{\p \xi^b}
\ee
satisfies
\[
\{f, g\}=-\{g, f\},
\]
\[
\label{jacobi_sheet}
\{f,\{ g, h\}\}+
\{h,\{ f, g\}\}+\{g,\{ h, f\}\}=0.
\]
\end{defi}
The second property is called the Jacobi identity. It puts restrictions on 
$\omega^{ab}(\xi)$ which can be seen noting that 
\[
\omega^{ab}(\xi)=\{\xi^a, \xi^b\}
\]
and evaluating the Jacobi identity on coordinate functions.

Given a Hamiltonian $H:M\times \R\longrightarrow \R$ the dynamics is
governed by
\[
\frac{d f}{d t}=\frac{\p f}{\p t}+\{ f, H\}
\]
and the Hamilton's equations generalising (\ref{caneq}) become
\be
\label{poisson_ham}
\dot{\xi}^a=\sum_{b=1}^m\omega^{ab}(\xi)\frac{\p H}{\p \xi^b}.
\ee

\begin{itemize}
\item {\bf Example}. Let $M=\R^3$ and 
$\omega^{ab}=\sum_{c=1}^3\varepsilon^{abc}\xi^c$,
where $\varepsilon^{abc}$ is the standard totally antisymmetric tensor.
Thus
\[
\{\xi^1, \xi^2\}=\xi^3, \quad \{\xi^3, \xi^1\}=\xi^2,
\quad \{\xi^2, \xi^3\}=\xi^1.
\]
This Poisson structure admits a Casimir - any function $f(r)$ where
\[r=\sqrt{(\xi^1)^2+(\xi^2)^2+(\xi^3)^2}\] Poisson commutes
with the coordinate functions
\[
\{ f(r), \xi^a\}=0.
\]
This is independent on the choice of the Hamiltonian. With a choice
\[
H=\frac{1}{2}\Big(\frac{(\xi^1)^2}{a_1}+\frac{(\xi^2)^2}{a_2}+
\frac{(\xi^3)^2}{a_3}\Big)
\]
where  $a_1, a_2, a_3$ are constants, the Hamilton's equations 
(\ref{poisson_ham})
become the equations of motion of a rigid body fixed at its centre of 
gravity
\[
\dot{\xi}^1=\frac{a_3-a_2}{a_2a_3}\xi^2\xi^3, 
\qquad \dot{\xi}^2=\frac{a_1-a_3}{a_1a_3}\xi^1\xi^3,  \qquad \dot{\xi}^3=\frac{a_2-a_1}{a_1a_2}\xi^1\xi^2.
\]
\end{itemize}
Assume that $m=2n$ is even and the matrix  $\omega$ is invertible
with $W_{ab}:=(\omega^{-1})_{ab}$. The Jacobi identity implies 
that the antisymmetric matrix
$W_{ab}(\xi)$ is closed, i.e.
\[
\p_{a} W_{bc} +\p_{c} W_{ab}+\p_{b} W_{ca}  =0, 
\qquad \forall a, b, c=1,\dots, m.
\]
In this case $W_{ab}$ is  called a symplectic structure. The Darboux theorem
states that in this case there locally
exists a coordinate system 
\[
\xi^1=q_1,\cdots, \xi^n=q_n, \xi^{n+1}=p_1,\cdots, \xi^{2n}=p_n
\]
such that 
\[
\omega=\left (
\begin{array}{cc}
0&1_n\\
-1_n&0
\end{array}
\right ) 
\]
and the Poisson bracket reduces to the standard form (\ref{poisson}).
A simple proof can be found in \cite{arnold}. One constructs a 
local coordinate system  $(p, q)$ by induction w.r.t half of the dimension 
of $M$. Choose
a function $p_1$, and find $q_1$ by solving the equation $\{q_1, p_1\}=1$.
Then consider a level set of $(p_1, q_1)$ in $M$ which is locally a symplectic 
manifold. Now look for $(p_2, q_2)$ etc.

\begin{itemize}
\item {\bf Example.} The Poisson structure in the last example is 
degenerate as the matrix $\omega^{ab}$ is not invertible. This
degeneracy always occurs if the phase space is odd dimensional or/and
there exists a non-tivial Casimir. Consider the restriction of
$\omega^{ab}=\sum_{c=1}^3\varepsilon^{abc}\xi^c$ to a two-dimensional 
sphere $r=C$. This gives a symplectic structure on the sphere
given by
\[
\{\xi^1, \xi^2\}=\sqrt{C^2-(\xi^1)^2-(\xi^2)^2}
\]
or
\[
W=\frac{1}{\sqrt{C^2-(\xi^1)^2-(\xi^2)^2}}
\left(\begin{array}{cc}
0&1\\
-1&0
\end{array}
\right).
\]
This of course has no Casimir functions apart from constants.
It is convenient to choose a different parametrisation of the sphere: if
\[
\xi^1=C\sin{\theta}\cos{\phi} , \quad \xi^2=C\sin{\theta}\sin{\phi}, \quad
\xi^3=C \cos{\theta}
\]
then in the local coordinates $(\theta, \phi)$ 
the symplectic structure is given by $\{\theta, \phi\}=\sin^{-1}{\theta}$ or
\[
W=\sin{\theta}
\left(\begin{array}{cc}
0&1\\
-1&0
\end{array}
\right)
\]
which is equal to the volume form on the two--sphere.
The radius $C$ is arbitrary. Therefore the Poisson phase
space $\R^3$ is foliated by symplectic phase spaces $S^2$ as there
is exactly one sphere centred at the origin through any point
of $\R^3$. This is a general phenomenon: fixing the values of the
Casimir functions on Poisson spaces gives the foliations by
symplectic spaces. The local Darboux coordinates on $S^2$ 
are given by $q=-\cos{\theta}, p
=\phi$ as then
\[
\{q, p\}=1. 
\]
\end{itemize}

The Poisson generalisation is useful to set up the Hamiltonian
formalism in the infinite--dimensional case. Formally
one can think of replacing the coordinates
on the trajectory $\xi^a(t)$ by a dynamical variable
$u(x, t)$. Thus the discrete index $a$ becomes the continuous independent
variable $x$  (think of $m$ points on a string versus
the whole string). The phase space $M=\R^{m}$ is replaced by
a space of smooth functions on a line with appropriate
boundary conditions (decay or periodic).
The whole formalism may be set up
making the following replacements 
\begin{eqnarray*}
\mbox{ODEs}&\longrightarrow&\mbox{PDEs}\\
\xi^a(t), a=1, \dots, m&\longrightarrow& u(x, t), x\in \R\\
\sum_a&\longrightarrow& \int_{\R} dx\\
\mbox{function}\,\,f(\xi)&\longrightarrow& \mbox{functional}\,\, F[u]\\
\frac{\p}{\p \xi^a}&\longrightarrow&\frac{\delta}{\delta u}.
\end{eqnarray*}
The functionals are given by integrals
\[
F[u]=\int_\R f(u, u_x, u_{xx}, \ldots ) dx
\]
(we could in principle allow the $t$ derivatives but we will not
for the reasons to become clear shortly). 
Recall that the functional derivative is
\[
\frac{\delta F}{\delta u(x)}=\frac{\p f}{\p u}-\frac{\p}{\p x}
\frac{\p f}{\p (u_x)}+\Big(\frac{\p}{\p x}\Big)^2\frac{\p f}{\p (u_{xx})}+\dots
\]
and
\[
\frac{\delta u(y)}{\delta u(x)}=\delta(y-x)
\]
where the $\delta$ on the RHS is the Dirac delta which satisfies
\[
\int_{\R}\delta(x)dx=1,\qquad \delta(x)=0 \,\,\mbox{for}\,\,x\neq 0.
\]
The presence of the Dirac delta will constantly remind us that
we have entered a territory which is rather slippery from a pure
mathematics perspective. We should rest reassured that the formal replacements
made above can nevertheless be given a solid functional-analytic foundation.
This will not be done in this course.

The analogy with finite dimensional situation 
(\ref{poisson_bracket_poisson}) suggests a following
definition of a Poisson bracket
\[
\{ F, G\}=\int_{\R^2}\omega(x, y, u)\frac{\delta F}{\delta u(x)}
\frac{\delta G}{\delta u(y)}dxdy
\]
where the Poisson structure $\omega(x, y, u)$ should be such that
the bracket is anti--symmetric and the Jacobi identity holds.
A canonical (but not the only) choice is
\[
\omega(x, y, u)=\frac{1}{2}\frac{\p}{\p x}\delta(x-y)-\frac{1}{2}\frac{\p}{\p y}\delta(x-y).
\]
This is analogous to the Darboux form in which $\omega^{ab}$
is a constant and antisymmetric matrix and the Poisson bracket
reduces to (\ref{poisson}). This is because
the differentiation operator $\p/\p x$ is  anti--self--adjoint
with respect to an inner product
\[
<u, v>=\int_\R u(x)v(x)dx
\]
which is analogous to a matrix being anti--symmetric.
With this choice
\be
\label{standard_bracket}
\{ F, G\}=\int_{\R}\frac{\delta F}{\delta u(x)}\frac{\p}{\p x}
\frac{\delta G}{\delta u(x)}dx
\ee
and the Hamilton's equations become
\begin{eqnarray}
\label{can_PDEs}
\frac{\p  u}{\p t}&=&\{u, H[u]\}=\int_{\R} 
\frac{\delta u(x)}{\delta u(y)} \frac{\p}{\p y}
\frac{\delta H}{\delta u(y)}dy\nonumber\\
&=& \frac{\p }{\p x}\frac{\delta H[u]}{\delta u(x)}.
\end{eqnarray}
\begin{itemize}
\item {\bf Example.} The KdV equation mentioned earlier is a Hamiltonian
system with the Hamiltonian given by the functional
\[
H[u]=\int_{\R}\Big(\frac{1}{2}u_x^2+u^3\Big)dx.
\]
It is assumed that $u$ belongs to the space of functions decaying 
sufficiently fast at when $x\rightarrow \pm \infty$.
\end{itemize}

\chapter{Soliton equations and Inverse Scattering Transform}
\label{chapter_sol}
The universally accepted definition of integrability does not exist
for partial differential equations. 
The phase space is infinite dimensional but having `infinitely many'
first integrals may not be enough - we could have missed every second one.
One instead focuses on properties of solutions and solutions generation 
techniques. We shall study solitons - 
solitary non-linear waves which preserve their shape (and other 
characteristics) in the evolution. These soliton solutions will be constructed
by means of an inverse problem: recovering 
a potential from the scattering data.

\section{History of two examples}
Soliton equations originate in the $19$th century. Some of them appeared
in the study of non-linear wave phenomena and other arose in differential
geometry of surfaces in $\R^3$
\begin{itemize}
\item The KdV equation
\be 
\label{kdv}
u_t-6uu_x+u_{xxx}=0, \qquad\mbox{where}\quad u=u(x,t)
\ee
has been written down, and solved in the simplest case,  
by Korteweg and de-Vires in 1895 to explain the following account of
J. Scott Russell. Russell  observed a soliton while ridding on horseback
beside a narrow barge channel. The following passage has been taken from
J . Scott Russell. Report on waves, 
Fourteenth meeting of the British Association for the 
Advancement of Science, 1844.
{\it`I was observing the motion of a boat which was rapidly drawn 
along a narrow channel by a pair of horses, when the boat suddenly stopped -  
not so the mass of water in the channel which 
it had put in motion; it accumulated  round the prow of the vessel 
in a state of violent agitation, then suddenly leaving  it behind, 
rolled forward with great velocity, assuming the form of a large  
solitary elevation, a rounded, smooth and well-defined heap of water, 
which continued its course along the channel apparently without 
change of form or diminution of speed. 
I followed it on horseback, and overtook it 
still rolling on at a rate of some 
eight or nine miles an hour, preserving 
its original figure some thirty feet long and 
a foot to a foot and a half in height. Its height gradually 
diminished, and after a chase of one or two miles I lost 
it in the windings of the channel. Such, in 
the month of August 1834, was my first chance interview 
with that singular and beautiful phenomenon which I have called 
the Wave of Translation'.}
\item
The Sine--Gordon equation 
\be
\label{sine_gordon}
\phi_{xx}-\phi_{tt}=\sin{\phi} \qquad\mbox{where}\quad \phi=\phi(x,t)
\ee
locally describes the isometric embeddings of surfaces with constant
negative Gaussian curvature in the Euclidean space $\R^3$. The function
$\phi=\phi(x, t)$ is the angle between two asymptotic directions
$\tau=(x+t)/2$ and $\rho=(x-t)/2$ on the surface 
along which  the second 
fundamental form is zero. 
If the first fundamental form of a surface parametrised by $(\rho, \tau)$ 
is 
\[
ds^2=d\tau^2+2\cos{\phi}\; d\rho d\tau+d\rho^2,\qquad
\mbox{where}\quad \phi=\phi(\tau, \rho)
\]
then the Gaussian curvature is constant and equal to $-1$ provided that
\[
\phi_{\tau\rho}=\sin{\phi}.
\]
which is (\ref{sine_gordon}).

The integrability of the Sine--Gordon equation 
have been used by Bianchi, B\"acklund, Lie and other
classical differential geometers to construct new embeddings.
\end{itemize}
\subsection{Physical derivation of KdV}
Consider the linear wave equation
\[
\Psi_{xx}-\frac{1}{v^2}\Psi_{tt}=0
\]
where $\Psi_{xx}=\p^2_x\Psi$ etc. which  describes a propagation
of waves travelling with a constant velocity $v$. Its derivation is 
based on three simplifying assumptions:
\begin{itemize}
\item
There is no dissipation i.e.
the equation is invariant with respect to time inversion
$t\rightarrow -t$. 
\item
The amplitude of oscillation is small and so the nonlinear
terms (like $\Psi^2$) can be omitted.
\item There is no dispersion, i.e. the group velocity is  constant.
\end{itemize}
In the derivation of the KdV we follow 
\cite{NMPZ} and  relax these assumptions.

The  general solution of the wave equation is a superposition
of two waves travelling in opposite directions
\[
\Psi=f(x-vt)+g(x+vt)
\]
where $f$ and $g$ are arbitrary  functions of one variable.
Each of these two waves is characterised by  a linear 1st order PDE, e.g.
\[
\Psi_{x}+\frac{1}{v}\Psi_t=0\qquad\longrightarrow\quad \Psi=f(x-vt).
\]
To introduce the dispersion consider a complex wave
\[
\Psi=e^{i(kx-\omega(k) t)}
\]
where $\omega(k)=vk$ and so the group velocity $d\omega/d k$ 
equals to the phase velocity $v$.
We change this relation by introducing the dispersion 
\[
\omega(k)=v(k-\beta k^3+\dots)
\]
where the absence of even terms in this expansion guarantees real dispersion 
relations. Let us assume that the dispersion is small and truncate this series
keeping only the first two terms. The equation satisfied by
\[
\Psi=e^{i(kx-v(kt-\beta k^3t))}
\]
is readily found to be 
\[
\Psi_x+\beta\Psi_{xxx}+\frac{1}{v}\Psi_t=0.
\]
This can be rewritten in a form of a conservation law
\[
\rho_t+j_x=0,
\]
where the density $\rho$ and the current $j$ are given by
\[
\rho=\frac{1}{v}\Psi, \qquad j=\Psi+\beta\Psi_{xx}.
\]
To introduce nonlinearity modify the current 
\[
j=\Psi+\beta\Psi_{xx}+\frac{\alpha}{2}\Psi^2.
\]
The resulting equation is
\[
\frac{1}{v}\Psi_t+\Psi_x+\beta\Psi_{xxx}+\alpha\Psi\Psi_x=0.
\]
The non--zero constants $(v, \beta, \alpha)$ can be eliminated by a simple change of variables $x\rightarrow x-vt$ and rescaling $\Psi$. This leads
to the standard form of the KdV equation
\[
u_t-6uu_x+u_{xxx}=0.                           
\]
The simplest 1--soliton solution found by Korteweg and de-Vires
is
\be
\label{1-soliton}
u(x, t)=-\frac{2\chi^2}{\cosh^2{\chi(x-4\chi^2 t-\phi_0)}}.
\ee
The KdV is not a linear equation therefore multiplying this
solution by a constant will not give another solution.
The constant $\phi_0$ determines the location of
the extremum at $t=0$. We should therefore think of a one--parameter
family of solutions labelled by $\chi\in \R$.

The one--soliton (\ref{1-soliton}) was the only regular solution of KdV
such that $u, u_x\rightarrow 0$ as $|x|\rightarrow\infty$
known until 1965 when Gardner, Green, Kruskal and Miura analysed KdV 
numerically. They took two waves with different amplitudes  as their initial 
profile. The computer simulations revealed that 
the initially separated waves approached each-other distorting their shapes,
but eventually the larger wave overtook the smaller wave and both
waves re-emerged with their sizes and shapes intact. The relative
phase shift was the only result of the non--linear interaction.
This behaviour resembles what we usually associate with particles and
not waves. Thus Zabruski and  Kruskal named these waves `solitons'
(like electrons, protons, barions and other particles ending with `ons').
In this Chapter we shall construct more general
$N$--soliton solutions describing the interactions of 1--solitons.

To this end we note that the existence of a stable 
solitary wave is a consequence
of cancellations of effects caused by non--linearity and dispersion.
\begin{itemize}
\item
If the dispersive term
were not present the equation would be
\[
u_t-6uu_x=0
\]
and the resulting solution would exhibit a discontinuity
of first derivatives at some $t_0>0$  (shock, or `breaking the wave').
This solution can be easily found using the method of characteristics.

\begin{center}
\includegraphics[width=10cm,height=5cm,angle=0]{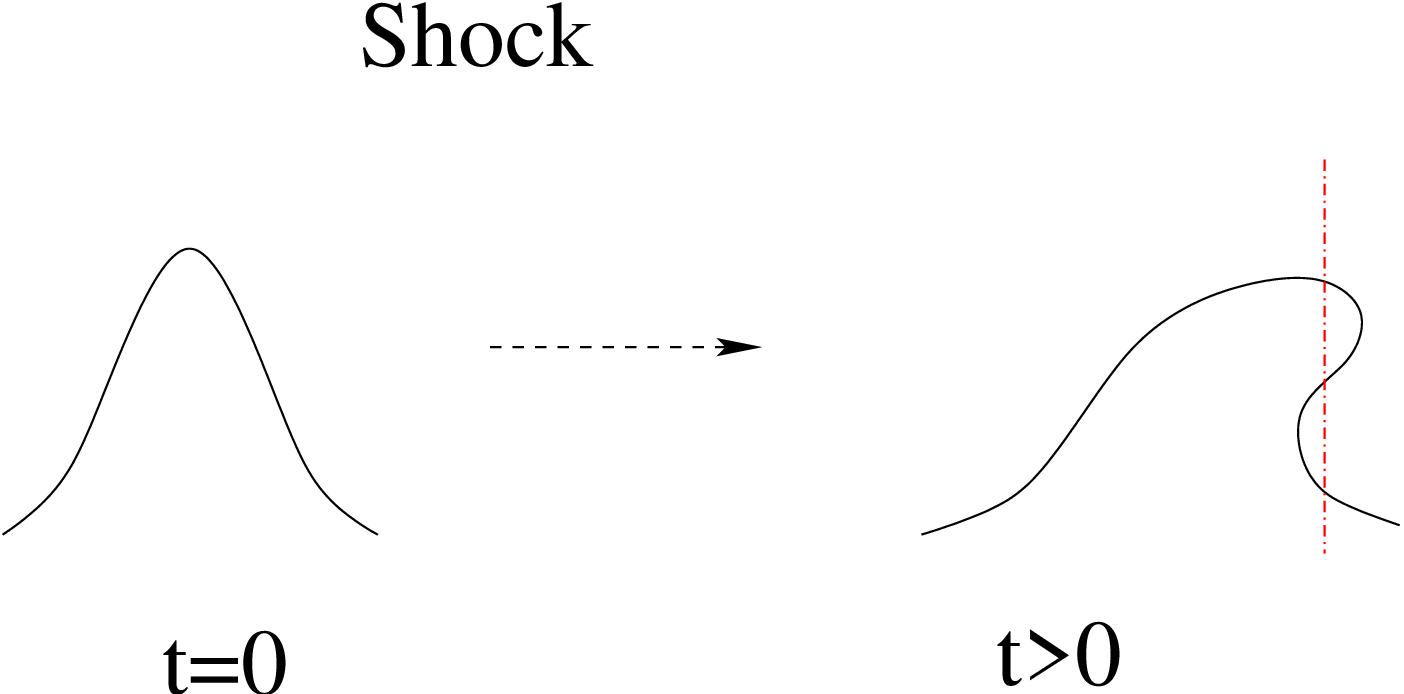}
\end{center}

\item If the nonlinear term were not present the initial wave profile would
disperse in the evolution $u_t+u_{xxx}=0$.
\begin{center}
\includegraphics[width=10cm,height=5cm,angle=0]{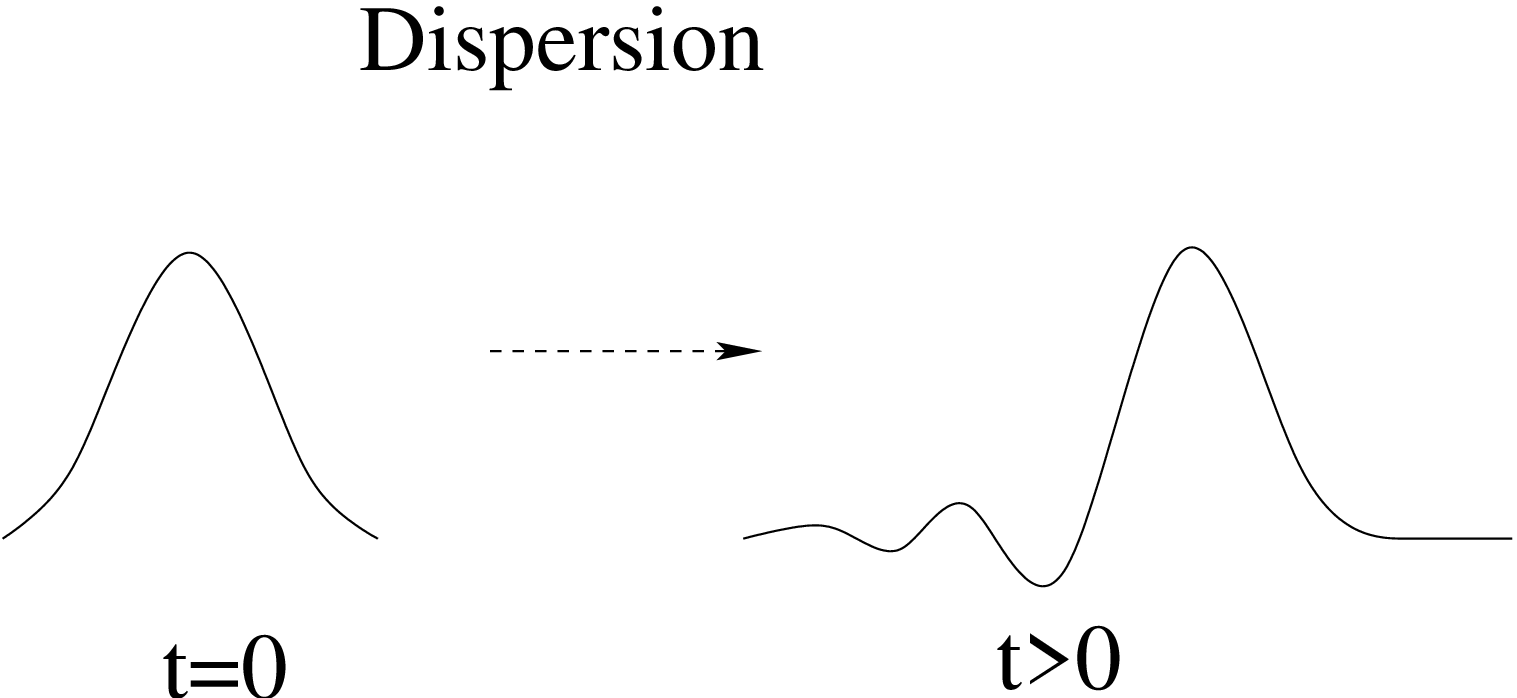}
\end{center}
\item
The presence of both terms allows smooth localised soliton solutions
\begin{center}
\includegraphics[width=10cm,height=5cm,angle=0]{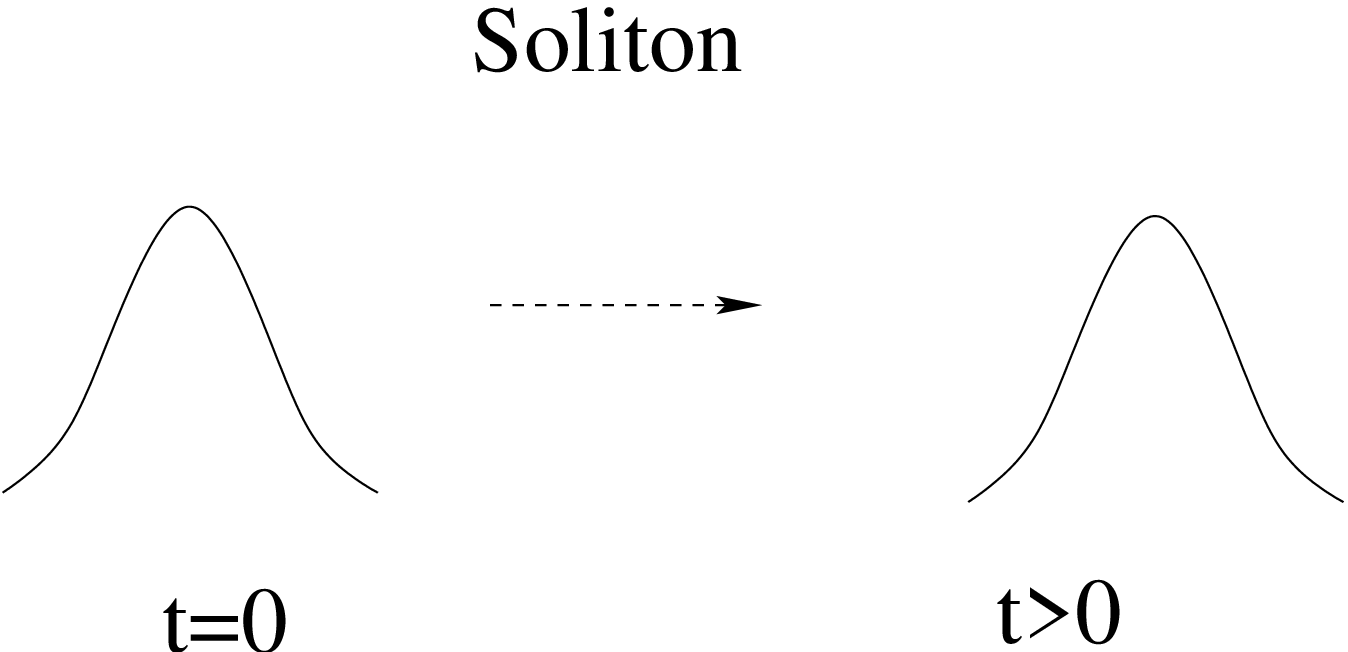}
\end{center}
of which (\ref{1-soliton}) is an example (the plot gives $-u(x, t)$).
\end{itemize}
\subsection{B\"acklund transformations for the Sine--Gordon equation}
\label{sec_backl}
\begin{figure}
\caption{Sine--Gordon Kink}
\label{IS_kink}
\begin{center}
\includegraphics[width=10cm,height=5cm,angle=0]{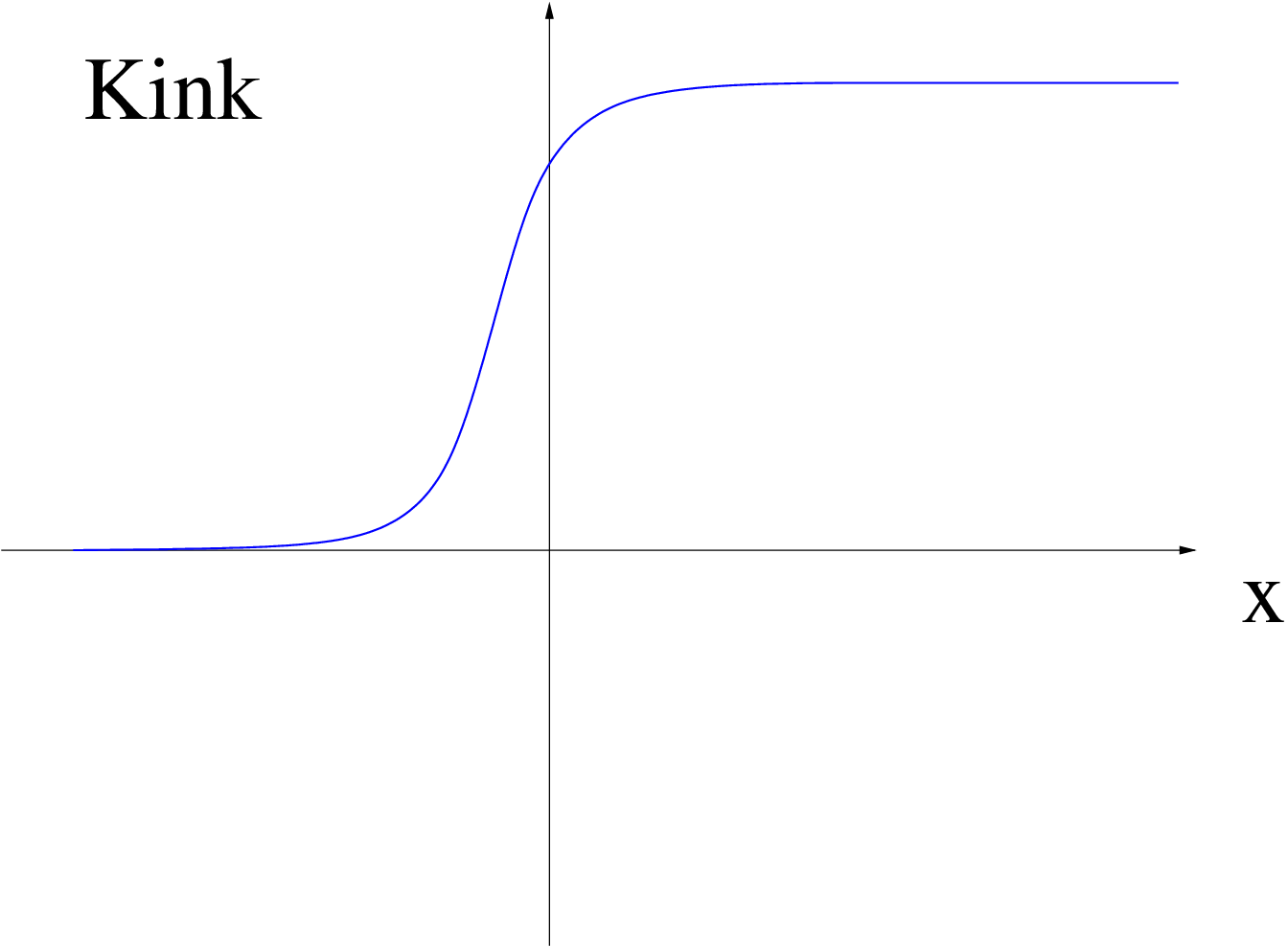}
\end{center}
\end{figure}

Let us consider the Sine--Gordon equation - the 
other soliton equation mentioned in the introduction
to this Chapter. 
The simplest solution generating technique is
the B\"acklund transformation. Set   
$\tau=(x+t)/2, \rho=(x-t)/2$ so that the equation (\ref{sine_gordon})  becomes
\[
\phi_{\tau\rho}=\sin{\phi}.
\]
Now define the B\"acklund relations
\[
\p_\rho(\phi_1-\phi_0)=2b\sin{\Big(\frac{\phi_1+\phi_0}{2}\Big)}, \qquad
\p_\tau(\phi_1+\phi_0)=2b^{-1}\sin{\Big(\frac{\phi_1-\phi_0}{2}\Big)}, \qquad 
b=\mbox{const}.
\]
Differentiating the first equation w.r.t $\tau$, and using the second
equation yields
\begin{eqnarray*}
\p_{\tau}\p_\rho(\phi_1-\phi_0)&=&2b \;\p_{\tau}
\sin{\Big(\frac{\phi_1+\phi_0}{2}\Big)}=
2 \sin{\Big(\frac{\phi_1-\phi_0}{2}\Big)}
\cos{\Big(\frac{\phi_1+\phi_0}{2}\Big)}\\
&=&
\sin{\phi_1}-\sin{\phi_0}.
\end{eqnarray*}
Therefore $\phi_1$ is a solution to the Sine--Gordon equation if $\phi_0$ is.
Given $\phi_0$ we can solve the first order B\"acklund  relations for $\phi_1$
and generate new solutions form the ones we know. The trivial solution 
$\phi_0=0$ yields 1--soliton solution of  Sine--Gordon 
\[
\phi_1(x, t)=4
\arctan{\Big(\exp{\Big(\frac{x-vt}{\sqrt{1-v^2}}-x_0\Big)}\Big)}
\]
where $v$ is a constant with $|v|<1$.
This solution is called a kink (Figure \ref{IS_kink}).
A static kink corresponds to a special case $v=0$.
\vskip10pt
One can  associate a topological charge 
\[
N=\frac{1}{2\pi}\int_{\R}d\phi=\frac{1}{2\pi}\Big(\phi(x=\infty, t)
-\phi(x=-\infty, t)\Big)
\]
with any solution of the Sine 
Gordon equation. It is an integral of 
a total derivative which depends only on boundary conditions. 
It is conserved if one insists on finiteness of the energy
\[
E=\int_{\R}\Big(\frac{1}{2}\Big(\phi_t^2+\phi_x^2\Big)+(1-\cos{(\phi)})\Big)dx.
\]
Note that the Sine--Gordon  
equations didn't enter the discussion at this stage.
Topological charges, like $N$, are in this sense 
different form first integrals like $E$ which  satisfy
$\dot{E}=0$ as a consequence of (\ref{sine_gordon}).  
For the given kink solution $N(\phi)=1$ and the  kink is stable as it 
would take infinite energy to change this solution into a constant 
solution $\phi=0$ with $E=0$.

There exist interesting solutions 
with $N=0$: a soliton--antisoliton 
pair  has $N=0$ but is non--trivial
\[
\phi(x, t)=
4\arctan\Big(
\frac
{v\cosh{\frac{x}{\sqrt{1-v^2}}}}
{\sinh{\frac{vt}{\sqrt{1-v^2}}}} 
\Big).
\]
At $t\rightarrow -\infty$ this solution represents widely separated pair of kink and
anti--kink approaching each-other with velocity $v$. A non--linear interaction takes place at $t=0$ and as $t\rightarrow \infty$ kink and anti-kink reemerge
unchanged.

\section{Inverse scattering transform for KdV}
One of the most spectacular methods of solving soliton equations
comes from quantum mechanics. It is quite remarkable, as the soliton
equations we have discussed so far have little to do with the quantum world.

Recall that the  mathematical arena of quantum mechanics  is the
infinite--dimensional complex vector space ${\cal H}$ of functions 
\cite{schieff_book}.  
Elements $\Psi$ 
of this space are referred to as wave functions, or state vectors.
In case
of one--dimensional quantum mechanics we have $\Psi:\R\rightarrow \C$,
$\Psi=\Psi(x)\in\C$. The space 
${\cal H}$ is equipped with a unitary inner product
\be
\label{L_2_inner}
(\Psi, \Phi)=\int_{\R}\overline{\Psi(x)}\Phi(x)dx.
\ee
The functions which are square integrable, i.e. $(\Psi, \Psi)<\infty$
like $\Psi=e^{-x^2}$, are called bound states. 
Other functions, like $e^{-ix}$, are called the scattering states.

Given a real valued function $u=u(x)$ called the potential,
the time independent
Schr\"odinger equation
\[
-\frac{\hbar^2}{2m}\frac{d^2\Psi}{dx^2}+u\Psi=E\Psi
\]
determines the $x$--dependence  of a wave function. 
Here $\hbar$ and $m$ are constants which we shall not worry about 
and $E$ is the energy
of the quantum system. The energy levels  can be discrete for bound
states or continuous for scattering states. This depends
on the potential $u(x)$.  We shall regard the Schr\"odinger equation as
an eigen--value problem and refer to $\Psi$ and $E$ as  eigenvector
and eigenvalue respectively. 

According to the Copenhagen interpretation of quantum mechanics the
probability density for the position  of a quantum  particle 
is given by $|\Psi|^2$, where $\Psi$ is a solution to the 
the Schr\"odinger  equation.  The time evolution of the
wave function is governed by a time dependent Schr\"odinger equation
\[
i\hbar\frac{\p \Psi}{\p t}=-\frac{\hbar^2}{2m}\frac{\p^2\Psi}{\p x^2}+u\Psi.
\]
This equation implies that for bound states 
the quantum--mechanical probability is  conserved in a sense
that
\[
\frac{d}{dt}\int_{\R}|\Psi|^2dx=0.
\]

The way physicists discover 
new elementary particles is by scattering
experiments. Huge accelerators collide particles through targets and,
by analysing the changes to momenta of scattered particles, a picture
of a target is built\footnote{These kind of experiments
will take place in the Large Hadron Collider LHC 
opened in September 2008 at CERN.
The LHC is located in a $27$km  long tunnel under 
the Swiss/French border outside Geneva. It is hoped that
the elusive Higgs particle and a whole bunch of other exotic form of matter
will be discovered.}. Given a potential $u(x)$ one can use the Schr\"odinger equation to find $\Psi$,
the associated energy levels and
the scattering data in
the form of so called reflection and transmission coefficients. 
Experimental needs are however different: the scattering data is measured
in the accelerator but the potential (which gives the internal structure
of the target) needs to be recovered. This comes down to the following 
mathematical problem
\begin{itemize}
\item Recover the potential from the scattering data.
\end{itemize}
This problem was solved in the 1950s
by the Gelfand, Levitan and Marchenko \cite{GLM_paper,Marchenko_ref}
who gave a linear algorithm for reconstructing $u(x)$.
Gardner, Green, Kruskal and Miura \cite{GGKM_paper} used this 
algorithm to solve
the Cauchy problem for the KdV equation. Their remarkable idea
was to regard the initial data in the solution of KdV as 
a potential in the Schr\"odinger equation.

Set $\hbar^2/(2m)=1$ and write the 1-dimensional Schr\"odinger equation as an
eigenvalue problem
\[
\Big(-\frac{d^2}{d x^2}+u(x)\Big)\Psi=E\Psi.
\]
We allow $u$ to depend on $x$ as well as $t$ which at this stage should be 
regarded as a parameter.

In the scattering theory one considers the beam of free 
particles incident from $+\infty$. Some of the particles will be reflected
by the potential (which is assumed to decay sufficiently fast as 
$|x|\rightarrow \infty$)
and some will be transmitted. There may also be a number of bound states with
discrete energy levels.
\begin{figure}
\caption{Reflection and Transmission  }
\label{IS_reflection}
\begin{center}
\includegraphics[width=10cm,height=6cm,angle=0]{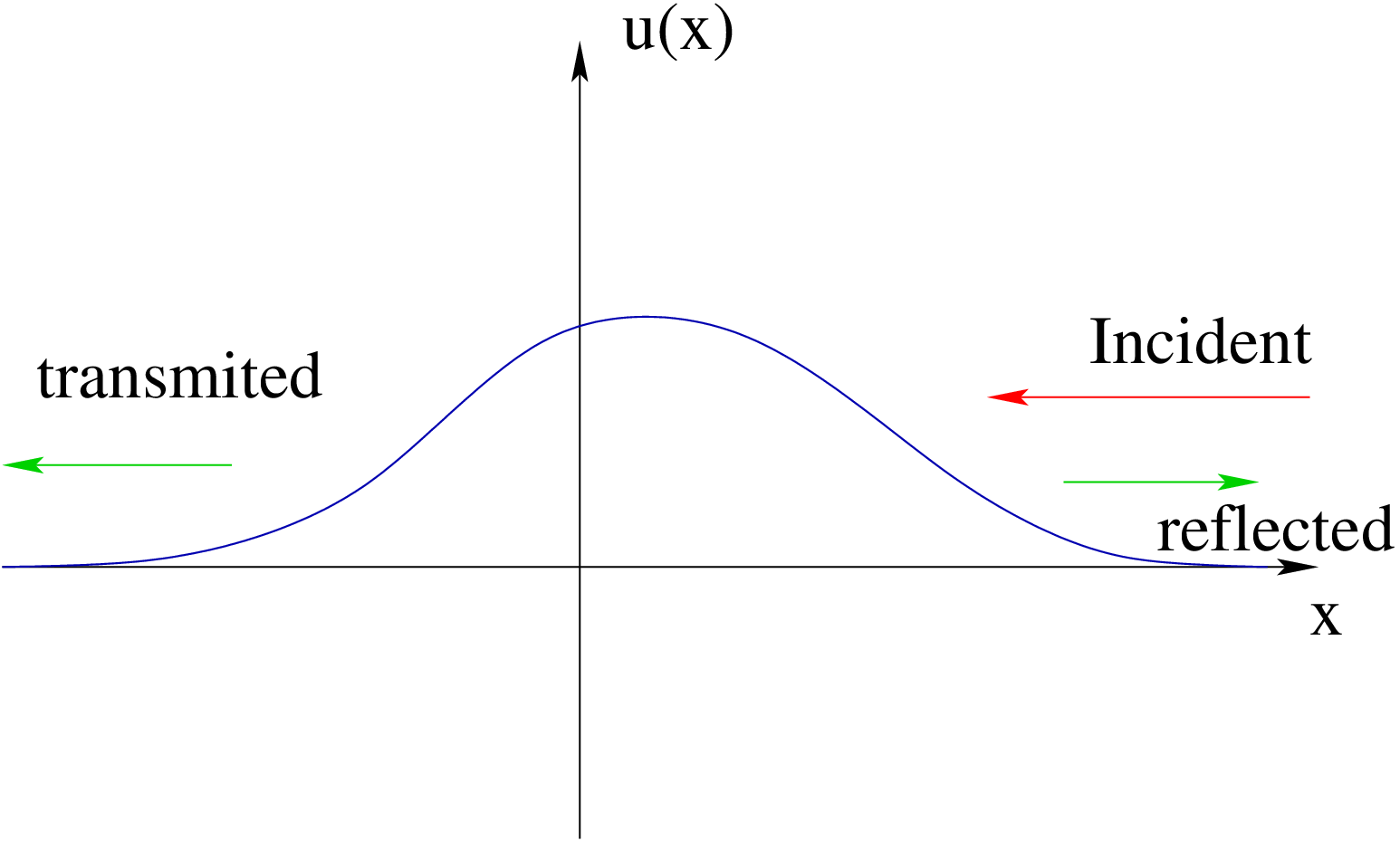}
\end{center}
\end{figure}
The Gelfand--Levitan--Marchenko theory shows that given
\begin{itemize}
\item energy levels $E$,
\item transmission probability $T$,
\item reflection probability $R$,
\end{itemize}
one can find the potential $u$: Given $u_0(x)$ one finds the scattering 
data at $t=0$. 
If $u(x, t)$ is a solution to the KdV equation (\ref{kdv}) 
with $u(x, 0)=u_0(x)$
then the scattering 
data $(E(t), T(t), R(t))$ satisfies simple linear ODEs determining 
their time evolution. In particular $E$ does not depend on $t$.
 Once this has been determined, $u(x, t)$ is recovered
by solving a linear integral equation.
The Gardner, Green, Kruskal and Miura scheme for
solving KdV is summarised in the following table
\begin{center}
\includegraphics[width=10cm,height=6cm,angle=0]{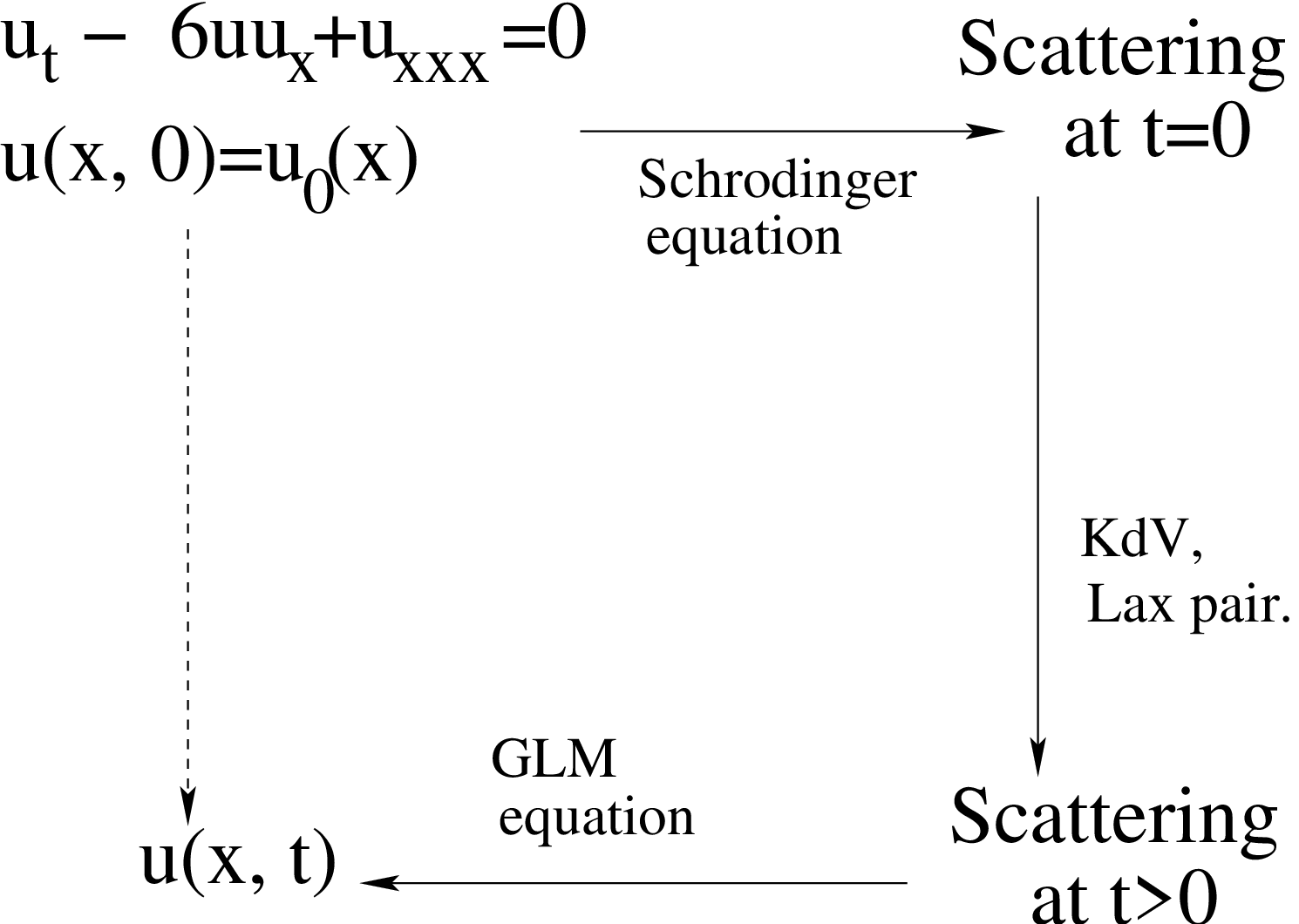}
\end{center}
We should stress that in this method the time evolution 
of the scattering data is governed by the KdV and not by 
the time dependent Schr\"dinger equation. In fact the time dependent
Schro\"dinger equation will not play any role in the following discussion.

\subsection{Direct scattering}
The following discussion summarises the basic one--dimensional quantum mechanics of a particle scattering on a potential \cite{schieff_book,NMPZ}.
\begin{itemize}
\item Set $E=k^2$ and rewrite the Schr\"odinger equation as
\be
\label{Sch}
Lf:=\Big(-\frac{d^2}{d x^2}+u(x)\Big)f=k^2f
\ee
where $L$ is called the  Schr\"odinger operator.
Consider the class of potentials $u(x)$ such that
\[
\int_{\R}(1+|x|) |u(x)|dx<\infty
\]
which  of course implies that $|u(x)|\rightarrow 0$ as 
$x\rightarrow \pm\infty$. This condition guaranties  that there exists
only  a finite number of discrete energy levels (thus it rules
out both the harmonic oscillator and the hydrogen atom).
\item At $x\rightarrow \pm\infty$ the problem 
(\ref{Sch})
reduces to a `free particle'
\[
f_{xx}+k^2f=0
\]
with the general solution
\[
f=C_1 e^{ikx}+C_2e^{-ikx}.
\]
The pair of constants $(C_1, C_2)$ is in general  different
at $+\infty$ and $-\infty$.
\item
For each $k\neq 0$ the set of solutions to (\ref{Sch}) forms a 
2--dimensional complex vector space $G_k$. The reality of $u(x)$ 
implies that if $f$ satisfies  (\ref{Sch}) then so does $\overline{f}$.

Consider two bases $(\psi, \opsi)$ and
$(\phi, \ophi)$ of $G_k$ determined by the asymptotic
\[
\psi(x, k)\cong e^{-ikx}, \quad \opsi(x, k)\cong e^{ikx}
\qquad{\mbox{as}}\quad x\longrightarrow \infty
\]
and
\[
\phi(x, k)\cong e^{-ikx}, \quad \ophi(x, k)\cong e^{ikx}
\qquad{\mbox{as}}\,\, x\longrightarrow -\infty.
\]
Any solution can be expanded in the first basis, so in particular
\[
\phi(x, k)=a(k)\psi(x, k)+b(k)\opsi(x, k).
\]
Therefore, if $a\neq 0$, we can write
\be
\label{phi_x}
\frac{\phi(x, k)}{a(k)}=
\left\{
\begin{array}{ll}
&\frac{e^{-ikx}}{a(k)},\qquad\mbox{for}\;x\rightarrow-\infty\\
&e^{-ikx}+\frac{b(k)}{a(k)}e^{ikx},\qquad\mbox{for}\;x\rightarrow\infty.
\end{array}
\right.
\ee
\item 
Consider a particle incident from $\infty$ with the wave function $e^{-ikx}$
(Figure \ref{IS_reflection}).
The transmission coefficient $t(k)$ and the reflection coefficient
$r(k)$ are given by
\[
t(k)=\frac{1}{a(k)}, \quad r(k)=\frac{b(k)}{a(k)}.
\]
They satisfy 
\be
\label{ttrr_1}
|t(k)|^2+|r(k)|^2=1
\ee
which is intuitively clear as the particle is `either reflected or transmitted'.
To prove it recall that given the  Wronskian
\[
W(f, g)=fg_x-gf_x
\]
of any two functions we have
\[
W_x=fg_{xx}-gf_{xx}=0
\]
if $f, g$ both satisfy the Schr\"odinger equation (\ref{Sch}). Thus
$W(\phi, \ophi)$ is a constant which can be calculated for
$x\rightarrow -\infty$ 
\[
W(\phi, \ophi)=e^{-ikx}(e^{ikx})_x-e^{ikx}(e^{-ikx})_x=2ik.
\]
Analogous calculation at $x\rightarrow \infty$  gives $W(\psi, \opsi)=2ik$. On the other hand
\begin{eqnarray*}
W(\phi, \ophi)&=&W(a\psi+b\opsi, \overline{a}\opsi+\overline{b}\psi)\\
&=&|a|^2W(\psi, \opsi)+a\overline{b}W(\psi, \psi)
+b\overline{a}W(\opsi, \opsi)-
|b|^2W(\psi, \opsi)\\
&=& 2ik(|a|^2-|b|^2).
\end{eqnarray*}
Thus 
$
|a(k)|^2-|b(k)|^2=1
$
or equivalently (\ref{ttrr_1}) holds.
\end{itemize}
\subsection{Properties of the scattering data}

There is a connection between scattering states and bound states. 
Assume that $k\in\C$, and by analytic continuation allow $a(k)$ and $b(k)$ to be complex valued. Set $k=i\chi$ for $\chi>0$. The a scattering state has
\be
\phi\sim
\left\{
\begin{array}{ll}
&a(i\chi) e^{\chi x}+b(i\chi) e^{-\chi x}
\qquad\mbox{for}\;x\rightarrow \infty\\
&e^{\chi x},\qquad\mbox{for}\;x\rightarrow-\infty
\end{array}
\right.
\ee
which yields a bound state with energy $E=-\chi^2$ if $a(i\chi)=0$. Therefore
the bound state energies correspond to zeroes of $a$ on the positive imaginary
axis.

In scattering theory (see e.g.\cite{NMPZ}) one proves
the following
\begin{itemize}
\item $a(k)$ is holomorphic in the upper half plane
$Im(k)>0$.
\item
$
\{Im(k)\geq 0, |k|\rightarrow \infty\}\quad\longrightarrow\quad
|a(k)|\rightarrow 1.
$
\item
Zeroes of $a(k)$ in the upper half plane lie on the imaginary axis.
The number of these zeroes is finite if
\[
\int_{\R}(1+|x|)|u(x)|<\infty.
\]
Thus $a(i\chi_1)=\cdots =a(i\chi_N)=0$ where $\chi_n\in\R$ can be ordered as
\[
\chi_1>\chi_2>\cdots>\chi_N>0.
\]
\item Consider the asymptotics of $\phi$ at these zeroes.
Formula (\ref{phi_x}) gives
\[
{\phi}(x, i\chi_n)=
\left\{ \begin{array}{ll}
&e^{-i(i\chi_n)x},\qquad\mbox{for}\;x\rightarrow-\infty\\
&a(i\chi_n)e^{-i(i\chi_n)x}+{b(i\chi_n)}e^{i(i\chi_n)},
\qquad\mbox{for}\;x\rightarrow\infty.
\end{array}
\right.
\]
Thus
\be
\label{new_formula_1}
{\phi}(x, i\chi_n)=\left\{ \begin{array}{ll}
&e^{\chi_nx},\qquad\mbox{for}\;x\rightarrow-\infty\\
&b_ne^{-\chi_nx},
\qquad\mbox{for}\;x\rightarrow\infty.
\end{array}
\right .
\ee
where  $b_n=b(i\chi_n)$. We should stress that this considerations apply to the discrete part of the spectrum so $b_n$ should be regarded as independent 
from  $b$ which appears in the reflection coefficient. One can show that
$b_n\in\R$ and that it satisfies
\[
b_n=(-1)^n|b_n|
\]
and $ia'(i\chi_n)$ has the same sign as $b_n$.

Moreover
\[
\Big(-\frac{d^2}{dx^2}+u(x)\Big)\phi(x, i\chi_n)=-\chi_n^2\phi(x, i\chi_n)
\]
so $\phi$ is square integrable with energy $E=-\chi_n^2$.
\end{itemize}
Let us consider a couple of examples
\begin{itemize}
\item{\bf Example.}
Consider $L_0=-{\p_x}^2+\chi^2$ which shifts the energy levels of $u=0$ by
  $\chi^2$. Solving $L_0 f=Ef$ gives
\[
f=\left\{ \begin{array}{ll}
            &C_1\sin{(\sqrt{E-\chi^2}x)}+C_2\cos{(\sqrt{E-\chi^2}x)},
              \qquad\mbox{for}\;E>\chi^2\\
&C_1x+C_2 \qquad\mbox{for}\; E=\chi^2\\
            &C_1e^{-\sqrt{\chi^2-E}x}+ C_2e^{\sqrt{\chi^2-E}x}
              \qquad\mbox{for}\; E<\chi^2 . 
\end{array}
\right .
\]
None of these is normalisable, so this potential only admits scattering states.
\item{\bf Example.}
Let $L_1=-{\p_x}^2+\chi^2+u(x)$
where
\[
u(x)=-\frac{2\chi^2}{\cosh^2{\chi x}}
\]
is the so called  P\"oschl--Teller potential.

We will see in \S\ref{onesolsec} that this potential forms the initial data
for the 1--soliton solution to KdV. We will show that
there is only one bound state with $E=-\chi^2$, and the reflection coefficient
vanishes.

The creation and annihilation operators
\[
  \alpha=\p_x+\chi \tanh(\chi x), \quad \alpha^{\dagger}=-\p_x+\chi \tanh{(\chi x)}
\]
satisfy $\alpha\alpha^{\dagger}=L_0$ and $\alpha^{\dagger}\alpha=L_1$. We look for bound states
satisfying $Lf=Ef$  with $E<0$, 
of $L=-{\p_x}^2+u=L_1-\chi^2$. Therefore
\[
  \alpha^{\dagger}\alpha f=(E+\chi^2)f
\]
and $<\alpha f, \alpha f>=(E+\chi^2)<f, f>$ is finite for a bound state. Hence, either
$\alpha f=0$ or $\alpha f$ is normalisable and is a bound state of $L_0$ as
\[
  \alpha\alpha^{\dagger}(\alpha f)=(E+\chi^2)\alpha f.
\]
However, in the previous example we have shown that $\alpha\alpha^{\dagger}$ does not have
any bound states. Therefore $\alpha f=0$ and 
\[
f=\sqrt{\frac{\chi}{2}}\frac{1}{\cosh{\chi x}}, \quad E=-\chi^2.
\]
Now look for the scattering states with energy $E=k^2>0$. Consider $
-{\p_x}^2 f=k^2 f$ and pick $f=e^{-ikx}$ corresponding to a wave incident from
$\infty$. Now $\alpha\alpha^{\dagger}f=(k^2+\chi^2)f$ so that
\[
  L(\alpha^{\dagger} f)=k^2 (\alpha^{\dagger} f).
\]
Therefore $\alpha^{\dagger} f$ is the scattering state of $L$ and all scattering
states are of this form. As $x\rightarrow\infty$ this gives
$\alpha^{\dagger} f\sim(ik+\chi) e^{-ikx}$. This has no $e^{ikx}$ term, and so 
(\ref{phi_x}) implies that $b(k)=r(k)=0$ and the potential is relfectionless.
We also conclude that $\lim_{x\rightarrow \infty} \alpha^{\dagger}f=0$ if $k=i\chi$.
Therefore $k=i\chi$ is a zero of $a$, and we recover a bound state
with the energy $E=-\chi^2$.
\end{itemize}
\subsection{Inverse Scattering}
We want to recover the potential $u(x)$ from the scattering data which 
consists of the reflection coefficients and the energy levels
\[
r(k), \{\chi_1, \dots, \chi_N\}
\]
so that $E_n=-\chi_n^2$ and
\[
\phi(x, i\chi_n)=\left\{ \begin{array}{ll}
&e^{\chi_nx},\qquad\mbox{for}\;x\rightarrow-\infty\\
&b_ne^{-\chi_nx},
\qquad\mbox{for}\;x\rightarrow\infty.
\end{array}
\right.
\]
The inverse scattering transform Gelfand--Levitan--Marchenko
consist of the following steps
\begin{itemize}
\item Set 
\be
F(x)=
\sum_{n=1}^N {b_n}^2e^{-\chi_n x} +\frac{1}{2\pi}\int_{-\infty}^{\infty}r(k)e^{ikx}dk.
\label{GLMF}
\ee
\item
Consider the GLM integral equation
\be
\label{GLMI}
K(x, y)+F(x+y)+\int_x^{\infty}K(x, z)F(z+y)dz=0
\ee 
and solve it for $K(x, y)$.
\item
Then
\be
\label{GLMu}
u(x)=-2\frac{d}{dx}K(x, x)
\ee
is the potential in the corresponding Schr\"odinger equation.
\end{itemize}
These formulae are given in the $t$--independent way,
but $t$ can be introduced as a parameter. 
If the time dependence
of the scattering data is known, the solution of the GLM integral equation
$K(x, y, t)$ will also depend on $t$ and so will the potential
$u(x, t)$.
\subsection{Lax formulation}
If the potential $u(x)$ in the Schr\"odinger equation depends on a parameter
$t$, its eigenvalues will in general change with $t$. 
The inverse scattering transform is an example of an isospectral problem,
when this does not happen
\begin{prop}
If there  exist a differential  operator $A$ such that
\be
\label{KdVlax}
\dot{L}=[L, A]
\ee
where
\[
L=-\frac{d^2}{dx^2}+u(x, t), 
\]
then the spectrum of $L$ does not depend on $t$.
\end{prop}
{\bf Proof.} Consider the eigenvalue problem
\[
Lf=Ef.
\]
Differentiating gives
\[
L_tf+Lf_t=E_tf+Ef_t.
\] 
Note that $ALf=EAf$ and use the representation (\ref{KdVlax}) to find 
\be
\label{lax_step}
(L-E)(f_t+Af)=E_tf.
\ee
Take the inner product (\ref{L_2_inner}) of this equation with $f$ and use the fact that
$L$ is self--adjoint
\[
E_t||f||^2=<f, (L-E)(f_t+Af)>=<(L-E)f, f_t+Af>=0.
\]
Thus $E_t=0$.
This derivation also implies that if 
$f(x, t)$ is an eigenfunction of $L$ with eigenvalue $E=k^2$ then so is 
$(f_t+Af)$.
\koniec

What makes the method applicable to KdV equation (\ref{kdv}) is that KdV 
is equivalent to  (\ref{KdVlax}) with
\be
\label{KdVlax1}
L=-\frac{d^2}{dx^2}+u(x, t), 
\quad A=4\frac{d^3}{dx^3}
-3\Big(u\frac{d}{dx}+\frac{d}{dx}u\Big).
\ee
To prove this statement it is enough to
compute both sides of (\ref{KdVlax}) on a function and verify that
$[L, A]$ is the multiplication by $6uu_x-u_{xxx}$ (also $\dot{L}=u_t$).
This is the Lax representation of KdV \cite{Lax_cite}. 
Such representations (for various 
choices of operators $L, A$) underlie integrability of PDEs and ODEs.

\subsection{Evolution of the scattering data}
We will now use the Lax representation to determine the time evolution
of the scattering data.
\begin{itemize}
\item{\bf Continuous spectrum.}
Assume that the potential $u(x, t)$ in the Schr\"odinger equation satisfies
the KdV equation (\ref{kdv}). Let $f(x, t)$ be an eigenfuction of the 
Schr\"odinger operator $Lf=k^2f$  defined  by its asymptotic behaviour
\[
f=\phi(x, k)\longrightarrow e^{-ikx}, \qquad \mbox{as}\,\,x\rightarrow -\infty.
\]
Equation
(\ref{lax_step}) implies 
that if $f(x, t)$ is an eigenfunction of $L$ with eigenvalue $k^2$ 
then so is $(f_t+Af)$. Moreover 
$u(x)\rightarrow 0$ as $|x|\rightarrow \infty$ therefore
\[
\dot{\phi}+A\phi\longrightarrow 4\frac{d^3}{dx^3}e^{-ikx}
=4ik^3e^{-ikx}\qquad\mbox{as}\,\,x\rightarrow -\infty.
\]
Thus $4ik^3\phi(x, k)$ and $\dot{\phi}+A\phi$ are eigenfunctions
of the Schr\"odinger operator with the same asymptotic and we deduce that 
they must be equal: Their difference is in the kernel of $L-k^2$ and so
must be a linear combination of $\psi$ and $\overline{\psi}$. But this 
combination vanishes at $\infty$ so, using the independence of
$\psi$ and $\overline{\psi}$, it must vanish everywhere.
Thus the ODE
\[
\dot{\phi}+A\phi=4ik^3\phi
\]
holds for all $x\in\R$.
We shall use this ODE and  the asymptotics at $+\infty$
to find ODEs for $a(k)$ and $b(k)$. Recall
that
\[
\phi(x, k)=a(k, t)e^{-ikx}+b(k, t)e^{ikx}\quad \mbox{as}\,\,x\rightarrow 
\infty.
\]
Substituting this to the ODE gives
\begin{eqnarray*}
\dot{a}e^{-ikx}+\dot{b}e^{ikx}&=&\Big(-4\frac{d^3}{dx^3}+4ik^3\Big)
({a}e^{-ikx}+{b}e^{ikx})\\
&=& 8ik^3be^{ikx}.
\end{eqnarray*}
Equating the exponentials gives
\[
\dot{a}=0, \qquad \dot{b}=8ik^3b
\]
and
\[
a(k, t)=a(k, 0), \qquad b(k, t)=b(k, 0)e^{8ik^3t}.
\]
\item{\bf Discrete spectrum.}
In the last Section we have shown  that the discrete energy levels are constant in time, i. e.
$E=-{\chi_n}^2, \chi_n(t)=\chi_n(0)$. As $x\rightarrow\infty$ the wave function
is
$f\sim b_n(t)e^{-\chi_n x}$. Another energy eigenstate with $E=-{\chi_n}^2$ is given by
$f_t+Af$. Non degeneracy of discrete energy levels in 1D implies that $f_t+Af=cf$, and $c=0$
from orthogonality. Therefore $f_t+Af=0$, and $A\sim 4{\p_x}^3$ as $x\rightarrow\infty$.
Therefore $b_n-4{\chi_n}^3 b_n=0$.
\end{itemize}
The evolution of the scattering
data is thus given by the following
\begin{eqnarray}
\label{scattering_data}
a(k, t)&=&a(k, 0),\nonumber\\
b(k, t)&=&b(k, 0)e^{8ik^3t},\nonumber\\
r(k, t)&=&\frac{b(k, t)}{a(k, t)}=r(k, 0)e^{8ik^3t},\nonumber\\
\chi_n(t)&=&\chi_n(0),\nonumber\\
b_n(t)&=&b_n(0)e^{4\chi_n^3t},\nonumber\\
\beta_n(t)&\equiv& {b_n}^2=\beta_n(0)e^{8\chi_n^3t}.
\end{eqnarray}
\section{Reflectionless potentials and solitons}
The formula (\ref{scattering_data}) implies that if 
the reflection coefficient is initially zero, it is zero for
all $t$. In this case the inverse scattering procedure can be carried
out explicitly. The resulting solutions are called $N$-solitons, where
$N$ is the number of zeroes $i\chi_1, \dots, i\chi_N$ of $a(k)$.
These solutions describe collisions of 1--solitons
(\ref{1-soliton}) without any non-elastic effects. The 1-solitons generated
after collisions are `the same' as those before the collision.
This fact was discovered numerically in the 1960s and boosted the interest
in the whole subject.

Assume $r(k, 0)=0$ so that (\ref{scattering_data}) implies
\[
r(k, t)=0.
\]
\subsection{One soliton solution}
\label{onesolsec}
We shall first derive the 1-soliton solution. The formula
(\ref{GLMF}) with $N=1$ gives
\[
F(x, t)=\beta(t) e^{-\chi x}.
\]
This depends on $x$ as well as $t$ because
$\beta(t)= \beta(0)e^{8\chi^3t}$ from 
 (\ref{scattering_data}). We shall suppress this explicit $t$ dependence
in the following calculation and regard $t$ as a parameter.
The GLM equation (\ref{GLMI}) becomes
\[
K(x, y)+\beta e^{-\chi(x+y)}+\int_x^\infty K(x, z)\beta e^{-\chi(z+y)}dz=0.
\]
Look for solutions in the form
\[
K(x, y)=K(x)e^{-\chi y}.
\]
This gives
\[
K(x)+\beta e^{-\chi x}+K(x)\beta\int_x^\infty e^{-2\chi z}dz=0,
\]
and after a simple integration
\[
K(x)=-\frac{\beta e^{-\chi x}}{1+\frac{\beta}{2\chi}e^{-2\chi x}}.
\]
Thus
\[
K(x, y)=
-\frac{\beta e^{-\chi (x+y)}}{1+\frac{\beta}{2\chi}e^{-2\chi x}}.
\]
This function also depends on $t$ because $\beta$ does.
Finally the formula (\ref{GLMu}) gives
\begin{eqnarray*}
u(x,t)&=& -2\frac{\p}{\p x}K(x, x)=-
\frac{4\beta\chi e^{-2\chi x}}{(1+\frac{\beta}{2\chi}e^{-2\chi x})^2}\\
&=&-\frac{8\chi^2}{\hat{\beta}^{-1}e^{\chi x}+\hat{\beta}e^{-\chi x}},\quad\mbox{where}\quad \hat{\beta}=\sqrt{\beta/(2\chi)}
\\
&=&-\frac{2\chi^2}{\cosh{(\chi(x-4\chi^2t-\phi_0))}^2}, \quad
\phi_0=\frac{1}{2\chi}\log{\Big(\frac{\beta_0}{2\chi}\Big)}
\end{eqnarray*}
which is the 1-soliton solution (\ref{1-soliton}).
 
 The energy of the corresponding solution to the Schr\"odinger equation 
determines the amplitude and the velocity of the soliton. The soliton is of the form $u=u(x-4\chi^2t)$ so it represents a wave moving to the right
with velocity $4\chi^2$ and phase $\phi_0$.
\subsection{N--soliton solution}
There are $N$ energy levels which we order
$\chi_1>\chi_2>\dots>\chi_N>0$. The function (\ref{GLMF}) is 
\[
F(x)=\sum_{n=1}^N\beta_ne^{-\chi_n x}
\]
and the GLM equation (\ref{GLMI}) becomes
\[
K(x, y)+\sum_{n=1}^N\beta_n e^{-\chi_n(x+y)}+\int_x^\infty K(x, z)
\sum_{n=1}^N\beta_n e^{-\chi_n(z+y)}dz=0.
\]
The kernel of this integral equation is degenerate in a sense that
\[
F(z+y)=\sum_{n=1}^Nk_n(z)h_n(y),
\]
so we seek solutions of the form
\[
K(x, y)=\sum_{n=1}^N K_n(x)e^{-\chi_n y}.
\]
After one integration this gives
\[
\sum_{n=1}^N(K_n(x)+\beta_ne^{-\chi_n x})e^{-\chi_n y}+\sum_{n=1}^N\Big(\beta_n\sum_{m=1}^N\frac{K_m(x)}{\chi_m+\chi_n} e^{-(\chi_n+\chi_m)x}\Big)e^{-\chi_n y} =0.
\]
The functions $e^{-\chi_n y}$ are linearly independent, so
\[
K_n(x)+\beta_ne^{-\chi_n x}+\sum_{m=1}^N\beta_nK_m(x)\frac{1}{\chi_m+\chi_n}
e^{-(\chi_n+\chi_m) x}=0.
\]
Define a matrix
\[
A_{nm}(x)=\delta_{nm}+\frac{\beta_n e^{-(\chi_n+\chi_m) x}}{\chi_n+\chi_m}.
\]
The linear system becomes
\[
\sum_{m=1}^NA_{nm}(x) K_m(x)=-\beta_ne^{-\chi_n x},
\]
or 
\[
AK+B=0,
\]
where $B$ is a column vector
\[
B=[\beta_1 e^{-\chi_1 x}, \beta_2 e^{-\chi_2 x}, \cdots,
\beta_n e^{-\chi_n x}]^T.
\]
The solution of this system is
\[
K=-A^{-1}B.
\]
Using the relation
\[
\frac{d A_{mn}(x)}{dx}=-B_me^{-\chi_n x}
\]
we can write
\begin{eqnarray*}
K(x, x)&=&\sum_{m=1}^Ne^{-\chi_m x}K_m(x)=-\sum_{m,n=1}^N e^{-\chi_m x}
(A^{-1})_{mn}B_n\\
&=&\sum_{m,n=1}^N(A^{-1})_{mn}\frac{d A_{nm}(x)}{dx}
=\mbox{Tr}\Big(A^{-1}\frac{d A}{dx}\Big)\\
&=&\frac{1}{\det{A}}\frac{d}{d x}{\det{A}}.
\end{eqnarray*}
Finally we reintroduce the explicit $t$-dependence to write
the N--soliton solution as
\be
\label{N-soliton}
u(x, t)=-2\frac{\p^2}{\p x^2}\ln{({\det{A(x)}})}\qquad\mbox{where}\qquad
A_{nm}(x)=\delta_{nm}+\frac{\beta_n e^{-(\chi_n+\chi_m) x}}{\chi_n+\chi_m}.
\ee
\subsection{Two-soliton asymptotics}
Let us analyse a two-soliton solution with $\chi_1>\chi_2$
in more detail. Set 
\[
\tau_k=\chi_k x-4\chi_k^3t,\qquad  k=1, 2
\]
and consider the determinant
\[
\det{A}=\Big(1+\frac{\beta_1(0)}{2\chi_1}e^{-2\tau_1}\Big)\Big(
1+\frac{\beta_2(0)}{2\chi_2}e^{-2\tau_2}\Big)
-\frac{\beta_1(0)\beta_2(0)}{(\chi_1+\chi_2)^2}e^{-2(\tau_1+\tau_2)}.
\]
We first analyse the case $t\rightarrow -\infty$.
In the limit $x\rightarrow -\infty$ we have
$\det{A}\sim e^{-2(\tau_1+\tau_2)}$ so
\[
\log{(\det{A})}\sim \co-2(\tau_1+\tau_2)
\]
and $u\sim 0$ which we already knew. 
Now move along the $x$ axis and  consider the leading term in $\det{A}$ 
when $\tau_1=0$ and then when $\tau_2=0$. We first reach the point
$\tau_1=0$ or
\[
x=4\chi_1^2t.
\]
In the neighbourhood of this point $\tau_2=4t\chi_2(\chi_1^2-\chi_2^2)\ll 0$
and 
\[
\det{A}\sim \frac{\beta_2(0)}{2\chi_2}e^{-2\tau_2}\Big(1+\frac{\beta_1(0)}{2\chi_1}
\Big(\frac{\chi_1-\chi_2}{\chi_1+\chi_2}\Big)^2e^{-2\tau_1}\Big).
\]
Differentiating the  logarithm of $\det{A}$ yields
\[
u\sim -2\frac{\p^2}{\p x^2}\Big(1+\frac{\beta_1(0)}{2\chi_1}
\Big(\frac{\chi_1-\chi_2}{\chi_1+\chi_2}\Big)^2
e^{-2\chi_1(x-4\chi_1^2t^2) }\Big)
\]
which looks like a one soliton solution with a phase
\[
(\phi_1)_-=\frac{1}{2\chi_1}\log{\Big(\frac{\beta_1(0)}{2\chi_1}
\Big(\frac{\chi_1-\chi_2}{\chi_1+\chi_2}\Big)^2\Big)}.
\]
 We now move along the $x$ axis
until  we reach $\tau_2=0$. Repeating the above analysis
shows that now $\tau_1=4\chi_1(\chi_2^2-\chi_1^2)t\gg 0$
and around the point $x=4\chi_2^2t$ we have
\[
\det{A}\sim 1+\frac{\beta_2(0)}{2\chi_2}e^{-2\tau_2}.
\]
Therefore the function $u$
looks like a one--soliton solution with a  phase
\[
(\phi_2)_-=\frac{1}{2\chi_2}\log{\Big(\frac{\beta_2(0)}{2\chi_2}\Big)}.
\]
As $t$ approaches $0$ the two solitons coalesce and the exact behaviour
depends on the ratio $\chi_1/\chi_2$. 

We perform analogous
analysis as $t\rightarrow \infty$. If $x\rightarrow \infty$ then
$\det{A}\sim 1$ and $u\sim 0$. We move along the $x$ axis to the left
until we reach $\tau_1=0$ where $\tau_2\gg 0$ and 
the profile of $u$ is given by
one--soliton with the phase
\[
(\phi_1)_+=\frac{1}{2\chi_1}\log{\Big(\frac{\beta_1(0)}{2\chi_1}\Big)}.
\]
Then we reach the point $\tau_2=0, \tau_1\ll 0$ where there is 
a  single soliton with
the phase
\[
(\phi_2)_+=\frac{1}{2\chi_2}\log{\Big(\frac{\beta_2(0)}{2\chi_1}\Big(\frac{\chi_1-\chi_2}{\chi_1+\chi_2}\Big)^2\Big)}.
\]
Thus the larger soliton has overtaken the smaller one.
This asymptotic analysis shows that the solitons have preserved their shape but their phases
have changed
\begin{eqnarray*}
\Delta \phi_1&=&(\phi_1)_+ -(\phi_1)_-
=-\frac{1}{\chi_1}\log{\frac{\chi_1-\chi_2}{\chi_1+\chi_2}},\\
\Delta \phi_2&=&(\phi_2)_+ -(\phi_2)_-
=-\frac{1}{\chi_2}\log{\frac{\chi_1-\chi_2}{\chi_1+\chi_2}}.
\end{eqnarray*}
The only result of the interaction can  be measured  by
\[
-\log{\frac{\chi_1-\chi_2}{\chi_1+\chi_2}}
\]
which is large if the difference between the velocities
$\chi_1$ and $\chi_2$ is small.

The figures 
show the two--soliton solution at $t=-1, t=0$ and $t=1$
(for the chosen parameters $t=-1$ is considered to be a large negative 
time when the two solitons are separated). 
It should be interpreted as a
passing collision of fast and slow soliton.
The larger, faster soliton  
has amplitude $8$, and the slower, smaller  soliton has  amplitude $2$. Its velocity is one half of that of the fast soliton. 
The solitons are separated at $t=-1$.
At $t=0$ the  collision takes place. The wave amplitude becomes 
smaller than the sum of the two waves. At $t=1$ the larger soliton has 
overtaken the smaller one. The amplitudes and shapes have not changed.
\newpage
\begin{center}
{\bf 2-soliton solution at $t=-1$.}
\includegraphics[width=15cm,height=9cm,angle=0]{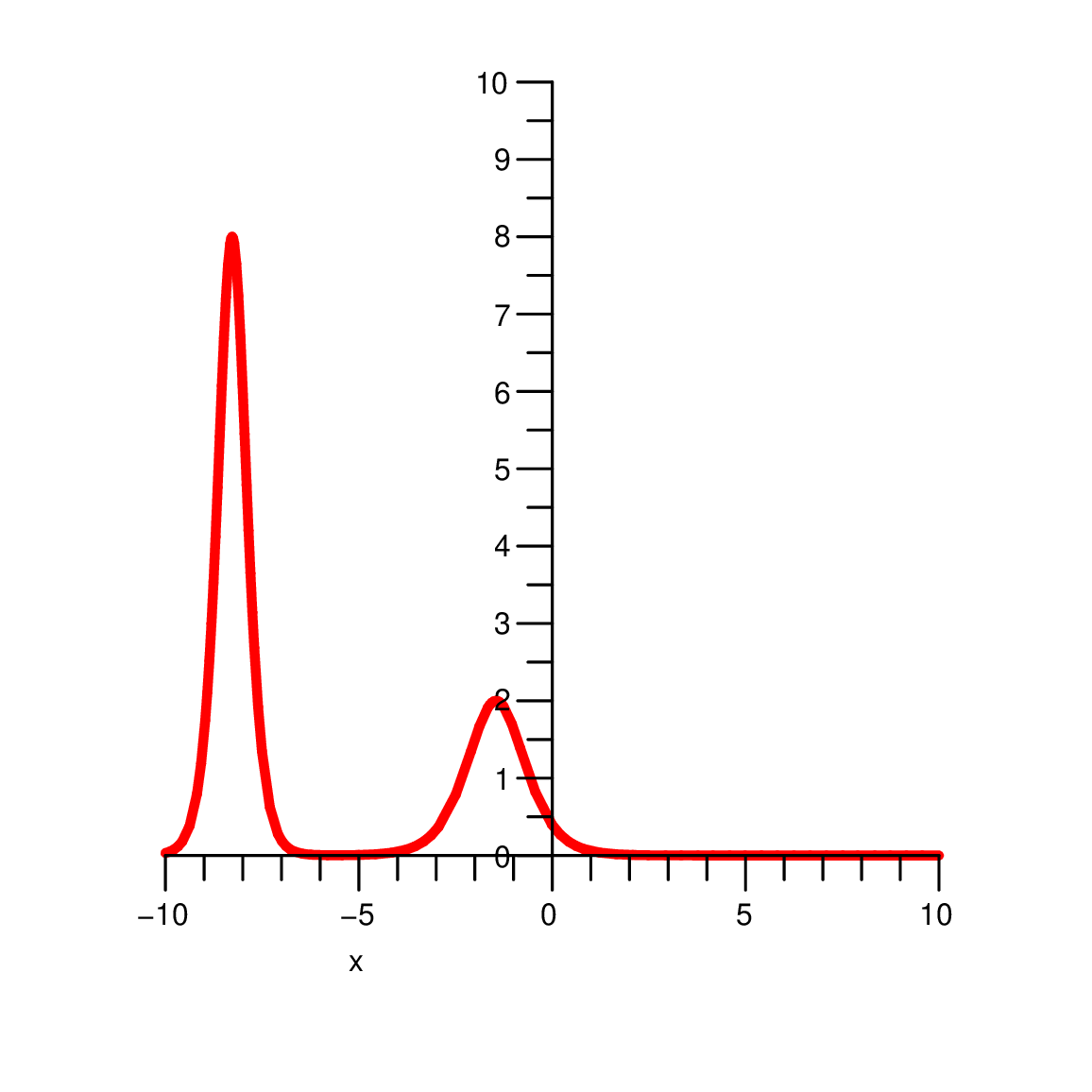}
\end{center}
{\bf 2-soliton solution at $t=0$. The total amplitude is smaller 
than the sum of the two amplitudes.}
\begin{center}
\includegraphics[width=15cm,height=9cm,angle=0]{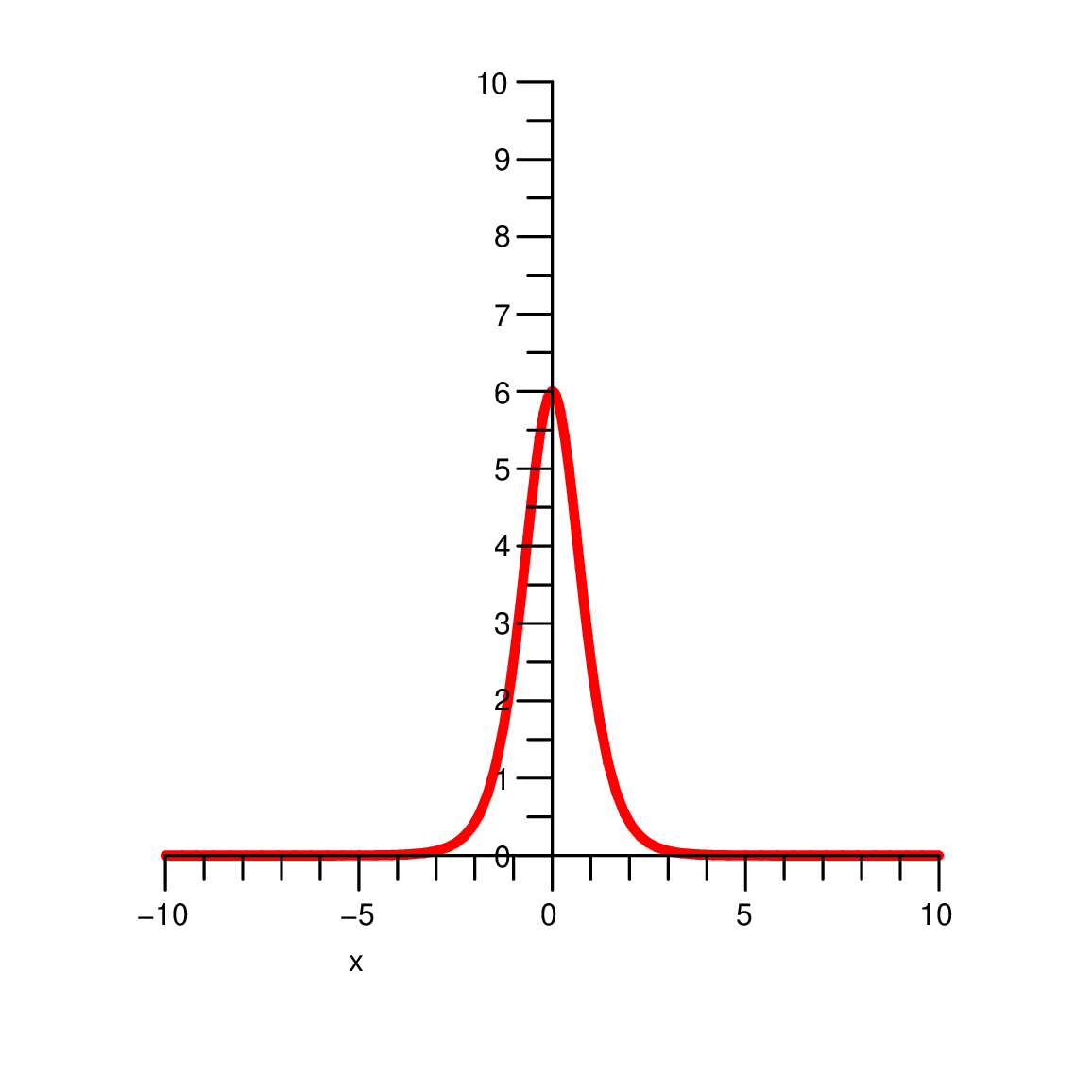}
\end{center}
\newpage
\begin{center}
{\bf 2-soliton solution at $t=1$. Amplitudes and shapes 
preserved by the collision.}
\includegraphics[width=15cm,height=9cm,angle=0]{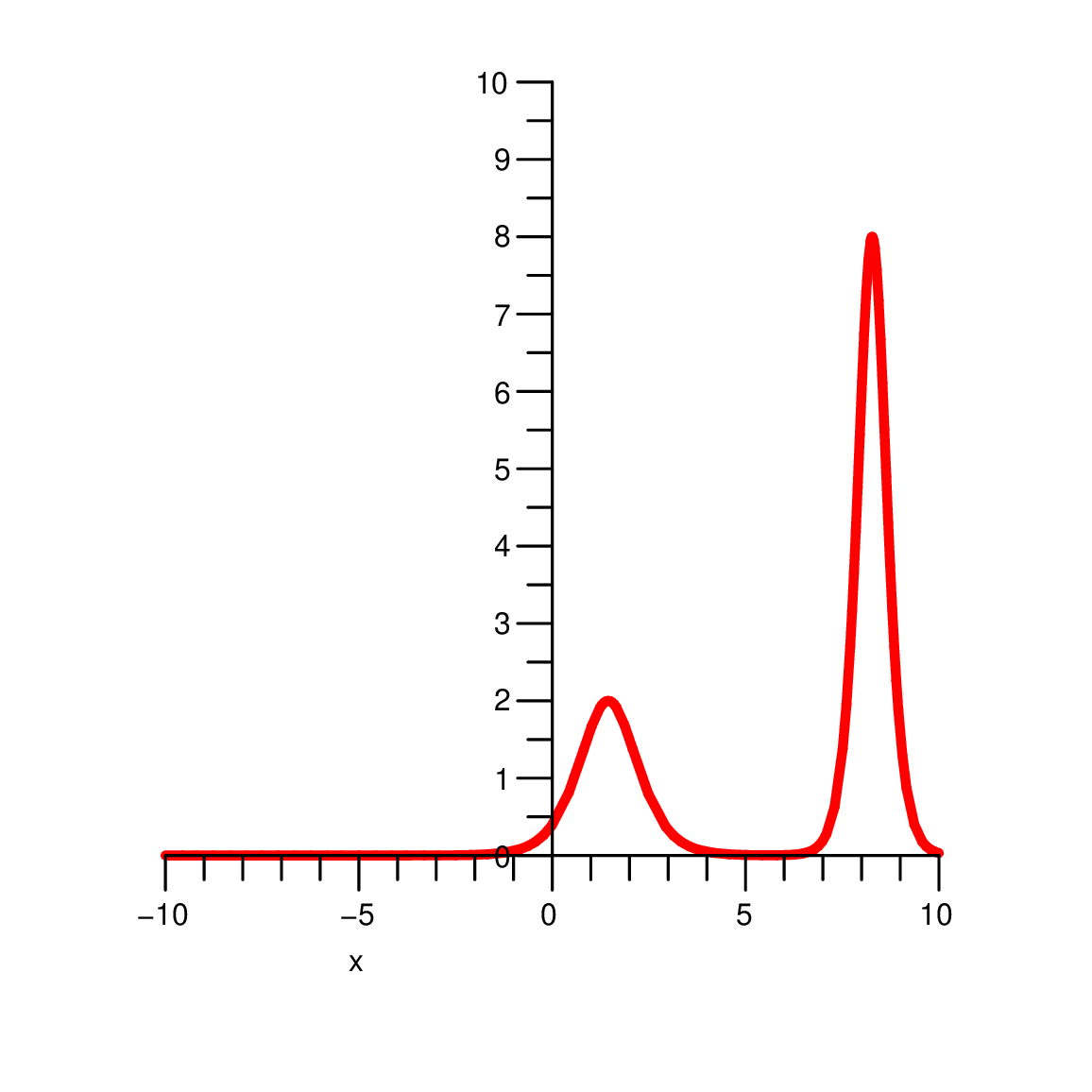}
\end{center}
This picture generalises to $N>2$. The general solution
(\ref{N-soliton}) asymptotically represents $N$ separate solitons
ordered accordingly to their speed. The tallest (and therefore fastest)
soliton is at the front, followed by the second tallest etc. 
At $t=0$ the `interaction' takes place and then the individual solitons
re-emerge in the opposite order as $t\rightarrow\infty$.
The total phase--shift is the sum of pairwise 
phase--shifts \cite{NMPZ}.
\subsection{Initial data}
The number of the discrete
eigenvalues $N$ in the Schr\"odinger operator is equal to the number of solitons
at $t\rightarrow \pm\infty$. This number is of course encoded in 
the initial conditions. To see it consider
\be
\label{PTinitial}
u(x, 0)=u_0(x)=-\frac{N(N+1)}{\cosh^2{(x)}}, \quad N\in \Z^+.
\ee
Substituting $\xi=\tanh{(x)}\in (-1, 1)$ in the Schr\"odinger equation
\[
-\frac{d^2 f}{dx^2}+u_0(x)f=k^2f
\]
yields the associated Legendre equation
\[
\frac{d}{d\xi}\Big((1-\xi^2)\frac{d f}{d \xi}\Big)+
\Big(N(N+1)+\frac{k^2}{1-\xi^2}\Big)f=0.
\]
Analysis of the power series solution shows that
the square integrable solutions exist if $k^2=-\chi^2$ and
$\chi=1, 2, \dots, N$. 
Therefore $F(x)$ in the GLM equation is given by
\[
F(x)=\sum_{n=1}^N\beta_ne^{-\chi_n x},
\]
and the earlier calculation applies leading to a particular case of 
the $N$--soliton solution (\ref{N-soliton}). See the more complete discussion
of this point in \cite{DJ}
\vskip 5pt
We have seen that the initial data for the one--soliton solution with $\chi=1$ is
$u(x, 0)=-2\cosh^{-2}(x)$. This potential admits one bound state. The two--soliton initial data
is the potential with two bound states $u(x, 0)=-6\cosh^{-2}(x)$. Consider instead the initial data
\[
u(x, 0)=-\frac{2.001}{\cosh^{2}(x)}.
\]
This also has one bound state, but unlike (\ref{PTinitial}) is not reflectionless. The corresponding solution to KdV can be found numerically. It gives one soliton but with a radiation tail.

\chapter{Hamiltonian formalism and the zero curvature representation}
\section{First integrals}
We shall make contact with the Definition \ref{definition_al}
of finite--dimensional integrable systems and 
show that KdV has infinitely many first integrals.
Rewrite the expression (\ref{phi_x})
\[
{\phi(x, k)}=
\left\{
\begin{array}{ll}
&{e^{-ikx}},\qquad\mbox{for}\;x\rightarrow-\infty\\
&a(k, t)e^{-ikx}+{b(k, t)}e^{ikx},\qquad\mbox{for}\;x\rightarrow\infty,
\end{array}
\right.
\]
when the time dependence of the scattering data has been determined using
the KdV equation.
The formula (\ref{scattering_data}) gives
\[
\frac{\p}{\p t}a(k, t)=0, \quad \forall k
\]
so the scattering data gives infinitely many first integrals provided
that they are non--trivial and independent. 
We aim to express these first 
integrals in the form
\[
I[u]=\int_{\R}P(u, u_x, u_{xx}, \dots)dx
\]
where $P$ is a polynomial in $u$ and its derivatives.

Set
\[
\phi(x, t, k)=e^{-ikx+\int_{-\infty}^x S(y, t,  k)dy}.
\]
For large $x$ the formula (\ref{phi_x}) gives
\[
e^{ikx}\phi\cong a(k)+b(k, t)e^{2ikx}.
\]
If we assume that $k$ is in the upper half plane $Im(k)>0$ the second term
on the RHS goes to 0 as $x\rightarrow \infty$. Thus
\begin{eqnarray}
\label{for_a}
a(k)&=&\lim_{x\rightarrow\infty} e^{ikx}\phi(x, t, k)=
\lim_{x\rightarrow\infty}e^{\int_{-\infty}^x S(y,t, k)dy}\nonumber\\
&=&e^{\int_{-\infty}^\infty S(y,t, k)dy},
\end{eqnarray}
where the above formula also holds in the limit $Im(k)\rightarrow 0$
because of the real analyticity.
Now we shall use the Schr\"odinger equation with $t$ regarded as a parameter
\[
-\frac{d^2\phi}{d x^2}+u\phi=k^2\phi
\]
to find an equation for $S$. Substituting
\[
\frac{d\phi}{d x}=(-ik+S(x, k))\phi, 
\quad \frac{d^2\phi}{d x^2}=\frac{d S}{d x}\phi+(-ik+S(x, k))^2\phi
\]
gives the Riccati type equation 
\be
\label{riccati}
\frac{d S}{d x}-2ik S+S^2=u,
\ee
(we stress that both $S$ and $u$ depend on $x$ as well as $t$).
Look for solutions of the form 
\[
S=\sum_{n=1}^\infty\frac{S_n(x, t)}{{(2ik)}^n}.
\]
Substituting this to (\ref{riccati})
yields a recursion relation
\be
\label{recursions_for_S}
S_1(x,t)=-u(x, t), \quad
S_{n+1}=\frac{dS_n}{d x}+\sum_{m=1}^{n-1}S_{m}S_{n-m}
\ee
which can be solved for the first few terms  
\[
S_2=-\frac{\p u}{\p x}, \quad S_3=-\frac{\p^2 u}{\p x^2}+u^2, \quad
S_4=-\frac{\p^3 u}{\p x^3}+2\frac{\p}{\p x}u^2, 
\]
\[
S_5=-\frac{\p^4 u}{\p x^4}+2\frac{\p^2}{\p x^2}u^2+
\Big(\frac{\p u}{\p x}\Big)^2+2\frac{\p^2 u}{\p x^2}u-2u^3.
\]
Now using the time independence (\ref{scattering_data}) of 
$a(k)$ for all $k$ and combining it with (\ref{for_a}) implies that 
\[
\int_{\R}S_n(x, t)dx
\]
are first integrals of the KdV equation.
Not all of these integrals are non-trivial. For example $S_2$ and $S_4$
given above are total $x$ derivatives so they integrate to $0$ 
(using the boundary conditions for $u$). The same is true for all
even terms $S_{2n}$.
To see it set
\[
S=S_R+iS_I
\]
where $S_R, S_I$ are real valued functions and substitute this to
(\ref{riccati}). Taking the imaginary part gives
\[
\frac{d S_I}{dx}+2S_RS_I-2kS_R=0
\]
which integrates to
\[
S_R=-\frac{1}{2}\frac{d}{dx}\log{(S_I-k)}.
\]
The even terms
\[
\frac{S_{2n}(x)}{{(2ik)}^{2n}}, \qquad n=1, 2, \ldots
\]
in the expansion of $a$ are real. Comparing this with the expansion
of $S_R$ in $k$ shows that $S_{2n}$ are all total derivatives
and therefore
\[
\int_{\R}S_{2n}dx=0.
\]
Let us now concentrate on the remaining non--trivial first integrals.
Set 
\be
\label{kdv_first_integrals}
I_{n-1}[u]=\frac{1}{2}\int_{\R}S_{2n+1}(x, t)dx, \qquad n=0, 1, 2, \cdots\,.
\ee
Our analysis shows
\[
\frac{d I_{n}}{d t}=0.
\]
The first of these is just the integral of $u$ itself. The next
two are known as momentum and energy respectively
\[
I_0=\frac{1}{2}\int_{\R} u^2dx, \quad I_1=-\frac{1}{2}\int_{\R}(u_x^2+2u^3)dx,
\]
where in the last integral we have isolated the total derivative in
\[
S_5=-\frac{\p^4}{\p x^4}u+2\frac{\p^2}{\p x^2} u^2+2\frac{\p}{\p x}
\Big(u\frac{\p u}{\p x}\Big)-\Big(\frac{\p u}{\p x}\Big)^2-2u^3
\]
and eliminated it using the integration by parts and boundary conditions.
These two first integrals are associated, via Noether's theorem,  
with the translational invariance of KdV: if $u(x, t)$ is a solution
then $u(x+x_0, t)$ and $u(x, t+t_0)$ are also solutions. The systematic
way of constructing such symmetries will be presented 
in Chapter \ref{chap_lie_ref}.

\section{Hamiltonian formalism}
We can now cast the KdV in the Hamiltonian form with the
Hamiltonian functional given by the energy integral $H[u]=-I_1[u]$.
First calculate.
\[
\frac{\delta I_1[u]}{\delta u(x)}=-3u^2+u_{xx}, \quad
\frac{\p }{\p x}\frac{\delta I_1[u]}{\delta u(x)}=-6uu_x+u_{xxx}.
\]
Recall that the Hamilton canonical equations
for PDEs take the form (\ref{can_PDEs})
\[
\frac{\p  u}{\p t}=\frac{\p }{\p x}\frac{\delta H[u]}{\delta u(x)}.
\]
Therefore
\be
\label{ham_kdv_eq}
\frac{\p  u}{\p t}=-\frac{\p }{\p x}\frac{\delta I_1[u]}{\delta u(x)},
\ee
is the KdV equation.
With some more work (see \cite{NMPZ}) it can be shown that
\[
\{ I_m, I_n\}=0
\]
where the Poisson bracket is given by (\ref{standard_bracket})
so that KdV is indeed integrable in the Arnold--Liouville sense.
For example
\begin{eqnarray*}
\{I_n, I_1\}&=&\int_{\R}\frac{\delta I_n}{\delta u(x)}\frac{\p}{\p x}
\frac{\delta I_1}{\delta u(x)}dx=-\int_{\R}\frac{\delta I_n}{\delta u(x)}u_t
\,dx\\
&=&-\frac{1}{2}\int_{\R}\sum_{k=0}^{2n}(-1)^k\Big(\Big(\frac{\p}{\p x}\Big)^k
\frac{\p S_{2n+1}}{\p u^{(k)}}\Big)u_t\, dx\\
&=&\frac{1}{2}\int_{\R}\sum_{k=0}^{2n}\frac{\p S_{2n+1}}{\p u^{(k)}}
\frac{\p}{\p t}u^{(k)}\, dx\\
&=&\frac{d}{d t}I_n[u]=0
\end{eqnarray*}
where we used integration by parts and the boundary conditions.
\subsection{Bi--Hamiltonian systems}
\label{sec_bi_ham}
Most systems integrable by the inverse scattering transform are Hamiltonian
in two distinct ways. This means that for a given  evolution equation
$u_t=F(u, u_x, \dots)$ there exist two Poisson structures
${\cal D}$ and ${\cal E}$ and two functionals 
$H_0[u]$ and $H_1[u]$ such that
\be
\label{bihamiltonian_system}
\frac{\p u}{\p t}={\cal D}\frac{\delta H_1}{\delta u(x)}={\cal E}\frac{\delta H_0}{\delta u(x)}.
\ee
One of these Poisson structures can be put in a form
${\cal D}=\p/\p x$ and corresponds to the standard Poisson bracket
(\ref{standard_bracket}), but the second structure ${\cal E}$ gives a new Poisson bracket.

In the finite--dimensional context discussed in Section \ref{act_ang_sec}
this would correspond to having two skew-symmetric matrices
$\omega, \Omega$ which satisfy the Jacobi Identity. 
The Darboux theorem implies the existence of a  local coordinate system
$(p, q)$ in which  one of these, say $\omega$,
is a constant skew-symmetric matrix.  The matrix components 
of second structure $\Omega$ will however be non--constant functions
of $(p, q)$. Using (\ref{poisson_ham}) we write the bi--Hamiltonian condition
as
\[
\omega^{ab}\frac{\p H_1}{\p \xi^a}=\Omega^{ab}\frac{\p H_0}{\p \xi^a},
\]
where $\xi^a, a=1, \dots, 2n$ are local coordinates on the phase space $M$, and
$H_0, H_1$ are two distinct functions on $M$. The matrix
valued function
\[
{R_a}^c=\Omega^{bc}(\omega^{-1})_{ab}
\]
is called a recursion operator. It should be thought of as an endomorphism $R=\Omega\circ \omega^{-1}$ acting on the tangent space $T_pM$, where $p\in M$. This endomorphism smoothly depends on a point $p$. 
The existence of such recursion operator is, 
under certain technical assumptions, equivalent
to Arnold--Liouville integrability  in a sense of Theorem \ref{al}.
This is because  given one first
integral $H_0$ the remaining $(n-1)$ 
integrals $H_1, \dots, H_{n-1}$ can be constructed recursively by
\[
\omega^{ab}\frac{\p H_i}{\p \xi^a}=
R^{i}\Big(\omega^{ab}
\frac{\p H_0}{\p \xi^a}\Big)
\qquad i=1, 2, \dots, n-1.
\]
The extension of this formalism to the infinite dimensional setting provides
a practical way of constructing first integrals.
In the case of KdV the first Hamiltonian formulation (\ref{ham_kdv_eq})
has ${\cal D}=\p/\p x$ and 
\[
H_1[u]=\int_{\R}\Big(\frac{1}{2}{u_x}^2+u^3\Big) dx.
\]
The second formulation
can be obtained taking
\[
H_0[u]=\frac{1}{2}\int_{\R} u^2 dx, \qquad {\cal E}=-\p_x^3+4u\p_x+2u_x.
\]
In general it is required that a pencil of Poisson structures ${\cal D}+c{\cal E}$
is also a Poisson structure (i.e. satisfies the Jacobi identity) for
any constant $c\in \R$. If this condition is satisfied, 
the bi--Hamiltonian formulation gives an effective way to construct first
integrals. The following result is proved in the book of Olver
\cite{olver}
\begin{theo}
Let {\em(\ref{bihamiltonian_system})} be a bi--Hamiltonian system,
such that the Poisson structure ${\cal D}$ is  
non--degenerate\footnote{A differential operator ${\cal D}$
is degenerate is there exists a non--zero differential
operator $\hat{\cal D}$ such that the operator $\hat{\cal D}\circ{\cal D}$ is identically zero.},
and let
\[
R= {\cal E}\circ {\cal D}^{-1}
\]
be the corresponding recursion operator. Assume that
\[
R^n\,\Big({\cal D}\frac{\delta H_0}{\delta u(x)}\Big)
\]
lies in the image of ${\cal D}$ for each $n=1, 2, \dots\,\,$.
Then there exists conserved functionals
\[H_1[u], H_2[u], \dots\]  which are in involution, i. e.
\[
\{H_m, H_n\}:=\int_{\R}\frac{\delta{H_m}}{\delta u(x)}{\cal D}\frac{\delta H_n}{\delta u(x)} dx=0.
\]
\end{theo}
The conserved functionals $H_n[u]$ are constructed recursively from $H_0$ by
\be
\label{recursive_formula_olver}
{\cal D}\frac{\delta H_n}{\delta u(x)}=R^n\Big({\cal D}\frac{\delta H_0}{\delta u(x)}\Big), \qquad n=1, 2, \dots\, \,.
\ee
In the case of the KdV equation the recursion operator is
\be
\label{kdv_rec_op}
R=-\p_x^2+4u+2u_x{\p_x}^{-1},
\ee
where ${\p_x}^{-1}$ is formally defined as integration with respect to $x$, and
formula (\ref{recursive_formula_olver}) gives an alternative way of constructing the first
integrals (\ref{kdv_first_integrals}).

\section{Zero curvature representation}
\label{sec_zcr}
We shall discuss a more geometric form of the Lax representation
where integrable systems arise as compatibility conditions of
overdetermined system of matrix PDEs. Let $U(\lambda)$ and
$V(\lambda)$ be matrix valued functions of $(\rho, \tau)$ depending on 
the auxiliary variable $\lambda$ called the spectral parameter. 
Consider a system of linear PDEs
\be
\label{linear_system_2D}
\frac{\p}{\p \rho} v=U(\lambda) v, \quad
\frac{\p}{\p \tau} v=V(\lambda) v
\ee
where $v$ is a column vector whose components
depend on $(\rho, \tau, \lambda)$.
This is an overdetermined system as there are twice as many equations
as unknowns. 
The compatibility conditions can be obtained by 
cross--differentiating and commuting the partial derivatives 
\[
\frac{\p}{\p \tau}\frac{\p}{\p \rho} v
-\frac{\p}{\p \rho}\frac{\p}{\p \tau}v=0
\]
which gives
\[
\frac{\p}{\p \tau}(U(\lambda) v)-\frac{\p}{\p \rho}(V(\lambda) v)=
\Big(
\frac{\p}{\p \tau}U(\lambda)-\frac{\p}{\p \rho}V(\lambda)+
[U(\lambda), V(\lambda)]\Big) v=0.
\]
This has to hold for all characteristic initial data
so the linear system 
(\ref{linear_system_2D}) is consistent iff the nonlinear equation
\be
\label{zcr}
\frac{\p}{\p \tau}U(\lambda)-\frac{\p}{\p \rho}V(\lambda)
+[U(\lambda), V(\lambda)]=0
\ee
holds. The whole scheme is known as the zero curvature 
representation\footnote{The terminology, 
due to Zaharov and Shabat, comes from 
differential geometry where (\ref{zcr}) means
that the curvature of a connection $Ud\rho+Vd\tau$ is zero.}.
Most non-linear integrable equation admit a zero--curvature representation 
analogous to (\ref{zcr}). 
\begin{itemize}
\item
{\bf Example.} If
\be
\label{sine_gord_case}
U=  
\frac{i}{2}
\left(\begin{array}{cc}
2\lambda&\phi_\rho\\
\phi_\rho&-2\lambda
\end{array}
\right),\quad
\quad V= 
\frac{1}{4i\lambda}
\left(\begin{array}{cc}
\cos{(\phi)}&-i\sin{(\phi)}\\
i\sin{(\phi)}&-\cos{(\phi)}
\end{array}
\right)
\ee
where $\phi=\phi(\rho, \tau)$ then (\ref{zcr}) is equivalent to the 
Sine--Gordon 
equation 
\[
\phi_{\rho\tau}=\sin{(\phi)}.
\]
\item {\bf Example.} Consider the zero curvature representation with
\begin{eqnarray}
\label{nls_lax}
U&=&  
i\lambda\left(\begin{array}{cc}
1&0\\
0&-1
\end{array}
\right)+i \left(\begin{array}{cc}
0&\ov{\phi}\\
\phi&0\end{array}\right),\\
V&=& 2i\lambda^2\left(\begin{array}{cc}
1&0\\
0&-1
\end{array}
\right)+2i\lambda
\left(\begin{array}{cc}
0&\ov{\phi}\\
\phi&0\end{array}\right)
+ \left(\begin{array}{cc}
0&\ov{\phi}_\rho\\
-\phi_\rho&0\end{array}\right)
-i\left(\begin{array}{cc}
|\phi|^2&0\\
0&-|\phi|^2\end{array}\right).\nonumber
\end{eqnarray}
The condition (\ref{zcr}) holds if
the  complex valued function $\phi=\phi(\tau, \rho)$
satisfies the nonlinear Schr\"odinger equation
\[
i\phi_{\tau}+\phi_{\rho\rho}+2|\phi|^2\phi=0.
\]
This is another famous soliton equation which can be solved by
inverse scattering transform.
\end{itemize}
There is a freedom in the matrices $U(\lambda), V(\lambda)$
known as the gauge invariance. 
Let $g=g(\tau, \rho)$ be an arbitrary invertible matrix.
The transformation
\be
\label{gauge_transformUV}
\widetilde{U}= gUg^{-1}+\frac{\p g}{\p \rho} g^{-1}, \quad
\widetilde{V}= gVg^{-1}+\frac{\p g}{\p \tau} g^{-1}
\ee
maps solutions to the zero curvature equation into new solutions: 
if the matrices $(U, V)$ satisfy (\ref{zcr}) then so do
the matrices $(\widetilde{U}, \widetilde{V})$. To see
it assume that $v(\rho, \tau, \lambda)$ is a solution
to the linear system (\ref{linear_system_2D}), and demand that
$\tilde{v}=g(\rho, \tau)v$ be another solution for some  
$(\widetilde{U}, \widetilde{V})$.
This leads to the gauge transformation (\ref{gauge_transformUV}).

 One can develop a version of inverse scattering transform
which recovers $U(\lambda)$ and $V(\lambda)$ from a linear scattering 
problem (\ref{linear_system_2D}). 
The  representation (\ref{zcr}) 
can also be an effective direct method of finding solutions
if we know $n$ linearly independent solutions 
$v_1, \dots, v_n$ to the linear system 
(\ref{linear_system_2D}) at the first place. 
Let $\Phi(\rho, \tau , \lambda)$ be a fundamental matrix solution to 
(\ref{linear_system_2D}). The columns of $\Phi$ are the 
$n$ linearly independent solutions  $v_1, \dots, v_n$. Then
(\ref{linear_system_2D}) holds with $v$ replaced by $\Phi$ and we can write
\[
U(\lambda)=\frac{\p \Phi}{\p \rho}\Phi^{-1}, \quad
V(\lambda)=\frac{\p \Phi}{\p \tau}\Phi^{-1}.
\]
In practice one assumes a simple $\lambda$ dependence in $\Phi$,
characterised by a finite number of poles with given multiplicities.
One general scheme of solving (\ref{zcr}), known as the dressing method, 
is based on the Riemann--Hilbert problem which we shall review next.
\subsection{The Riemann--Hilbert problem}
\label{RH_subsec}
Let $\lambda\in \overline{\C}=\C+\{\infty\}$ and let $\Gamma$ be a closed
contour in the extended complex plane. In particular we can
consider $\Gamma$ to be a real line $-\infty<\lambda<\infty$ regarded
as a circle in $\overline{\C}$ passing through $\infty$.
Let $G=G(\lambda)$ be a matrix valued function on the contour $\Gamma$.
The Riemann--Hilbert problem  is to construct two matrix valued functions
$G_+(\lambda)$ and $G_-(\lambda)$ holomorphic respectively inside and outside 
the contour such that on $\Gamma$
\be
\label{Riemann_problem}
G(\lambda)=G_+(\lambda)G_-(\lambda).
\ee
In the case when $\Gamma$ is the real axis $G_+$ is required to be holomorphic
in the upper half--plane and  $G_-$  is required to be holomorphic
in the lower half--plane. If $(G_+, G_-)$ is a solution of the 
Riemann--Hilbert
problem, then
\[
\widetilde{G}_+=G_+\, g^{-1},\qquad \widetilde{G}_-=g\,G_-
\]
will also be a solution for any constant 
invertible matrix $g$. This ambiguity can be avoided by fixing a
values of $G_+$ or $G_-$ at some point in their domain,  for example by
setting $G_-(\infty)=I$. If the matrices $G_\pm$ are everywhere
invertible then this normalisation guarantees that the solution to
(\ref{Riemann_problem}) is unique.

Solving a Riemann--Hilbert problem comes down to an integral equation.
Choose a normalisation $G_+(\lambda_0)=I$ and set
$G_-(\lambda_0)=g$ for some $\lambda_0\in\C$. 
Assume that the Riemann--Hilbert problem
has a solution of the form
\[
(G_+)^{-1}=h+\oint_\Gamma\frac{\Phi(\xi)}{\xi-\lambda}d\xi
\]
inside the contour $\Gamma$, and
\[
G_-=h+\oint_\Gamma\frac{\Phi(\xi)}{\xi-\lambda}d\xi
\]
outside $\Gamma$,
where $h$ is determined by  the normalisation condition to be
\[
h=g-\oint_\Gamma\frac{\Phi(\xi)}{\xi-\lambda_0}d\xi.
\]
The Plemelj  formula  \cite{Ablowitz_Fokas} can be used to determine 
$(G_+)^{-1}$ and $G_-$ on the contour: If $\lambda\in\Gamma$ then
\begin{eqnarray*}
(G_+)^{-1}(\lambda)&=&h+\oint_\Gamma\frac{\Phi(\xi)}{\xi-\lambda}d\xi+
\pi i\Phi(\lambda)\\
G_-(\lambda)&=&h+\oint_\Gamma\frac{\Phi(\xi)}{\xi-\lambda}d\xi-
\pi i\Phi(\lambda),
\end{eqnarray*}
where the integrals are assumed to be  defined by the principal value.
Substituting these expressions to  (\ref{Riemann_problem}) yields
the integral equation for $\Phi=\Phi(\lambda)$. 
If the normalisation is canonical, so that $h=g=1$, the equation is
\[
\frac{1}{\pi i}\Big(\int_\Gamma\frac{\Phi(\xi)}{\xi-\lambda}d\xi+I\Big)
+\Phi(\lambda)(G+I)(G-I)^{-1}=0.
\]
\vskip 10pt
The simplest case is the  scalar Riemann--Hilbert problem 
where $G, G_+, G_-$ are ordinary functions.
In this case
the solution can be 
written down explicitly as 
\begin{eqnarray*}
G_+&=&
\exp{\Big(-\Big(\frac{1}{2\pi i}\int_{-\infty}^{\infty} \frac{\log{G(\xi)}}{\xi-\lambda}d\xi\Big)\Big)},\quad\mbox{Im}(\lambda)>0 \\
G_-&=&
\exp{\Big(\frac{1}{2\pi i}\int_{-\infty}^{\infty} \frac{\log{G(\xi)}}{\xi-\lambda}d\xi\Big)},\quad\mbox{Im}(\lambda)<0.
\end{eqnarray*}
This is verified by taking a logarithm of (\ref{Riemann_problem})
\[
\log{G}=\log{ (G_-)}-\log{(G_+)^{-1}}
\]
and   applying  the Cauchy integral formulae. 
\subsection{Dressing method}
We shall assume that the matrices $(U, V)$ in the zero curvature
representation (\ref{zcr}) have rational dependence 
on the spectral parameter $\lambda$. The complex analytic data for each of
these matrices consist of a set of poles (including poles at $\lambda=\infty$) 
with the corresponding multiplicities.
Define the divisors to be the  sets
\[
S_U=\{\a_i, n_i, n_\infty\}, \quad 
S_V=\{\beta_j, m_j,  {m}_\infty\}, \qquad i=1, \ldots, n, \quad
\qquad j=1, \ldots, {m}
\]
so that
\begin{eqnarray}
\label{pole_structure}
U(\rho, \tau, \lambda)&=&\sum_{i=1}^n\sum_{r=1}^{n_i}
\frac{U_{i, r}(\rho,\tau)}{(\lambda-\a_i)^r}+\sum_{k=0}^{n_\infty}\lambda^k U_k(\rho, \tau)\nonumber\\
V(\rho, \tau, \lambda)&=&\sum_{j=1}^m\sum_{r=1}^{m_i}\frac{V_{j, r}
(\rho, \tau)}{(\lambda-\beta_j)^r}+\sum_{k=0}^{m_\infty}\lambda^k V_k(\rho, \tau).
\end{eqnarray}

The zero curvature condition (\ref{zcr}) is a system of non--linear PDEs
on coefficients \[U_{i, r},\quad U_k,\quad V_{j, r},\quad V_k\] 
of $U$ and $V$. Consider
a trivial solution to  (\ref{zcr})
\[
U=U_0(\rho, \lambda), \qquad V=V_0(\tau, \lambda)
\] 
where $U_0, V_0$ are any two commuting matrices with divisors
$S_U$ and $S_V$ respectively. 

Let $\Gamma$ be a contour in the extended complex plane which does not contain
any points from $S_U\cup S_V$, and let $G(\lambda)$ be a smooth matrix--valued
function defined on $\Gamma$. The dressing method \cite{ZM} is a way
of constructing a non--trivial solution with analytic
structure specified by divisors $S_U, S_V$ out of the data 
\[
(U_0,\, V_0,\, \Gamma,\, G).
\]
It consists of the following steps
\begin{enumerate}
\item Find a fundamental matrix solution to a linear system
of equations
\be
\label{linear_system_2D_0}
\frac{\p}{\p \rho} \Psi_0=U_0(\lambda) \Psi_0, \quad
\frac{\p}{\p \tau} \Psi_0=V_0(\lambda) \Psi_0.
\ee
This overdetermined system is compatible as $U_0, V_0$ satisfy
(\ref{zcr}).
\item Define a family of smooth functions $G(\rho, \tau. \lambda)$
parametrised by $(\rho, \tau)$ on $\Gamma$ 
\be
\label{family_of_functions}
G(\rho, \tau, \lambda)=\Psi_0(\rho, \tau, \lambda)G(\lambda)
{\Psi_0}^{-1}(\rho, \tau, \lambda).
\ee

This family admits a factorisation
\be
\label{RHGpm}
G(\rho, \tau, \lambda)=G_+(\rho, \tau, \lambda)G_-(\rho, \tau, \lambda)
\ee
where $G_+(\rho, \tau, \lambda)$ and $G_-(\rho, \tau, \lambda)$ are
solutions to the Riemann--Hilbert problem described in the last subsection, and are
holomorphic respectively inside and outside  the contour $\Gamma$.
\item Differentiate (\ref{RHGpm}) with respect
to $\rho$ and use  (\ref{linear_system_2D_0}) and 
(\ref{family_of_functions}). This yields
\[
\frac{\p G_+}{\p \rho}G_- +G_+\frac{\p G_-}{\p \rho}=U_0 G_+G_- -G_+ G_- U_0.
\]
Therefore we can define
\[
U(\rho, \tau, \lambda):=\Big(\frac{\p G_-}{\p\rho}+G_- U_0\Big){G_-}^{-1}
=-{G_+}^{-1}\Big(\frac{\p G_+}{\p\rho}- U_0G_+\Big)
\]
which is holomorphic in $\overline{\C}/S_U$. 
The Liouville Theorem (stating that every bounded holomorphic function is 
constant)
applied to the extended complex plane
implies that 
$U(\rho, \tau, \lambda)$ is
rational in $\lambda$ and has the same pole structure as $U_0$.

Analogous argument leads to 
\[
V(\rho, \tau, \lambda):=\Big(\frac{\p G_-}{\p\tau}+G_- V_0\Big){G_-}^{-1}
=-{G_+}^{-1}\Big(\frac{\p G_+}{\p\tau}- V_0G_+\Big)
\]
which has the same pole structure as $V_0$.
\item Define two matrix valued functions
\[
\Psi_+={G_+}^{-1}\Psi_0, \qquad \Psi_-={G_-}^{-1}\Psi_0.
\]

Equations (\ref{linear_system_2D_0}) and the definitions of
$(U, V)$  imply that these matrices both satisfy the overdetermined system
\[
\frac{\p}{\p \rho} \Psi_\pm=U(\lambda) \Psi_\pm, \quad
\frac{\p}{\p \tau} \Psi_\pm=V(\lambda) \Psi_\pm.
\]
We can therefore deduce that $U(\rho, \tau, \lambda)$ and
$V(\rho, \tau, \lambda)$ are of the form (\ref{pole_structure}) and  satisfy the zero curvature relation
(\ref{zcr}).
\end{enumerate}
This procedure is called `dressing' as the bare, trivial solution $(U_0, V_0)$
has been dressed by an application of a Riemann--Hilbert 
problem to a non-trivial $(U, V)$. Now, given another matrix valued
function  $G=G'(\lambda)$ on the contour we could repeat the whole procedure
and apply it to $(U, V)$ instead of $(U_0, V_0)$. This would lead
to another solution $(U', V')$ with the same pole structure. Thus dressing
transformations act on the space of solutions to (\ref{zcr}) and form a group.
If $G=G_+G_-$ and $G'={G'}_+{G'}-$ then
\[
(G\circ G')=G_+{G'}_+{G'}_-G_-.
\]
The solution to the Riemann--Hilbert problem 
(\ref{RHGpm}) is not unique. If $G_\pm$ give a factorisation of 
$G(\rho, \tau, \lambda)$
then so do 
\[
\widetilde{G}_+=G_+ g^{-1},\qquad \widetilde{G}_- = gG_-
\]
where $g=g(\rho, \tau)$ is a matrix valued function. 
The corresponding solutions
$(\widetilde{U}, \widetilde{V})$ are related to $(U, V)$ by the 
gauge transformation
(\ref{gauge_transformUV}). Fixing the gauge is therefore equivalent to 
fixing the value
of $G_+$ or $G_-$ at one point of the extended complex plane, say 
$\lambda=\infty$.
This leads to a 
unique solution of the Riemann--Hilbert problem 
with $G_\pm(\infty)=G(\infty)=I$.
\vskip10pt

The dressing  method leads to a general form of $U$ and $V$ with prescribed 
singularities,
but more work is required to make contact with specific integrable models
when additional algebraic constraints need to be imposed on $U$ and $V$. 
For example in the 
Sine--Gordon case (\ref{sine_gord_case}) the matrices are anti-Hermitian. 
The anti-Hermticity condition gives certain constraints
on the contour $\Gamma$ and the function $G$. Only 
if these constraints hold, the
matrices resulting from the dressing procedure will be given (in some gauge) 
in terms of the  solution to the Sine--Gordon equation.

\subsection{From Lax representation to zero curvature}
\label{section_lax_zero}
The zero curvature representation (\ref{zcr}) is more general than the scalar
Lax representation but there is a connection between the two. 
First similarity is that the Lax equation (\ref{KdVlax}) 
also arises as a compatibility
condition for two overdetermined PDEs. To see it 
take $f$ to be an eigenfunction of $L$ with a simple eigenvalue $E=\lambda$
and
consider the
relation (\ref{lax_step}) which follows from the Lax equations.
If $E=\lambda$ is a simple eigenvalue then
\[
\frac{\p f}{\p t}+Af=C(t) f
\]
for some function $C$ which depends on $t$ but not on $x$. Therefore
one can use an integrating factor to find a function
$\hat{f}=\hat{f}(x, t, \lambda)$ such that 
\be
\label{overdet_lax}
L\hat{f}=\lambda \hat{f}, \qquad \frac{\p \hat{f}}{\p t}+A\hat{f}=0,
\ee
where $L$ is the Schr\"odinger operator and $A$ is some differential operator
(for example given by (\ref{KdVlax1})). 
Therefore the Lax relation
\[
\dot{L}=[L, A]
\]
is a compatibility of an overdetermined system (\ref{overdet_lax}).

Consider a general scalar Lax pair
\begin{eqnarray*}
L&=&
\frac{\p^n~}{\p x^n}+u_{n-1}(x, t)\frac{\p^{n-1}}{\p x^{n-1}}+
\dots+  u_1(x, t)\frac{\p}{\p x}+u_0(x, t)  \\
A&=&
\frac{\p^m}{\p x^m}+v_{m-1}(x, t)\frac{\p^{m-1}~}{\p x^{m-1}}+
\dots  +v_{1}(x, t)\frac{\p}{\p x}+v_0(x, t)  
\end{eqnarray*}
given by differential operators  with coefficients depending on $(x, t)$.
The Lax equations 
\[
\dot{L}=[L, A]
\]
(in general there will be more than one)
are non--linear PDEs for the coefficients 
\[
(u_0, \dots, u_{n-1}, v_0, \dots, v_{m-1}).
\] 
The linear $n$th order scalar
PDE
\be
\label{generalised_lax}
L\hat{f}=\lambda \hat{f}
\ee
is equivalent to the first order matrix PDE
\[
\frac{\p F}{\p x}=U_L F
\]
where $U_L=U_L(x, t, \lambda)$ is an $n$ by $n$ matrix
\[
U_L=\left (
\begin{array}{cccccc}
0&1&0& \dots&0&0\\
0&0&1& \dots&0&0\\
...&...&...&\dots&...&...\\
...&...&...&\dots&...&...\\
0&0&0& \dots&0&1\\
\lambda-u_0&-u_1&-u_2& \dots&-u_{n-2}&-u_{n-1}
\end{array}
\right )
\]
and $F$ is a column vector
\[
F=(f_0, f_1, \dots, f_{n-1})^T,\qquad \mbox{where}\quad 
f_k=\frac{\p^k \hat{f}}{\p x^k}.
\]
Now consider the second equation in (\ref{overdet_lax})
\[
\frac{\p \hat{f}}{\p t}+A\hat{f}=0
\]
which is compatible with  (\ref{generalised_lax}) if the Lax equations hold.
We differentiate this equation with respect to  
$x$ and use (\ref{generalised_lax})
to express $\p_x^n \hat{f}$ in terms of $\lambda$ and lower order  derivatives.
Repeating this process $(n-1)$ times gives an action of $A$ on components
of the vector $F$. We write it as
\[
\frac{\p F}{\p t}=V_A F
\]
using the method described above. This leads to a pair of 
first order linear matrix equations with the zero curvature 
compatibility conditions
\[
\frac{\p U_L}{\p t}-\frac{\p V_A}{\p x}+
[U_L, V_A]=0.
\]
These conditions hold if the operators $(L, A)$ satisfy the Lax relations $\dot{L}=[L, A]$.
\vskip5pt
\begin{itemize}
\item
{\bf Example.} Let us apply  this procedure to  the KdV Lax pair
(\ref{KdVlax1}). Set \[
f_0=\hat{f}(x, t, \lambda),\qquad f_1=\p_x \hat{f}(x, t, \lambda).\]
The eigenvalue problem $L\hat{f}=\lambda \hat{f}$ gives
\[
(f_0)_x=f_1, \qquad (f_1)_x=(u-\lambda) f_0.
\]
The  equation $\p_t \hat{f}+A\hat{f}=0$ gives
\begin{eqnarray*}
(f_0)_t&=&-4(f_0)_{xxx}+6u f_1+ 3u_x f_0\\
&=&-u_x (f_0)+(2u+4\lambda )f_1.
\end{eqnarray*}
We differentiate this equation with respect to $x$ and
eliminate the second derivatives of $\hat{f}$ to get
\[
(f_1)_t=((2u+4\lambda)(u-\lambda)-u_{xx})f_0+u_x f_1.
\]
We now collect the equations in the matrix form
$\p_x F=U_L F, \p_t F=V_A F$ where $F=(f_0, f_1)^T$ and
\be
\label{kdv_matrix}
U_L=  
\left(\begin{array}{cc}
0&1\\
u-\lambda&0
\end{array}
\right), \quad
V_A= 
\left(\begin{array}{cc}
-u_x&2u+4\lambda\\
2u^2-u_{xx}+2u\lambda-4\lambda^2&u_x
\end{array}
\right).
\ee
We have therefore obtained a zero curvature representation for KdV.
\end{itemize}
\section{Hierarchies and finite gap
solutions.}
\label{Section_finite_gap}
We shall end our discussion of the KdV equation with a description of
KdV hierarchy.  Recall that KdV is a Hamiltonian system 
(\ref{ham_kdv_eq}) with the Hamiltonian given by the first integral $-I_1[u]$.
Now choose a (constant multiple of) 
a different first integral $I_n[u]$ as
a Hamiltonian and consider the equation
\be
\label{nKdV}
\frac{\p u}{\p t_n}=(-1)^n\frac{\p}{\p x}\frac{\delta I_n[u]}{\delta u(x)}
\ee
for a function $u=u(x, t_n)$. This leads to an infinite set of equations
known as higher KdVs. The first three equations are
\begin{eqnarray*}
u_{t_0}&=&u_x,\\
u_{t_1}&=&6uu_x-u_{xxx},\\
u_{t_2}&=&10uu_{xxx}-20 u_xu_{xx}-30 u^2u_x-u_{xxxxx}.
\end{eqnarray*}
Each of these equations can be solved by inverse scattering method
we have discussed, and the functionals $I_{k}, k=-1, 0, \cdots$ are
first integrals regardless which one of them is chosen as a Hamiltonian.
In the associated Lax representation $L$ stays unchanged, but $A$ is 
replaced by a differential operator of degree $(2n+1)$.
One can regard the higher KdVs as a system of overdetermined PDEs
for
\[
u=u(t_0=x, t_1=t, t_2, t_3, \cdots),
\]
where we have identified $t_0$ with $x$ using the first equation in 
(\ref{nKdV}).

This system is called a hierarchy and
the coordinates $(t_2, t_3, \cdots)$ are known as higher times.
The equations of the hierarchy 
are consistent as the flows generated by time translations 
commute
\begin{eqnarray*}
\frac{\p}{\p t_m}\frac{\p}{\p t_n}u
-\frac{\p}{\p t_n}\frac{\p}{\p t_m}u&=&
(-1)^n\frac{\p}{\p t_m}\frac{\p}{\p x}\frac{\delta I_n[u]}{\delta u(x)}
-(-1)^m\frac{\p}{\p t_n}\frac{\p}{\p x}\frac{\delta I_m[u]}{\delta u(x)}\\
&=&\{u, \p_m I_n-\p_n I_m+\{I_m, I_n\}\}=0
\end{eqnarray*}
where we used the Jacobi identity and the fact that $I_n[u]$
Poisson commute.

The concept of the hierarchy leads to a beautiful method of finding
solutions to KdV with periodic initial data, i.e.
\[
u(x, 0)=u(x+X_0, 0)
\]
for some period $X_0$.
The method is based on the concept of stationary
(i.e. time independent) solutions, albeit applied to a combination of 
the higher times.

Consider the first $(n+1)$ higher KdVs
and take $(n+1)$ real constants $c_0, \dots, c_{n}$.
Therefore
\[
\sum_{k=0}^n c_k\frac{\p u}{\p t_k}
=\sum_{k=0}^n(-1)^kc_k\frac{\p}{\p x}\frac{\delta I_k[u]}{\delta u(x)}.
\]
The stationary solutions correspond to $u$ being independent on
the combination of higher times on the LHS.
This leads to an ODE
\be
\label{finite_gap_ode}
\sum_{k=0}^n(-1)^k c_k\frac{\delta I_k[u]}{\delta u(x)}=c_{n+1},\qquad
c_{n+1}=\mbox{const}.
\ee
The recursion relations (\ref{recursions_for_S})
for $S_k$s imply that this ODE is of order $2n$. Its general
solution depends on $2n$ constants of integration as
well as $(n+1)$ parameters $c$ as we can always divide 
(\ref{finite_gap_ode}) by $c_n\neq 0$.
Altogether one has $3n+1$ parameters. The beauty of this method
is that the ODE is integrable in the sense
of Arnold--Liouville theorem and its solutions can be constructed by
hyper--elliptic functions. The corresponding solutions to KdV are known
as finite--gap solutions. Their description in terms
of a spectral data is rather involved and uses Riemann surfaces
and algebraic geometry - see Chapter 2 of \cite{NMPZ}.

 We shall now present  the construction of the first integrals
to equation (\ref{finite_gap_ode})
(we stress that (\ref{finite_gap_ode})
is an ODE in $x$ so the first integrals are functions
of $u$ and its derivatives which do not depend on $x$ when
(\ref{finite_gap_ode}) holds).
The higher KdV equations (\ref{nKdV}) admit a zero curvature representation
\[
\frac{\p}{\p t_n}U-\frac{\p}{\p x}V_n+
[U, V_n]=0
\]
where 
\[
U=  
\left(\begin{array}{cc}
0&1\\
u-\lambda&0
\end{array}
\right)
\]
is the matrix obtained for KdV in section (\ref{section_lax_zero}) and 
$V_n=V_n(x, t, \lambda)$ are traceless 2 by 2 matrices
analogous to $V_A$ which can be obtained 
using  (\ref{nKdV}) and the recursion relations for (\ref{riccati}).
The components
of $V_n$ depend on $(x, t)$ and are polynomials in $\lambda$
of degree $(n+1)$.
Now set
\[
\Lambda=c_0 V_0+\dots +c_n V_n
\]
where $c_k$ are constants and consider
solutions to
\[
\frac{\p}{\p T}U-\frac{\p}{\p x}\Lambda+
[U, \Lambda]=0, \qquad 
\]
such that
\[
\frac{\p }{\p T}U=0, \qquad\mbox{where}\qquad \frac{\p}{\p T}=
c_0\frac{\p}{\p t_0}+\dots +c_n\frac{\p}{\p t_n}.
\]
This gives rise to the ODE
\[
\frac{d}{dx}\Lambda=[U, \Lambda]
\]
which is the Lax representation of (\ref{finite_gap_ode}).
This representation reveals existence of many first integrals
for  (\ref{finite_gap_ode}) . We have
\[
\frac{d}{d x}\mbox{Tr}(\Lambda^p)=\mbox{Tr}(p[U, \Lambda]\Lambda^{p-1})=
p\mbox{Tr}(-\Lambda U \Lambda^{p-1}+U\Lambda^p)=0, \quad p=2, 3, \dots
\]
by the cyclic property of trace. 
Therefore all the coefficients of the polynomials 
$\mbox{Tr}(\Lambda(\lambda)^p)$ 
for all $p$ are conserved (which implies that the
whole spectrum of $\Lambda(\lambda)$ is constant in $x$).
It turns out  \cite{NMPZ} that one can find $n$ independent non-trivial integrals
in this set which  are in involution thus guaranteeing
the integrability of (\ref{finite_gap_ode}) is a sense of 
the Arnold--Liouville theorem \ref{al}.

The resulting solutions to KdV are known as `finite gap' potentials.
 Let us justify this terminology.
The spectrum of $\Lambda$ does not depend on $x$ and
so the coefficients of the characteristic polynomial
\[
\det{({\bf 1}\mu -\Lambda(\lambda))}=0
\] 
also do not depend on $x$. Using the fact that $\Lambda(\lambda)$
is trace free we can rewrite this polynomial
as
\be
\label{riemann_surface}
\mu^2+R(\lambda)=0
\ee
where 
\begin{eqnarray*}
R(\lambda)&=&\lambda^{2n+1}+a_1\lambda^{2n}+\dots +a_{2n}\lambda+a_{2n+1}\\
&=&(\lambda-\lambda_0)\dots (\lambda-\lambda_{2n}).
\end{eqnarray*}
Therefore the coefficients $a_1, \dots, a_{2n+1}$ 
(or equivalently $\lambda_0, \dots, \lambda_{2n}$)
do not depend on $x$. However 
$(n+1)$ of those coefficients can be expressed in terms
of the constants $c_k$ and thus the corresponding first integrals
are trivial. This leaves us with $n$ first integrals for
an ODE (\ref{finite_gap_ode}) of order $2n$.

It is possible to show \cite{NMPZ} that
\begin{itemize}
\item  All solutions to the KdV equation with periodic initial data
arise from   (\ref{finite_gap_ode}).
\item For each $\lambda$ the corresponding eigenfunctions of the Schr\"odinger operator $L\psi=\lambda\psi$
can be expanded in a basis $\psi_{\pm}$ such that
\[
\psi_\pm(x+X_0)=e^{\pm ipX_0}\psi_{\pm}(x)
\]
for some $p=p(\lambda)$ ($\psi_\pm$ are called  Bloch functions).
The set of real $\lambda$ for which $p(\lambda)\in\R$ is
called the permissible zone.
The roots of the polynomial $R(\lambda)$ are the end-points
of the permissible zones
\[
(\lambda_0, \lambda_1), \quad (\lambda_2, \lambda_3),\quad \dots,
\quad (\lambda_{2n-2}, \lambda_{2n-1}), \quad
(\lambda_{2n}, \infty).
\]
\end{itemize}
The equation (\ref{riemann_surface}) defines a Riemann surface $\Gamma$ of 
genus $n$. The number of forbidden zones (gaps) is therefore 
finite for the periodic solutions as the Riemann surface 
(\ref{riemann_surface}) has  finite genus. This
justifies the name `finite gap' potentials.

\begin{itemize}
\item {\bf Example.} If $n=0$ the Riemann surface $\Gamma$ has topology of the sphere and the corresponding solution to the KdV is a constant. If $n=1$ 
 $\Gamma$ is called an elliptic curve  (it has topology of a two-torus)
and the ODE (\ref{finite_gap_ode}) is solvable by elliptic functions:
the stationary condition
\[
c_0\frac{\p u}{\p t_0}+c_1\frac{\p u}{\p t_1} =0
\]
yields
\[
c_0u_x+c_1(6uu_x-u_{xxx})=b, \qquad b=\co
\]
where we used the fist two equations of the hierarchy.  We can 
set $c_1=1$ redefining the other two constants. This ODE can be integrated and 
the general solution is a Weierstrass elliptic function
\[
\int\frac{du}{\sqrt{2u^3+ c_0u^2-2bu+d}}=x-x_0.
\]
The stationary condition implies that $u=u(x-c_0t)$
where we have identified $t_0=x, t_1=t$. Thus $x_0=c_0t$.
These solutions are called cnoidal waves because the corresponding elliptic
function is often denoted `cn'.
\end{itemize}
If $n>1$ the Riemann surface $\Gamma$ is a hyper-elliptic curve and the 
corresponding KdV potential is given in terms of Riemann's Theta function
\cite{NMPZ}.

\chapter{Lie symmetries and reductions}
\label{chap_lie_ref}
\section{Lie groups and Lie algebras}
Phrases like `the unifying role of symmetry in \ldots' 
feature prominently in the
popular science literature. Depending on the subject, the symmetry may be
`cosmic', `Platonic', `perfect', `broken' or even 
`super'\footnote{Supersymmetry
is a symmetry between elementary particles known as bosons and fermions. It
is  a symmetry of equations underlying the current physical theories.
Supersymmetry predicts that each elementary particle
has its supersymmetric partner. No one has yet observed supersymmetry.
Perhaps it will be found in the LHC. See a footnote on page 26.}.

The mathematical framework used to define and describe the symmetries is group
theory. Recall that a group is a set $G$ with a
map \[G\times G\rightarrow G, \quad (g_1, g_2)\rightarrow g_1g_2\] called the group multiplication 
which satisfies the following properties:
\begin{itemize}
\item Associativity
\[
(g_1g_2)g_3=g_1(g_2g_3)\qquad \forall g_1, g_2, g_3\in G.
\]
\item There exist an identity element $e\in G$ such that
\[
eg=ge=g, \qquad \forall g\in G.
\]
\item For any $g\in G$ there exists an inverse element $g^{-1}\in G$ such that
\[
g g^{-1}=g^{-1} g=e.
\]
\end{itemize}
A group $G$ acts on a set
$X$ if there exists a map $G\times X\rightarrow X, \, (g, p)\rightarrow g(p)$ such that
\[
e(p)=p,\quad g_2(g_1(p))=(g_2 g_1)(p)
\]
for all $p\in X$, and $g_1, g_2 \in G$.
The set Orb$(p)=
\{ g(p),\, g\in G \}\subset X$ is called the orbit of $p$.
Groups acting on sets are often called groups of transformations.

In this Chapter
we shall explore the groups which act on solutions to differential equations.
Such group actions occur both for integrable and non-integrable systems
so the methods we shall study are quite universal\footnote{It is however
the case that integrable systems admit `large' groups of symmetries and
non-integrable  systems usually do not.}. In fact all the techniques of integration
of differential equations (like separation of variables, integrating factors,
homogeneous equations, \ldots)  students have encounter in their education 
are special cases of the symmetry approach. 
See \cite{olver} for a very complete treatment of this subject and
\cite{Hydon} for an elementary introduction at an undergraduate level.

The symmetry programme goes back to a 19th century Norwegian mathematician Sophus
Lie who developed a theory of continuous transformations now known as Lie groups.
One of most important of Lie's discoveries was that a continuous group $G$ of transformations is easy to
describe by infinitesimal transformations characterising group elements
close (in a sense of Taylor's theorem) to the identity element. These infinitesimal
transformations are elements of Lie algebra $\g$. For example a general
element of the rotation group $G=SO(2)$
\[
g(\ve) =\left(\begin{array}{cc}
\cos{\ve}&-\sin{\ve}\\
\sin{\ve}&\cos{\ve}
\end{array}
\right)
\]
depends on one parameter $\ve$. The group $SO(2)$ is a Lie group as $g$, its inverse and the group multiplication depend on $\ve$ in a continuous and 
differentiable way. This Lie group is one--dimensional as one parameter - the angle of rotation - is sufficient to describe any rotation
around the origin in $\R^2$. A rotation in $\R^3$ depends on three such parameters - the Euler
angles used in classical dynamics -  so $SO(3)$  is a three dimensional Lie group.
Now consider the Taylor series
\[
g(\ve)= \left(\begin{array}{cc}
1&0\\
0&1
\end{array}
\right)
+\ve  \left(\begin{array}{cc}
0&-1\\
1&0
\end{array}
\right)
+O(\ve^2).
\]
The antisymmetric matrix 
\[
A=\left(\begin{array}{cc}
0&-1\\
1&0
\end{array}
\right)
\]
represents an infinitesimal rotation  as $A {\bf x}=(-y, x)^T$ are components of the vector
tangent to the orbit of ${\bf x}$ at ${\bf x}$. The one dimensional vector space
spanned by $A$ is called a Lie algebra of $SO(2)$.

The following definition is not quite correct 
(Lie groups should be defined as  manifolds  -  see 
the Definition \ref{defi_of_lie_group_app}
in Appendix \ref{appendix_mdf}) 
but it is sufficient for our purposes. 
\begin{defi}
An $m$--dimensional Lie group is a group whose elements depend continuously of
$m$ parameters such that the maps $(g_1, g_2)\rightarrow g_1g_2$ and $g\rightarrow g^{-1}$
are smooth (infinitely differentiable) functions of these parameters.
\end{defi}
The infinitesimal description of Lie groups is given by Lie algebras.
\begin{defi} A Lie algebra  is a vector space $\g$  with an anti--symmetric bilinear 
operation called a Lie bracket  
$[\, ,\, ]_\g:\g\times\g\rightarrow \g$ which satisfies the Jacobi 
identity 
\[
[A, [B, C]]+[C, [A, B]]+ [B, [C, A]]=0, \qquad
\forall A, B, C \in \g.
\] 
\end{defi}
If the vectors 
$A_1, \ldots, A_{\dim{\g}}$ span $\g$, 
the algebra structure is determined by the structure constants
$f_{\alpha\beta}^\gamma$ such that
\[
[A_\alpha, A_\beta]_{\g}=\sum_{\gamma}f_{\alpha\beta}^\gamma A_\gamma, \qquad \alpha, \beta, \gamma =1, \ldots, \dim{\g}. 
\]
The Lie bracket is related to non-commutativity of the group operation 
as the following argument
demonstrates. Let $a, b\in G$.  Set 
\[
a=I+\ve A+ O(\ve^2), \qquad b=I+\ve B+ O(\ve^3)
\]
for some $A, B$ and calculate

\[
aba^{-1}b^{-1}=(I+\ve A+\dots)(I+\ve B+\dots)
(I-\ve A+\dots)(I-\ve B+\dots)=I+\ve^2[A, B]+O(\ve^2)
\]
where  $\dots$ denote terms of higher order in $\ve$ and  
we used the fact $(1+\ve A)^{-1}=1-\ve A+\dots$
which follows from the Taylor series. Some care
needs to be taken with the above argument
as we have neglected the second order terms in the
group elements but not in the answer.
The readers should convince themselves
that these terms indeed cancel
out.
\begin{itemize}
\item {\bf Example.} Consider the group of special orthogonal transformations $SO(n)$ which
consist of $n$ by $n$ matrices $a$ such that
\[
aa^T=I, \qquad \det{a}=1.
\]
These conditions imply that only $n(n-1)/2$  matrix components are independent
and $SO(n)$ is a Lie group of dimension $n(n-1)/2$.
Setting $a=I+\ve A +O(\ve^2)$ shows that
infinitesimal version of the orthogonal condition
is anti-symmetry
\[
A+A^T=0.
\]
Given two anti-symmetric matrices their commutator
is also anti-symmetric as
\[
[A, B]^T=B^TA^T-A^T B^T
=-[A, B].
\]
Therefore the vector space of antisymmetric matrices
is a Lie algebra with a Lie bracket defined to be the matrix commutator. 
This Lie algebra, called
$\mathfrak{so}(n)$, is a vector space of dimension
$n(n-1)/2$. This is equal to the dimension
(the number of parameters) of
the corresponding Lie group $SO(n)$.

\item {\bf Example.} An example of a three--dimensional 
Lie group is given by the group of $3$ by $3$ upper
triangular matrices
\be
\label{heisenberg_group}
g(m_1, m_2, m_3)=\left(\begin{array}{ccc}
1&m_1 &m_3\\
0&1&m_2\\
0&0&1
\end{array}
\right).
\ee
Note that $g={\bf 1} +\sum_{\alpha}m_\alpha T_\alpha$,
where the matrices $T_\alpha$ are
\be
\label{matrices_M}
T_1=\left(\begin{array}{ccc}
0&1 &0\\
0&0&0\\
0&0&0
\end{array}
\right), \quad
T_2=\left(\begin{array}{ccc}
0&0 &0\\
0&0&1\\
0&0&0
\end{array}
\right), \quad
T_3=\left(\begin{array}{ccc}
0&0 &1\\
0&0&0\\
0&0&0
\end{array}
\right).
\ee
This Lie group is called {\bf Nil}, as the  matrices $T_\alpha$ are all
nilpotent. These matrices span the Lie algebra of the group ${\bf Nil}$ and
have the commutation relations
\be
\label{heisenberg_la}
[T_1, T_2]=T_3, \quad [T_1, T_3]=0, \quad [T_2, T_3]=0.
\ee
This gives the structure constants $f_{12}^3=-f_{21}^3=1$
and all other constants vanish.

A three-dimensional Lie algebra with these structure constants is called the Heisenberg algebra because of its connection with Quantum
Mechanics - think of $T_1$ and $T_2$ as position
and momentum  operators respectively, and
$T_3$ as $i\hbar$ times the identity operator.
\end{itemize}

In the above example the Lie algebra of a Lie group was represented by matrices.
If the group acts on a subset $X$ of $\R^n$, its Lie algebra is represented by 
vector fields\footnote{The structure
constants $f_{\alpha\beta}^{\gamma}$ do not depend on which of these representation is used.} on $X$. This approach underlies the application of Lie groups to
differential equations so we shall study it next.
\section{Vector fields and one parameter groups of transformations}
Let $X$ be an open set in $\R^n$ with local coordinates $x^1, \dots, x^n$
and let
$\gamma:[0, 1]\longrightarrow X$ be a parametrised curve, so that
$\gamma(\ve)=(x^1(\ve), \ldots, x^n(\ve))$. 
The tangent vector 
$V|_p$ to this curve at a point $p\in X$ has components
\[
V^a=\dot{x}^a|_p, \quad a=1, \ldots, n, \qquad\mbox{where}\;\;\; 
\dot=\frac{d}{d\ve}.
\]
The collection of all tangent vectors to all possible curves through $p$ is an $n$-dimensional vector space called the tangent space $T_pX$. 
The collection of all tangent spaces as $x$ varies in $X$ is called a tangent 
bundle $TX=\cup_{x\in M}T_xX$. The tangent bundle is a manifold of 
dimension $2n$ (see Appendix \ref{appendix_mdf}). 

A vector field $V$ on $X$ assigns a tangent vector $V|_p\in T_pX$ to each point in $X$.
Let $f:X\longrightarrow \R$ be a function on $X$. The rate of change of 
$f$ along
the curve is measured by a derivative 
\begin{eqnarray*}
\frac{d}{d \ve}f(x(\ve))|_{\ve=0}&=&V^a\frac{\p f}{\p x^a}\\
&=&V(f)
\end{eqnarray*}
where 
\[
V=V^1\frac{\p}{\p x^1}+\ldots +V^n\frac{\p}{\p x^1}.
\]
Thus vector fields can be thought of as first order differential operators.
The derivations $\{ \frac{\p}{\p x^1}, \ldots,  \frac{\p}{\p x^n}\}$
at the point $p$ denote the elements of the basis of $T_p X$.

An integral curve $\gamma$ of a vector field $V$ is defined by
$\dot{\gamma}(\ve)=V|_{\gamma(\ve)} $or equivalently
\be
\label{ODEsFLOW}
\qquad
\frac{d x^a}{d \ve}=V^{a}(x).
\ee
This system of ODEs has a unique solution for each initial data, and the
integral curve passing through $p$ with coordinates
$x^a$ is called a flow  $\tilde{x}^a(\ve, x^b)$. 
The vector field $V$ is called a generator of the flow, as
\[
\tilde{x}^{a}(\ve, x)=x^{a}+\ve V^{a}(x)+O(\ve^2).
\]
Determining the flow of a given vector field comes down to solving a system of
ODEs (\ref{ODEsFLOW}).
\begin{itemize}
\item
{\bf Example.} Integral curves of the vector field
\[
V=x\frac{\p}{\p x}+\frac{\p}{\p y}
\]
on $\R^2$ are found by solving a pair of ODEs
$
\dot{x}=x, \, \dot{y}=1.
$
Thus 
\[
({x}(\ve), 
{y}(\ve))=(x(0)e^{\ve}, y(0)+\ve).
\]
There is one integral curve passing through each point in $\R^2$.
\end{itemize}
The flow is an example of 
one--parameter group of
transformations, as
\[
\tilde{x}(\ve_2, \tilde{x}(\ve_1, x))=\tilde{x}(\ve_1+\ve_2, x), \qquad
\tilde{x}(0, x)=x.
\]
An invariant of a flow is a function $f(x^a)$ such that 
 $f(x^a)=f(\tilde{x}^a)$ or equivalently
\[
V(f)=0
\]
where $V$ is the generating vector field.
\begin{itemize}
\item {\bf Example.}
The one parameter group $SO(2)$ of rotations on the plane 
\[
(\tilde{x}, \tilde{y})=(x\cos{\ve}-y\sin{\ve}, x \sin{\ve}+y \cos{\ve})
\]
is generated by 
\begin{eqnarray*}
V&=&\Big(\frac{\p \t{y}}{\p \ve}|_{\ve=0}\Big)\frac{\p}{\p y}
+\Big(\frac{\p \t{x}}{\p \ve}|_{\ve=0}\Big)\frac{\p}{\p x}\\
&=& x\frac{\p}{\p y}-y\frac{\p}{\p x}.
\end{eqnarray*}
The function $r=\sqrt{x^2+y^2}$ is an invariant of $V$.
\end{itemize}
A Lie bracket of two vector fields $V, W$ is a vector field $[V, W]$
defined by its action on functions
\be
\label{liebrac}
[V, W](f):=V(W(f))-W(V(f)).
\ee
The components of the Lie bracket are
\[
[V, W]^{a}=V^{b}\frac{\p W^{a}}{\p x^{b}}-W^{b}
\frac{\p V^{a}}{\p x^{b}}.
\]
From its definition the Lie bracket is bi-linear, antisymmetric and it
satisfies the Jacobi identity
\be
\label{jacobi_bracket}
[V, [W, U]]+[U,  [V, W]]+[W, [U, V]]=0.
\ee
A geometric interpretation of the Lie bracket is the infinitesimal
commutator of two flows.
To see it consider  
$\t{x}_1(\ve_1, x)$ and $\t{x}_2(\ve_2, x)$
which are the  flows of vector fields $V_1$ and $V_2$ respectively. For any 
$f:X\rightarrow \R$ define
\[
F(\ve_1, \ve_2, x):=f(\t{x}_1(\ve_1, (\t{x}_2(\ve_2, x))))-
f(\t{x}_2(\ve_2, (\t{x}_1(\ve_1, x)))).
\]
Then
\[
\frac{\p^2}{\p\ve_1\p\ve_2} F(\ve_1, \ve_2, x)|_{\ve_1=\ve_2=0}=[V_1, V_2](f).
\]
\begin{itemize}
\item {\bf Example.} Consider the 
three--dimensional Lie group {\bf Nil} of   $3$ by $3$ upper
triangular matrices
\[
g(m_1, m_2, m_3)=\left(\begin{array}{ccc}
1&m_1 &m_3\\
0&1&m_2\\
0&0&1
\end{array}
\right)
\]
acting  
on $\R^3$ by matrix multiplication
\[
\tilde{\bf {x}}=g(m_1, m_2, m_3){\bf x}=(x+m_1 y+m_3 z, y+m_2z, z).
\]
The corresponding vector fields\footnote{Note that
the lower index labels the vector fields while
the upper index labels the components. Thus 
$V_\alpha=V_\alpha^a\p/\p x^a$} $V_1, V_2, V_3$ are
\[
V_\alpha=\Big(\frac{\p\tilde{x}}{\p m_\alpha}\frac{\p}{\p \tilde{x}}
+\frac{\p\tilde{y}}{\p m_\alpha}\frac{\p}{\p \tilde{y}}+
\frac{\p\tilde{z}}{\p m_\alpha}\frac{\p}{\p \tilde{z}}\Big)|_{(m_1, m_2, m_3)=(0, 0, 0)}
\]
which gives
\[
V_1=y\frac{\p}{\p x},\quad V_2=z\frac{\p}{\p y}, \quad V_3=z\frac{\p}{\p x}.
\]
The Lie brackets of these vector fields are
\[
[V_1, V_2]=-V_3, \quad [V_1, V_3]=0, \quad [V_2, V_3]=0.
\]
Thus we have obtained the representation of the Lie algebra
of {\bf Nil} by vector fields on $\R^3$. Comparing this with the commutators of the matrices (\ref{matrices_M}) we see that
the structure constants only differ by an overall sign. 
The Lie algebra spanned by the vector fields $V_\alpha$ is isomorphic
to Lie algebra spanned by the matrices $M_\alpha$.
\item {\bf Example.} A driver of a car has two 
transformation at his disposal. These are generated by vector fields
\[
\mbox{STEER}=\frac{\p}{\p \phi}, \qquad \mbox{DRIVE}=
\cos{\th}\frac{\p}{\p x}+\sin{\th}\frac{\p}{\p y}+\frac{1}{L}
\tan{\phi}\frac{\p}{\p\theta}, \qquad 
L=\const.
\]
where $(x, y)$ are coordinates of the center of the rear axle, $\th$ specifies
the direction of the car, and $\phi$ is the angle between the front wheels and the direction of the car. These two flows don't commute, and 
\[
[\mbox{STEER}, \mbox{DRIVE}]=\mbox{ROTATE},
\]
where the vector field 
\[
\mbox{ROTATE}= \frac{1}{L\cos^2{\phi}}\frac{\p}{\p \theta}
\]
generates the manoeuvre 
steer, drive, steer back, drive back. This manoeuvre alone doesn't guarantee 
that the driver parks his car in a tight space. The commutator
\[
[\mbox{DRIVE}, \mbox{ROTATE}]=\frac{1}{L\cos^2{\phi}}
\Big(\sin{\th}\frac{\p}{\p x}-\cos{\th}\frac{\p}{\p y}\Big)=\mbox{SLIDE}
\]
is the key to successful parallel parking. One needs to perform
the following sequence
steer, drive, steer back, drive, steer, drive back, steer back, drive back!
\end{itemize}
In general  the Lie bracket is a closed operation in a set of the vector fields generating a group. The vector space of vector 
fields generating the group action
gives a representation of the corresponding Lie algebra.
The structure constants $f_{\alpha\beta}^{\gamma}$ do not depend on which of the representations (matrices or vector fields) is used.
\section{Symmetries of differential equations}
\label{sec_sym_DEs}
Let $u=u(x,t)$ be a solution to the KdV equation (\ref{kdv}). Consider a vector field
\[
V=\xi(x, t, u)\frac{\p}{\p x}+\tau(x, t, u)\frac{\p}{\p t}
+\eta(x, t, u)\frac{\p}{\p u}
\]
on the space of dependent and independent variables 
$\R\times \R^2 $. This vector field generates a one--parameter
group of transformations
\[
\tilde{x}=\tilde{x}(x, t, u, \ve), \qquad
\tilde{t}=\tilde{t}(x, t, u, \ve), \qquad
\tilde{u}=\tilde{u}(x, t, u, \ve).
\]
This group is called a symmetry of the KdV equation if 
\[
\frac{\p \tilde{u}}{\p \tilde{t}}-
6\tilde{u}\frac{\p \tilde{u}}{\p \tilde{x}}
+\frac{\p^3 \tilde{u}}{\p \tilde{x}^3}=0.
\]
The common  abuse of terminology is to refer to  the 
corresponding vector field as a symmetry,
although the term infinitesimal symmetry is more appropriate.
\begin{itemize}
\item {\bf Example.} An example of a symmetry of the KdV is given by
\[
\tilde{x}=x, \qquad \tilde{t}=t+\ve, \qquad  \tilde{u}=u.
\]
It is a symmetry as there is no explicit time dependence in the KdV.
Its generating  vector field is
\[
V=\frac{\p}{\p t}.
\]
\end{itemize}
Of course there is nothing special about KdV in this definition and
the concept of a symmetry applies generally to PDEs and ODEs. 
\begin{defi}
\label{defnition_of_Lie_point}
Let $X=\R^n\times \R$ be the space of independent and dependent
variables in a PDE.  A one--parameter group  of transformations
of this space
\[
\tilde{u}=\tilde{u}(x^a, u, \ve),\qquad
\tilde{x}^b=\tilde{x}^b (x^a, u, \ve)
 \]
is  called a Lie point symmetry (or symmetry for short) 
group of a PDE
\be
\label{PDEs}
F[u, \frac{\p u}{\p x^a}, \frac{\p^2 u}{\p x^a\p x^b}, \ldots]=0
\ee
if  its action transforms solutions  to other solutions
i.e.
\[
F[\tilde{u}, \frac{\p \tilde{u}}{\p \tilde{x}^a}, 
\frac{\p^2 \tilde{u}}{\p \tilde{x}^a\p \tilde{x}^b}, \ldots]=0.
\]
\end{defi}
This definition naturally extends to multi--parameter groups of transformation.
A Lie group $G$ is a symmetry of a PDE if any of its one--parameter subgroups
is a symmetry in a sense of Definition \ref{defnition_of_Lie_point}.

A knowledge of  Lie point symmetries is useful for the following reasons
\begin{itemize}
\item It allows to use known solutions to construct new solutions.

{\bf Example.} The Lorentz group 
\[
(\tilde{x}, \tilde{t})=\Big(
\frac{x-{\varepsilon}t}{\sqrt{1-{\varepsilon}^2}}, 
\frac{t-{\varepsilon}x}{\sqrt{1-{\varepsilon}^2}}\Big), 
\qquad \varepsilon\in(-1, 1)
\]
is the symmetry group of the Sine--Gordon equation (\ref{sine_gordon}).   
Any $t$--independent solution $\phi_S({x})$
to (\ref{sine_gordon}) can be used to 
obtain a time dependent solution
\[
\phi({ x}, t)
=\phi_S\Big(\frac{x-{\varepsilon}t}{\sqrt{1-{\varepsilon}^2}}\Big), \qquad
{\varepsilon}\in (-1, 1).
\]
In physics this procedure is known as `Lorentz boost'.
The parameter $\varepsilon$ is usually denoted by $v$ and called velocity.
For example the Lorentz boost of a static kink is a moving kink. 
\item
For ODEs each  symmetry reduces the order by 1. So a knowledge of a
sufficiently many symmetries allows a construction of the most general
solution. 

{\bf Example.} 
An ODE 
\[
\frac{d u}{d x}=F\Big(\frac{u}{x}\Big)
\]
admits a scaling symmetry 
\[
(x, u)\longrightarrow  (e^\ve x,  e^\ve u), \qquad \ve\in \R.
\]
This one--dimensional group is generated by a vector field
\[
V=x\frac{\p}{\p x}+u\frac{\p}{\p u}.
\]
Introduce the invariant coordinates
\[
r=\frac{u}{x}, \qquad s=\log{|x|}
\]
so that
\[
V(r)=0, \qquad V(s)=1.
\]
If $F(r)=r$ the general solution
is $r=\co$. Otherwise
\[
\frac{ds}{dr}=\frac{1}{F(r)-r}
\]
and  the general implicit solution is
\[
\log{|x|}+c=\int^{\frac{u}{x}}\frac{dr}{F(r)-r}.
\]
\item 
For PDEs the knowledge of the symmetry group 
is not sufficient to construct the most
general solution, but it can be used to find special solutions which
admit symmetry. 

{\bf Example.} Consider the one--parameter group of
transformations
\[
(\tilde{x}, \tilde{t}, \tilde{u})=(x+c\varepsilon, t+\varepsilon, u)
\]
where $c\in\R$ is a constant. It is straightforward to verify that
this group is a Lie point symmetry of the KdV equation (\ref{kdv}).
It is generated by the vector field
\[
V=\frac{\p}{\p t}+c\frac{\p}{\p x}
\]
and the corresponding invariants are $u$ and $\xi= x-ct$.
To find the group invariant solutions 
assume that a solution of the KdV equation 
is of the form 
\[
u(x, t)=f(\xi).
\]
Substituting this to the KdV yields a third order ODE which
easily integrates to
\[
\frac{1}{2}\Big(\frac{d f}{d\xi}\Big)^2=f^3+\frac{1}{2}cf^2+\alpha f+\beta
\]
where $(\alpha, \beta)$ are arbitrary constants. This
ODE is solvable in terms of an elliptic integral, which gives
all group invariant solutions in the implicit form
\[
\int\frac{df}{\sqrt{f^3+\frac{1}{2}cf^2+\alpha f+\beta}}=\sqrt{2}\xi.
\]
Thus we have recovered the cnoidal wave which in Section 
\ref{Section_finite_gap}
arose from the finite gap integration.
In fact the one--soliton solution (\ref{1-soliton}) falls into this 
category:
if $f$ and its first two derivatives tend to zero 
as $|\xi|\rightarrow\infty$ then $\alpha, \beta$
are both zero and the elliptic integral reduces
to an elementary one. Finally we obtain
\[
u(x, t)=-\frac{2\chi^2}{\cosh^2{\chi(x-4\chi^2 t-\phi_0)}}
\]
which is the one--soliton solution (\ref{1-soliton}) to the KdV equation.
\end{itemize}
\subsection{How to find symmetries}
Some of them can be guessed. For example if there 
is no explicit dependence of 
independent coordinates in the equation then the translations
$\tilde{x}^a={x}^a+c^a$ are symmetries. All translations form an 
$n$--parameter abelian group generated by $n$ vector fields
$\p/\p x^a$.

In the general case of (\ref{PDEs}) we could substitute
\[
\tilde{u}=u+\varepsilon\eta(x^a, u)+O(\varepsilon^2),\qquad
\tilde{x}^b={x}^b+\varepsilon\xi^b (x^a, u)+O(\varepsilon^2)
\]
into the equation (\ref{PDEs}) and keep the terms linear in $\varepsilon$.
A more systematic method is given by the {\em prolongation} of vector field.
Assume that the space of independent variables is coordinatised by $(x, t)$
and 
the equation $(\ref{PDEs})$ is of the form
\[
F(u, u_x, u_{xx}, u_{xxx}, u_t)=0.
\]
(for example KdV is of that form). The prolongation of the vector field
\[
V=\xi(x, t, u)\frac{\p}{\p x}+\tau(x, t, u)\frac{\p}{\p t}
+\eta(x, t, u)\frac{\p}{\p u}
\]
is
\[
\pr(V)=V+\eta^t\frac{\p}{\p u_t}+\eta^x\frac{\p}{\p u_x}
+\eta^{xx}\frac{\p}{\p u_{xx}}+\eta^{xxx}\frac{\p}{\p u_{xxx}},
\]
where $(\eta^t, \eta^x, \eta^{xx}, \eta^{xxx})$ are certain functions
of $(u, x, t)$ which can be determined algorithmically in terms of 
$(\xi, \tau, \eta)$ and their derivatives
(we will do it in the next Section). The prolongation $\pr(V)$ generates a one--parameter group of 
transformations on the 7--dimensional space
with coordinates
\[
(x, t, u, u_t, u_x, u_{xx}, u_{xxx}).
\]
(This is an example of a jet space. The symbols 
$(u_t, u_x, u_{xx}, u_{xxx})$ should be regarded as independent coordinates
and not as derivatives of $u$.). The vector field $V$ is a symmetry of 
the PDE if
\be
\label{criterion}
\pr(V)(F)|_{F=0}=0.
\ee
This condition gives a linear system of PDEs for $(\xi, \tau, \eta)$.
Solving this system yields the most general symmetry of a given PDE.
The important point is that (\ref{criterion}) is only required to hold
when (\ref{PDEs}) is satisfied (`on shell' as a physicist would put it).
\subsection{Prolongation formula}
The first step in implementing the prolongation procedure is
to determine the functions \[\eta^t, \eta^x, \eta^{xx}, \dots\] 
in the prolonged vector field.
For simplicity we shall assume that we want to determine
a symmetry of $N$th order ODE 
\[
\frac{d^N u}{d x^N}=F\Big(x, u, \frac{d u}{d x}, \cdots,  
\frac{d^{N-1} u}{d x^{N-1}}\Big).
\]
Consider a vector field
\[
V=\xi\frac{\p}{\p x}+\eta\frac{\p}{\p u}.
\]
Its prolongation
\[
\pr(V)=V+\sum_{k=1}^N\eta^{(k)}\frac{\p}{\p u^{(k)}}
\]
generates a one--parameter transformation group 
\[
\tilde{x}=x+\varepsilon\xi +O(\ve^2), \qquad
\tilde{u}=u+\varepsilon\eta+O(\ve^2), \qquad \tilde{u}^{(k)}
=u^{(k)}+\varepsilon\eta^{(k)}+O(\ve^2)
\]
of the $(N+2)$ dimensional jet space with coordinates 
$(x, u, u', \dots, u^{N})$.

The prolongation is an algorithm for the calculation of the 
functions $\eta^{(k)}$. Set
\[
D_x=\frac{\p}{\p x}+u'\frac{\p}{\p u}+u^{''}\frac{\p}{\p u'}+\cdots
+u^{(N)}\frac{\p}{\p u^{(N-1)}}.
\]
The chain rule gives
\[
\tilde{u}^{(k)}=\frac{d\tilde{u}^{(k-1)}}{d \tilde{x}}=
\frac{D_x\tilde{u}^{(k-1)}}{D_x\tilde{x}},
\]
so
\[
\tilde{u}^{(1)}=\frac{D_x\tilde{u}}{D_x\tilde{x}}=
\frac{\frac{d u}{dx}+\varepsilon D_x(\eta) +\dots}{1+\varepsilon D_x(\xi)+\dots}
=\frac{d u}{dx}+\varepsilon(D_x\eta-\frac{d u}{dx} D_x\xi)+O(\varepsilon^2).
\]
Thus
\[
\eta^{(1)}=D_x\eta-\frac{d u}{dx} D_x\xi.
\]
The remaining prolongation coefficients can now be constructed recursively:
The relation
\[
\tilde{u}^{(k)}= 
\frac{u^{(k)}+\varepsilon D_x(\eta^{(k-1)}) }{1+\varepsilon D_x(\xi)}
\]
yields the general prolongation formula
\be
\label{prolongation_formula}
\eta^{(k)}=D_x(\eta^{(k-1)})- \frac{d^k u}{d x^k}D_x\xi.
\ee
The procedure is entirely analogous
for PDEs, where $u=u(x^a)$ but one has to keep track of the index $a$ labelling
the independent variables. Set
\[
D_a=\frac{\p}{\p x^a}+(\p_a u)\frac{\p}{\p u}+(\p^2_au)\frac{\p}{\p (\p_a u)}
+\cdots+(\p^N_au)\frac{\p}{\p (\p^{N-1}_au)},
\]
where 
\[
\p_a^k u=\frac{\p^k u}{\p (x^a)^k}.
\]
The first prolongation is
\[
\eta^{(a)}=D_a\eta-\sum_{b=1}^n(D_a\xi^b)\frac{\p u}{\p x^b}
\]
and the higher prolongations are given recursively by the formula
\[
\eta^{A, a}=D_a\eta^A-\sum_{b=1}^n(D_a\xi^b)\frac{\p u^A}{\p x^b}
\]
where $A=(a_1, \dots, a_k)$ is a multi-index and 
\[
u^A=\frac{\p^k u}{\p x^{a_1}\p x^{a_2}\ldots\p x^{a_k}}.
\]
\begin{itemize}
\item 
{\bf Example.}
Let us follow the prolongation procedure 
to find the most general Lie--point symmetry of 
the second order ODE
\[
\frac{ d^2 u}{d x^2}=0.
\]

We first need to compute the second prolongation
\[
\pr(V)=\xi\frac{\p}{\p x}+\eta\frac{\p}{\p u}+\eta^x\frac{\p}{\p {u_x}}
+\eta^{xx}\frac{\p}{\p {u_{xx}}}.
\]
This computation does not depend on the details of the equation but only
on the prolongation formulae (\ref{prolongation_formula}).
The result  is
\[
\eta^{x}=\eta_x+(\eta_u-\xi_x)u_x-\xi_u u_x^2,
\]
\[
\eta^{xx}=\eta_{xx}+(2\eta_{xu}-\xi_{xx})u_x+(\eta_{uu}-2\xi_{xu})u_x^2
-\xi_{uu}u_x^3+(\eta_u-2\xi_x)u_{xx}-3\xi_u u_x u_{xx}.
\]
Now we substitute this, and the ODE to the symmetry criterion 
(\ref{criterion})
\[
\pr(V)(u_{xx})=\eta^{xx}=0.
\]
Thus
\[
\eta_{xx}+(2\eta_{xu}-\xi_{xx})u_x+(\eta_{uu}-2\xi_{xu})u_x^2
-\xi_{uu}u_x^3=0
\]
where we have used the ODE to set $u_{xx}=0$.
In the second order equation the value of 
$u_x$ can be prescribed in an arbitrary way
at each point (initial condition). Therefore the coefficients of $u_x, u_x^2$
and $u_{x}^3$ all vanish 
\[
\eta_{xx}=0, \quad 2\eta_{xu}-\xi_{xx}=0, \quad
\eta_{uu}-2\xi_{xu}=0, \quad \xi_{uu}=0.
\]
The general solution of these linear PDEs is
\begin{eqnarray*}
\xi(x, u)&=& \ve_1x^2+\ve_2xu+\ve_3x+\ve_4u+\ve_5, \\
\eta(x, u)&=&\ve_1xu+\ve_2u^2+\ve_6x+\ve_7u+\ve_8.
\end{eqnarray*}
Therefore the trivial ODE in our example admits an eight dimensional
group of symmetries. 

Let $V_\alpha, \alpha=1, \dots, 8$ be the corresponding vector fields
obtained by setting $\ve_{\alpha}=1$ and $\ve_{\beta}=0$ if $\beta\neq\alpha$
\[
V_1=x^2\frac{\p}{\p x}+xu\frac{\p}{\p u}, \quad
V_2=xu\frac{\p}{\p x}+u^2\frac{\p}{\p u}, \quad
V_3=x\frac{\p}{\p x}, \quad 
V_4=u\frac{\p}{\p x},
\]
\[
V_5=\frac{\p}{\p x}, \quad V_6=x\frac{\p}{\p u}, \quad
V_7=u\frac{\p}{\p u},  \quad V_8=\frac{\p}{\p u}.
\]
Each of the eight  vector fields generates a one--parameter 
group of transformations. Calculating the Lie brackets of these vector 
fields verifies that they form a Lie algebra 
of $PGL(3, \R)$.

It is possible to show that the Lie point symmetry group of a general second order ODE has dimension at
most 8. If this dimension is 8 then the ODE is equivalent
to $u_{xx}=0$ by a coordinate transformation
$u\rightarrow U(u, x), x\rightarrow X(u, x)$.
\end{itemize}
This example shows that the process of prolonging the vector fields
and writing down the linear PDEs characterising the symmetries
is tedious but algorithmic. It is worth doing a few examples 
by hand to familiarise oneself with the method
but in practice it is best to use computer
programmes like MAPLE or MATHEMATICA to do symbolic computations.
\begin{itemize}
\item {\bf Example.}  Lie point symmetries of KdV.
The vector fields
\[
V_1=\frac{\p}{\p x}, \quad V_2=\frac{\p}{\p t}, \quad
V_3=\frac{\p}{\p u}-6t\frac{\p}{\p x}, \quad
V_4=x \frac{\p}{\p x}+3t\frac{\p}{\p t}
-2 u\frac{\p}{\p u}
\]
generate a  four--parameter symmetry group of KdV. The group is
non--abelian as the structure constants of the Lie algebra
spanned by $V_\alpha$ are non--zero:
\[
[V_2, V_3]=-6V_1, \quad [V_1, V_4]=V_1, \quad [V_2, V_4]=3V_2, \quad
[V_3, V_4]=-2V_3
\]
and all other Lie brackets vanish.

One can  use the prolongation procedure to show that 
this is in fact the most general symmetry group of KdV. 
One needs to find a third prolongation of
a general vector field on $\R^3$ - this can be done
`by hand' but it is best to use {\tt MAPLE} package
{\tt liesymm} with the command {\tt determine}.
Type {\tt help(determine);} and take it from there.
\end{itemize}
\section{Painlev\'e equations}
In this Section 
we shall consider ODEs in complex domain. 
This means that both the dependent and 
independent variables are complex numbers. Let us first discuss the 
linear ODEs of the form
\be
\label{linear_ODE}
\frac{d ^N w}{d z^N}+p_{N-1}(z)\frac{d ^{N-1} w}{d z^{N-1}}
+\cdots +p_{1}(z)\frac{d  w}{d z}+p_0(z)w=0
\ee
where $w=w(z)$. If the functions $p_0, \dots, p_{N-1}$ are analytic at 
$z=z_0$, then $z_0$ is called a regular point and for a given initial data 
there exist
a unique analytic solution in the form of a power series
\[
w(z)=\sum_k a_k(z-z_0)^k.
\]
The singular points of the ODE (\ref{linear_ODE}) can be located
only at the singularities of $p_k$. Thus the singularities are fixed -- their 
location does not depend on the initial conditions.
Nonlinear ODEs lose this property. 
\begin{itemize}
\item {\bf Example.} Consider
a simple nonlinear ODE and its general solution
\[
\frac{d w}{d z}+w^2=0, \qquad w(z)=\frac{1}{z-z_0}.
\]
The location of the singularity depends on the constant of 
integration $z_0$.
This is a movable singularity.
\end{itemize}
A singularity of a nonlinear ODE can be a pole (of arbitrary order), 
a branch point or an essential singularity. 
\begin{itemize}
\item
{\bf Example.} The ODE with the general solution
\[
\frac{d w}{dz}+w^3=0, \quad w(z)=\frac{1}{\sqrt{2(z-z_0)}}.
\]
has a movable singularity which is a branch point.
Another example with a movable logarithmic branch point is
\[
\frac{d w}{dz}+e^w=0, \quad w(z)={\ln{(z-z_0)}}.
\]
\end{itemize}
\begin{defi}
\label{painleve_prop}
The ODE
\[
\frac{d ^N w}{d z^N}=F\Big(\frac{d ^{N-1} w}{d z^{N-1}}, \cdots,
\frac{d  w}{d z}, w, z\Big)
\]
where $F$ is rational in $w$ and its derivatives
has Painlev\'e property if its movable singularities are at worst poles.
\end{defi}
In 19th century Painlev\'e, Gambier and Kowalewskaya 
aimed to  classify all second order ODEs with
Painlev\'e property up to a change of variables
\[
\tilde{w}(w, z)=\frac{a(z)w+b(z)}{c(z)w+d(z)}, \quad \tilde{z}(z)=\phi(z)
\]
where the functions $a, b, c, d, \phi$ are analytic in $z$.
There exist 50 canonical types  44  of which are solvable in terms
of `known' functions (sinus, cosinus, elliptic functions or in general
solutions to linear ODEs) \cite{Ince}. The remaining 6 equations define 
new transcendental functions
\begin{eqnarray}
\label{painleve_equations}
\frac{d^2 w}{d z^2}&=&6w^2+z \quad \mbox{PI},\\
\frac{d^2 w}{d z^2}&=&2w^3+wz+\alpha \quad \mbox{PII},\nonumber\\
\frac{d^2 w}{d z^2}&=&\frac{1}{w}\Big(\frac{d w}{d z}\Big)^2  -\frac{1}{z}\frac{d w}{d z} +\frac{\alpha w^2+\beta}{z}+\gamma w^3 
+\frac{\delta}{w} \quad \mbox{PIII},\nonumber\\
\frac{d^2 w}{d z^2}&=&\frac{1}{2w}\Big(\frac{d w}{d z}\Big)^2 +\frac{3}{2}w^3
+4zw^2+2(z^2-\alpha)w+\frac{\beta}{w} \quad \mbox{PIV},\nonumber\\
\frac{d^2 w}{d z^2}&=&\left(\frac{1}{2w}
+\frac{1}{w-1}\right)\Big(\frac{d w}{d z}\Big)^2-\frac{1}{z}\frac{d w}{d z}
+\frac{(w-1)^2}{z^
2}\left(\alpha w +\frac{\beta}{w}\right)\nonumber\\
&&+\frac{\gamma w}{z}+\frac{\delta 
w(w+1)}{w-1} \quad \mbox{PV},\nonumber\\
\frac{d^2 w}{d z^2}&=& \frac{1}{2}\left(\frac{1}{w}+\frac{1}{w-1}
+\frac{1}{w-z}\right)\Big(\frac{d w}{d z}\Big)^2
-\left(\frac{1}{z}+\frac{1}{z-1}+\frac{1}{w-z}\right)
\frac{d w}{d z}\nonumber\\
& &+\frac{w(w-1)(w-z)}{z^2(z-1)^2}\left(\alpha +\beta\frac{z}{w^2}+
\gamma\frac{z-1}{(w-1)^2}+\delta\frac{z(z-1)}{(w-z)^2}\right)\quad 
\mbox{PVI}.\nonumber
\end{eqnarray}
Here $\alpha, \beta, \gamma, \delta$ are constants. Thus PVI belongs to
a four--parameter family of ODEs but PI is rigid up to coordinate 
transformations.

How to check the Painlev\'e property for a given ODE? 
If a second order equation
possesses the Painlev\'e property then it is either linearisable
or can be put into one of the six Painlev\'e types by appropriate
coordinate transformation. Exhibiting such transformation
is often the most straightforward way of establishing the Painlev\'e property.

Otherwise, especially if we suspect that the equation does not have
Painlev\'e property,  the singular point analysis may be performed.
If a general $N$th order ODE possesses the PP then the general
solution admits a Laurent expansion with a finite number of terms with 
negative powers. This expansion must contain $N$ arbitrary constants so that
the initial data consisting of $w$ and its first $(N-1)$
derivatives can be specified at any point.
Assume that a leading term in the expansion of the 
solution is of the form
\[
w(z)\sim a(z-z_0)^p, \qquad a\neq 0, \quad a, p\in \C
\]
as $z\rightarrow z_0$. Substitute this to the ODE and require the maximal
balance condition. This means that two (or more) terms must be of equal maximally small order as $(z-z_0)\rightarrow 0$. This should determine
$a$ and $p$ and finally the form of a solution around $z_0$. If $z_0$ is
a singularity we should also be able to determine if it is movable or
fixed.
\begin{itemize}
\item
{\bf Example.}  Consider the ODE
\[
\frac{d w}{d z}=w^3+z.
\]
The maximal balance condition gives
\[
ap(z-z_0)^{p-1}\sim a^3(z-z_0)^{3p}.
\]
Thus $p=-1/2, a=\pm i\sqrt{2}^{-1}$ and
\[
w(z)\sim \pm i\frac{\sqrt{2}}{2}(z-z_0)^{-1/2}
\]
possesses a movable branch point as $z_0$ depends on the initial conditions.
The ODE does not have  Painlev\'e property.
\vskip5pt
\item {\bf Example.} Consider the first Painlev\'e equation
\[
\frac{d^2 w}{d z^2}=6w^2+z.
\]
The orders of the three terms in this equations
are
\[
p-2, \qquad 2p, \qquad 0.
\]
Balancing the last two terms gives $p=0$ but this is not the maximal balance
as the first term is then of order $-2$. Balancing the first and last terms
gives $p=2$. This is a maximal balance and the corresponding solution
is analytic around $z_0$. Finally balancing the first two terms
gives $p=-2$ which again is the maximal balance: the `balanced' terms
behave like $(z-z_0)^{-4}$ and the remaining term is of order $0$.
Now we find that $a=a^2$ and so $a=1$
and
\[
w(z)\sim\frac{1}{(z-z_0)^2}.
\]
Thus the movable singularity is a second order pole.
\end{itemize}

This singular point analysis is good to rule out  Painlev\'e property,
but does not give sufficient conditions (at least not in the heuristic
form in which we presented it), as some singularities may have been missed or the Laurent series may be divergent.
The analysis of sufficient conditions is tedious and complicated - 
we shall leave it out.
\vskip10pt

The Painlev\'e property guarantees that the solutions of six
Painlev\'e equations are single valued thus giving rise to proper functions.
The importance of the Painlev\'e equations is that they define new
transcendental functions in the following way: Any sufficiently smooth 
function can be defined as a solution to certain DE. For example
we can define the exponential function as a general solution to
\[
\frac{d w}{d z}=w
\]
such that $w(0)=1$.
Similarly  we define the function PI from the general solution
of the first Painlev\'e equation. From this point of view
the exponential and PI functions are on equal footing. Of course
we know more about the exponential as it possesses
simple properties and arises in a wide range of problems
in natural sciences. We know much less about PI, but it is largely because
we have not yet been bothered to reveal its properties.

The irreducibility of the Painlev\'e equations is a more subtle issue. It roughly
means the following. One can define a field of classical functions by
starting off with the rational functions $Q[z]$  and adjoining those 
functions which arise as solutions of algebraic or linear differential 
equations with coefficients in $Q[z]$. For example the  
exponential, Bessel function, hyper-geometric function  are all solutions of linear DEs, and thus are classical. A function is called irreducible 
(or transcendental) if it is
not classical. Painlev\'e himself anticipated that the Painlev\'e equations
define irreducible functions but the rigorous proofs  for PI and PII 
appeared only recently. They use a far reaching extension of Galois theory
from the number fields to differential 
fields of functions. The irreducibility problem is analogous
to existence of non--algebraic numbers (numbers which are not
roots of any polynomial equations with rational coefficients).
Thus the the appearance of Galois theory is not that surprising.
\subsection{Painlev\'e test}
How to determine whether a given PDE is integrable? It is to large extend
an open problem as the satisfactory definition of integrability of PDEs 
is still missing. The following algorithm is based on the observation 
of Ablowitz, Ramani and Segur \cite{ARS} (see also \cite{AC92})
that
PDEs integrable by inverse scattering transform reduce
(when the solutions are required to be invariant under some Lie
symmetries) to ODEs with Painlev\'e property. 
\begin{itemize}
\item
{\bf Example}. Consider a Lie point symmetry 
\[
(\tilde{\rho}, \tilde{\tau})=(c\rho, \frac{1}{c}\tau), \qquad c\neq 0
\]
of the 
Sine--Gordon equation
\[
\frac{\p^2 \phi}{\p \rho\p\tau}=\sin{\phi}.
\]
The  group invariant solutions are of 
the form $\phi(\rho, \tau)= F(z)$ where $z=\rho\tau$ is an invariant
of the symmetry. Substituting $w(z)=\exp{(i F(z))}$ to the Sine--Gordon yields 
\[
\frac{d^2 w}{d z^2}=\frac{1}{w}\Big(\frac{d w}{d z}\Big)^2  -\frac{1}{z}\frac{d w}{d z} +\frac{1}{2}\frac{w^2}{z} -\frac{1}{2z},
\]
which is the  third Painlev\'e equation PIII with the special values
of parameters
\[
\alpha=\frac{1}{2}, \quad\beta=-\frac{1}{2}, \quad
\gamma=0, \quad \delta=0.
\]
\item {\bf Example.}
Consider the modified KdV equation
\[
v_t-6v^2v_x+v_{xxx}=0,
\]
and look for  a Lie point symmetry  of the form
\[
(\tilde{v}, \tilde{x}, \tilde{t})=(c^\alpha v, c^{\beta} x, c^{\gamma} t),
\qquad c\neq 0.
\]
The symmetry condition will hold if all three terms in the equation
have equal weight
\[
\alpha-\gamma=3\alpha-\beta=\alpha-3\beta.
\]
This gives $\beta=-\alpha, \gamma=-3\alpha$ where $\alpha$ can be chosen 
arbitrarily. The corresponding symmetry group depends on one parameter
$c^\alpha$ and is generated by
\[
V=v\frac{\p}{\p v}-x\frac{\p}{\p x}-3t\frac{\p}{\p t}.
\]
This has two independent invariants
\[
z= (3)^{-1/3}\,x\,t^{-1/3}, \qquad
w=(3)^{1/3}\,v\, t^{1/3}
\]
where the constant factor $(3)^{-1/3}$ has been added for
convenience. The group invariant solutions are of the form $w=w(z)$
which gives
\[
v(x, t)=(3t)^{-1/3} w(z).
\]
Substituting this to the modified KdV equation leads to 
a third order ODE for $w(z)$
\[
w_{zzz}-6w^2w_z-w-zw_z=0.
\]
Integrating this ODE once shows that  $w(z)$ satisfies 
the second Painlev\'e equation PII with general value of the parameter
$\alpha$.
\end{itemize}
The general  Painlev\'e test comes down to the following algorithm:
Given a PDE
\begin{enumerate}
\item Find all its Lie point symmetries.
\item Construct ODEs characterising the group invariant solutions.
\item Check for Painlev\'e property.
\end{enumerate}
This procedure only gives necessary conditions for integrability. 
If all reduction possess Painlev\'e property the PDE does not have
to be integrable in general.
\chapter{Exercises}
\section{Example sheet 1}
\begin{enumerate}
\item {\bf Jacobi identity.}
Assume that $(p_j, q_j)$ satisfy the Hamilton equations and show that any
function $f=f(p, q, t)$ satisfies
\[
\frac{d f}{d t}=\frac{\p f}{\p t}+\{f, H\},
\]
where $H$ is the Hamiltonian.

Show that the Jacobi identity 
\be
\label{jacobi_sheet_1}
\{f_1,\{ f_2, f_3\}\}+
\{f_3,\{ f_1, f_2\}\}+\{f_2,\{ f_3, f_1\}\}=0
\ee
holds for Poisson brackets.

Deduce that if functions $f_1$ and $f_2$ which do not explicitly depend on 
time are first integrals of 
a Hamiltonian system then so is $f_3=\{f_1, f_2\}$.
\item {\bf Canonical transformations.}
\begin{itemize}
\item Find the canonical transformation
generated by
\[
S=\sum_{k=1}^nq_kP_k.
\]
\item Show that the canonical transformations preserve volume in the 
two--dimensional phase space, i.e.
\[
\frac{\p (P, Q)}{\p (p, q)}=1.
\]
[This result also holds in phase spaces of arbitrary dimension.]
\item
Show that the transformation
\[
Q=\cos{(\beta)}q-\sin{(\beta)}p, \quad
P=\sin{(\beta)}q+\cos{(\beta)}p
\]
is canonical for any constant $\beta\in\R$. Find the corresponding generating
function. Is it defined for all $\beta?$

\end{itemize}
\item {\bf Action variables for the Kepler problem.}
Consider the four--dimensional phase space coordinatised by 
\[q_1=\phi,\quad q_2=r,\quad p_1=p_\phi,\quad p_2=p_r\] equipped with a Hamiltonian 
\[
H=\frac{{p_\phi}^2}{2r^2}+\frac{{p_r}^2}{2}-\frac{\alpha}{r}
\]
where $\alpha>0$ is a constant. Use the fact
that $\p_\phi H=0$ to show the existence of two first integrals in involution and deduce 
that this system is integrable in a sense of the Arnold--Liouville theorem.

Construct the action  variables.  Express the Hamiltonian 
in terms of the action variables to show that the frequencies associated to the corresponding angles are equal.

[Hint:  $\phi$ and one function of $(r, p_r)$ parametrise $M_f$. Varying $\phi$
and fixing the other coordinate gives one cycle $\Gamma_\phi\subset M_f$.
To find the second action coordinate fix $\phi$ (on top of $H$ and $p_\phi$).]

\item {\bf Poisson Structures.}
A Poisson on structure on $\R^{2n}$ is an anti--symmetric matrix $\omega^{ab}$ whose
components depend on the coordinates $\xi^a\in\R^{2n}, a=1,\cdots, 2n$
and such that the Poisson bracket
\[
\{f, g\}=\sum_{a,b=1}^{2n}\omega^{ab}(\xi)\frac{\p f}{\p \xi^a}\frac{\p g}{\p \xi^b}
\]
satisfies the Jacobi identity 
(\ref{jacobi_sheet_1}). 

Show that
\[
\{fg, h\}=f\{g, h\}+\{f, h\}g.
\]

Assume that the matrix $\omega$ is invertible
with $W:=(\omega^{-1})$ and show that the antisymmetric matrix
$W_{ab}(\xi)$ satisfies 
\be
\label{sheet_closure_1}
\p_{a} W_{bc} +\p_{c} W_{ab}+\p_{b} W_{ca}  =0.
\ee
[Hint: note that $\omega^{ab}=\{\xi^a, \xi^b\}$.]
Deduce that if $n=1$ then any antisymmetric invertible matrix
$\omega(\xi^1, \xi^2)$ gives rise to a Poisson structure (i.e. show that
the Jacobi identity holds automatically in this case).

[In differential geometry the invertible antisymmetric
 matrix $W$ which satisfies (\ref{sheet_closure_1}) is called 
a symplectic structure. We have therefore deduced that symplectic structures 
are special cases of Poisson structures.]

\item {\bf KdV.} 
Verify that the equation
\[
\frac{1}{v}\Psi_t+\Psi_x+\beta\Psi_{xxx}+\alpha\Psi\Psi_x=0.
\]
where $\Psi=\Psi(x, t)$ and  $(v, \beta, \alpha)$
are non--zero constants is equivalent to the KdV equation
\be
\label{kdv_sheet_1}
u_t-6uu_x+u_{xxx}=0, \qquad u=u(x,t)
\ee
after a suitable change of dependent and independent variables.
\item {\bf 1--soliton solution.}
Assume that a solution of the KdV equation (\ref{kdv_sheet_1})
is of the form 
\[
u(x, t)=f(\xi), \qquad \mbox{where}\quad \xi=x-ct
\]
for some constant $c$. Show that the function $f(\xi)$ satisfies the ODE
\[
\frac{1}{2}(f')^2=f^3+\frac{1}{2}cf^2+\alpha f+\beta
\]
where $(\alpha, \beta)$ are arbitrary constants. Assume that $f$ and its
first two derivatives tend to zero as $|\xi|\rightarrow\infty$ and solve
the ODE to construct 
the one--soliton solution to the KdV equation.

\item{\bf Sine--Gordon soliton from Backlund transformations.}

The Sine--Gordon equation is 
\[
\phi_{xx}-\phi_{tt}=\sin{(\phi)}, \qquad \phi=\phi(x, t).
\]
Set $\tau=(x+t)/2, \rho=(x-t)/2$ and consider the B\"acklund transformations 
\[
\p_\rho(\phi_1-\phi_0)=2b\sin{\Big(\frac{\phi_1+\phi_0}{2}\Big)}, \qquad
\p_\tau(\phi_1+\phi_0)=2b^{-1}\sin{\Big(\frac{\phi_1-\phi_0}{2}\Big)}, 
\]
where $b=\mbox{const}$ and $\phi_0, \phi_1$ are functions of $(\tau, \rho)$.
Take $\phi_0=0$ and construct the 1-soliton (kink) solution
$\phi_1$.

Draw the graph of $\phi_1(x, t)$ for a fixed value of $t$. What happens
when $t$ varies?
\item {\bf Miura transformation}. Let $v=v(x, t)$ 
satisfy the modified KdV equation
\[
v_t-6v^2v_x+v_{xxx}=0.
\]
Show that the function $u(x, t)$ given by
\be
\label{Miura_1}
u=v^2+v_x
\ee
satisfy the  KdV equation. Is it true that any solution $u$ to the KdV equation gives rise, via 
(\ref{Miura_1}), to a solution of the modified KdV equation?
\end{enumerate}

\section{Example sheet 2}
\begin{enumerate}

\item{\bf Lax pair.} 
Consider a one-parameter family of self--adjoint operators
$L(t)$  in some complex inner product space 
such that \[
L(t)=U(t)L(0)U(t)^{-1}\] 
where $U(t)$ is a unitary
operator, i.e. $U(t)U(t)^*=1$ where $U^*$ is the adjoint of $U$.

Show that $L(t)$ and $L(0)$ have the same eigenvalues. Show that
there exist an anti-self-adjoint operator $A$ such that
$U_t=-A U$ and
\[
L_t=[L, A].
\] 
\item {\bf Lax pair for KdV.} Show that the
The KdV equation  is equivalent to
\[
L_t=[L, A]
\]
where the Lax operators are 
\[
L=-\frac{d^2}{dx^2}+u, \quad A=4\frac{d^3}{dx^3}
-3\Big(u\frac{d}{dx}+\frac{d}{dx}u\Big), \qquad u=u(x, t).
\]
\item {\bf Dispersionless limit.}
Let $(L, A)$ be the KdV Lax pair.  Set $u(x, t)=U(X, T)$.
Substitute
\[
\frac{\p}{\p x}=\varepsilon\frac{\p}{\p X}, \quad
\frac{\p}{\p t}=\varepsilon\frac{\p}{\p T},
\]
and consider the operators acting on functions of the form
\[
\psi(x, t)=\exp{(\varepsilon^{-1}\Psi(X, T))}.
\]
Show that in the limit  $\varepsilon\longrightarrow 0$
the commutators of differential 
operators are replaced by the Poisson brackets 
according to the relation
\[
\frac{\p^k}{\p x^k}\psi\longrightarrow (\Psi_X)^k\psi, \qquad
[{ L}, { A}]\longrightarrow
\frac{\p {\cal L}}{\p p}\frac{\p {\cal A}}{\p x}-\frac{\p {\cal L}}{\p x}\frac{\p {\cal A}}{\p p}
=-\{{\cal L}, {\cal A}\}, 
\]
where $p=\Psi_X$ and 
\[
{\cal L}=-p^2+U(X, T), \quad {\cal A}=4p^3-6U(X, T)p.
\]
Deduce that dispersionless limit of the Lax representation is
\[
{\cal L}_T+\{{\cal L}, {\cal A}\}=0,
\]
and that $U=U(X, T)$ satisfies the dispersionless KdV equation
\[
U_T=6UU_X.
\]
[The functions $({\cal L}, {\cal A})$ are called symbols
of operators $({ L}, { A})$.
This method of taking the dispersionless limit is analogous to the WKB
approximation in quantum mechanics.]
Obtain the same limit by making the substitutions directly in the KdV equation.

\item {\bf Dispersionless KdV and the gradient catastrophe.}
Use a chain rule to verify that the implicit solution to the dispersionless KdV
is given by 
\[
U(X, T)=F(X+6TU(X, T)),
\]
where $F$ is an arbitrary differentiable function of one variable.
[This solution is obtained by the method of characteristics. Read about it
if you want to, or see if your supervisor is willing to explain it to you.]

Assume that 
\[
U(X, 0)=-\frac{2}{\cosh^2{(X)}}
\]
and show that $U_X$ is unbounded, i.e. that
for any $M>0$ there exists $T>0$ so
that $U_X(X_0, T)>M$ for some $X_0$. Deduce that
$U_X$ becomes infinite after finite time.
This is called
gradient a catastrophe, or a shock. Draw a graph
of $U(X, T)$ illustrating this situation. Compare it with
the one--soliton solution to the KdV equation  with the same initial
condition.

\item {\bf Lax representation of ODEs}. Let $L(t), A(t)$
be complex valued $n$ by $n$ matrices such that
\[
\dot{L}=[L, A].
\]
Deduce that $\mbox{Trace}{(L^p)}, p\in \Z$ does not depend on $t$.

[It is possible to show that systems integrable in a sense of
Arnold--Liouville's theorem can be put in this form, with
the Poisson commuting first integrals given by 
traces of powers of $L$].

Assume that
\begin{eqnarray*}
L&=&(\Phi_1+i \Phi_2)+2\Phi_3\lambda-(\Phi_1
-i \Phi_2)\lambda^2,\\
A&=&-i \Phi_3+i (\Phi_1-i \Phi_2)\lambda
\end{eqnarray*}
where $\lambda$ is a parameter and find the system of ODEs
satisfied by matrices $\Phi_j(t), j=1, 2, 3$.

[ The Lax relations should hold for any value of the parameter
$\lambda£$.
The system you are asked to find known as  Nahm's equations. 
It underlies the   construction of 
non--abelian magnetic monopoles.]

Now take
$\Phi_j(t)=-i\sigma_jw_j(t)$ (no summation) 
where $\sigma_j$ are  matrices 
\[
\sigma_1=\frac{1}{2}\left(\begin{array}{cc}
0&1\\
1&0
\end{array}
\right), \quad
\sigma_2=\frac{1}{2}\left(\begin{array}{cc}
0&-i\\
i&0
\end{array}
\right), \quad
\sigma_3=\frac{1}{2}\left(\begin{array}{cc}
1&0\\
0&-1
\end{array}
\right)
\]
which satisfy
$
[\sigma_j, \sigma_k]=i\sum_{k=1}^3\varepsilon_{jkl}\sigma_l.
$
Show that the system reduces  to the Euler equations
\[
\dot{w_1}=w_2w_3,\qquad \dot{w_2}=w_1w_3,\qquad \dot{w_3}=w_1w_2.
\]
Use $\mbox{Trace}{(L^p)}$ to construct first integrals
of this system.

\item {\bf 2--soliton solution.} Assume that the scattering data
consists of two energy levels $E_1=-\chi_1^2, E_2=-\chi_2^2$ where
$\chi_1>\chi_2$ and a vanishing reflection coefficient. 
Solve  the Gelfand--Levitan--Marchenko 
equation to find the 2-soliton solution.

[Follow the derivation of the 1--soliton in the Notes
but try not to look at the N--soliton unless you really get stuck.]

\item {\bf Asymptotics of 2-soliton solution.}
Consider the 2-soliton solution with $\chi_1>\chi_2$ and
$t\rightarrow -\infty$. Set 
\[
\tau_k=(x-4\chi_k^2t)\chi_k,\qquad  k=1, 2
\]
and show that $u\rightarrow 0$ as $x\rightarrow -\infty$.
Now move along the $x$ axis: consider the leading term in $\det{A}$ 
(see the notes) when $\tau_1=0$ and then when $\tau_2=0$. Deduce
that at $t=-\infty$ the $2$-soliton solution is a superposition
of two $1$-solitons. 
What happens as $t\rightarrow \infty?$ 
[Fix values of the parameters in your solution and
use a computer to produce the animation
of two colliding solitons.]

\item {\bf Integral equation.} Let $L\psi=k^2\psi$
where $L=-\p_x^2+u$. Consider $\psi$ of the form
\[
\psi(x)=e^{ikx}+\int_{x}^{\infty} K(x, z)e^{ikz}dz
\]
where $K(x, z), \p_z K(x, z)\rightarrow 0$ as $z\rightarrow \infty$
for any fixed $x$. Use integration by parts to
show 
\[
\psi=e^{ikx}\Big( 1+\frac{i \hat{K}}{k}-\frac{\hat{K}_z}{k^2}\Big)-
\frac{1}{k^2}\int_{x}^{\infty} K_{zz}e^{ikz} dz,
\]
where $\hat{K}=K(x, x)$ and $\hat{K}_z=(\p _zK)|_{z=x}$.
Deduce that the Schr\"odinger equation is satisfied if
\[
u(x)=-2(\hat{K}_x+\hat{K}_z), \qquad\mbox{and}
\]
\[
K_{xx}-K_{zz}-u K=0\quad \mbox{for}\quad z>x.
\]
\item{\bf First integrals for KdV.}
Consider the Riccati equation 
\[
\label{riccati1}
\frac{d S}{d x}-2ik S+S^2=u.
\]
for the first integrals of KdV.
Assume that
\[
S=\sum_{n=1}^{\infty}\frac{S_n(x)}{{(2ik)}^n}.
\]
and find the recursion relations
\[
S_1(x,t)=-u(x, t), \quad
S_{n+1}=\frac{d S_n}{d x}+\sum_{m=1}^{n-1}S_{m}S_{n-m}.
\]
Solve the first few relations to show that
\[
S_2=-\frac{\p u}{\p x}, \quad S_3=-\frac{\p^2 u}{\p x^2}+u^2, \quad
S_4=-\frac{\p^3 u}{\p x^3}+2\frac{\p}{\p x}u^2
\]
and find $S_5$.
Use the KdV equation to verify directly that
\[
\frac{d}{dt}\int_{\R} S_3 dx=0, \quad \frac{d}{dt}\int_{\R} S_5 dx=0.
\]
\end{enumerate}

\section{Example sheet 3}

\begin{enumerate}

\item {\bf Gauge invariance of zero curvature equations.}
Let $g=g(\tau, \rho)$ be an arbitrary invertible matrix.
Show that the transformation
\[
\widetilde{U}= gUg^{-1}+\frac{\p g}{\p \rho} g^{-1}, \quad
\widetilde{V}= gVg^{-1}+\frac{\p g}{\p \tau} g^{-1}
\]
maps solutions to the zero curvature equation into new solutions: 
if the matrices $(U, V)$ satisfy
\[
\frac{\p}{\p \tau}U(\lambda)-\frac{\p}{\p \rho}V(\lambda)
+[U(\lambda), V(\lambda)]=0
\]
then so do $\widetilde{U}(\lambda), \widetilde{V}(\lambda)$. 
What is the relationship between
the solutions of the associated linear problems?
\item{\bf Finite gap integration.} 
Consider solutions to the KdV hierarchy which are stationary with respect to
\[
c_0\frac{\p}{\p t_0}+c_1\frac{\p}{\p t_1}
\]
where the $k$th KdV flow is generated by the Hamiltonian $(-1)^kI_k[u]$ and $I_k[u]$ are the first integrals constructed in lectures.

Show that the  resulting solution to KdV is
\[
F(u)=c_1x-c_0t,
\]
where $F(u)$ is given by an integral which should be determined
and $t_0=x, t_1=t$.

Find the zero curvature representation for the ODE characterising the
stationary solutions.

\item {\bf Nonlinear Schr\"odinger equation.}
Consider the zero curvature representation with
\begin{eqnarray*}
U&=&  
i\lambda\left(\begin{array}{cc}
1&0\\
0&-1
\end{array}
\right)+i \left(\begin{array}{cc}
0&\ov{\phi}\\
\phi&0\end{array}\right),\\
V&=& 2i\lambda^2\left(\begin{array}{cc}
1&0\\
0&-1
\end{array}
\right)+2i\lambda
\left(\begin{array}{cc}
0&\ov{\phi}\\
\phi&0\end{array}\right)
+ \left(\begin{array}{cc}
0&\ov{\phi}_\rho\\
-\phi_\rho&0\end{array}\right)
-i\left(\begin{array}{cc}
|\phi|^2&0\\
0&-|\phi|^2\end{array}\right)
\end{eqnarray*}
and show that complex valued function $\phi=\phi(\tau, \rho)$
satisfies the nonlinear Schr\"odinger equation
\[
i\phi_{\tau}+\phi_{\rho\rho}+2|\phi|^2\phi=0.
\]
[This is another famous soliton equation which can be solved by
inverse scattering transform.]




\item {\bf From group action to vector fields.}
Consider three one--parameter groups of transformations
of $\R$
\[
x\rightarrow x+\varepsilon_1, \quad
x\rightarrow e^{\varepsilon_2} x, \quad x\rightarrow \frac{x}{1-\varepsilon_3 x},
\]
and find the vector fields $V_1, V_2, V_3$ generating these
groups. Deduce
that these vector fields
generate a three-parameter group of transformations
\[
x\rightarrow\frac{ ax+b}{cx+d}, \qquad ad-bc=1.
\]
Show that
\[
[V_\alpha, V_\beta]=\sum_{\gamma=1}^3 f_{\alpha\beta}^\gamma V_\gamma, 
\qquad \alpha, \beta=1, 2, 3
\]
for some constants  $f_{\alpha\beta}^\gamma$  which should be determined.

\item{\bf ODE with symmetry.}  Consider a vector field
\[
V=x\frac{\p}{\p x}-u\frac{\p}{\p u}
\]
and find the corresponding one parameter group of transformations
of $\R^2$. Sketch the integral curves of this vector field.

Find the invariant coordinates, i.e. functions 
$s(x, u), g(x, u)$ such that 
\[
V(s)=1, \qquad V(g)=0
\]
[These are not unique. Make sure that
that $s, g$ are functionally independent in a domain
of $\R^2$ which you should specify.]

Use your results to integrate
the ODE
\[
x^2\frac{du}{dx}=F(xu)
\]
where $F$ is arbitrary function of one variable.


\item{\bf Lie point symmetries of KdV.} Consider
the vector fields
\[
V_1=\frac{\p}{\p x}, \quad V_2=\frac{\p}{\p t}, \quad
V_3=\frac{\p}{\p u}+\alpha t\frac{\p}{\p x}, \quad
V_4=\beta x \frac{\p}{\p x}+\gamma t\frac{\p}{\p t}
+\delta u\frac{\p}{\p u}
\]
where $(\alpha, \beta, \gamma, \delta)$ are constants
and find the corresponding one parameter groups of transformations of $\R^3$ with coordinates  $(x, t, u)$. 

Find $(\alpha, \beta, \gamma, \delta)$ such that these are symmetry groups of KdV and deduce
the existence of a four--parameter symmetry group.

Determine the structure constants of the corresponding Lie 
algebra of vector fields.

\item {\bf Painlev\'e II from modified KdV.}
Consider the modified KdV equation
\[
v_t-6v^2v_x+v_{xxx}=0.
\]
Find a Lie point symmetry of this equation of the form
\[
(\tilde{v}, \tilde{x}, \tilde{t})=(c^\alpha v, c^{\beta} x, c^{\gamma} t),
\qquad c\neq 0
\]
for some $(\alpha, \beta, \gamma)$ which should be found,
and find the corresponding vector field generating this group.

Consider the group invariant solution of the form
\[
v(x, t)=(3t)^{-1/3} w(z), \quad\mbox{where}\quad z= x(3t)^{-1/3}
\]
and obtain a third order ODE for $w(z)$. Integrate this ODE once
to show that  $w(z)$ satisfies the second Painlev\'e equation.

\item{\bf Symmetry reduction of Sine--Gordon.} 
Show that the transformation
\[
(\tilde{\rho}, \tilde{\tau})=(c\rho, \frac{1}{c}\tau), \qquad c\neq 0
\]
is a one--parameter symmetry of the Sine--Gordon
equation and find its generating vector field.

Consider the  group invariant solutions of the form $\phi(\rho, \tau)= F(z)$
where $z=\rho\tau$. Substitute $w(z)=\exp{(i F(z))}$
and demonstrate that the ODE arising from a symmetry reduction
is one of the Painlev\'e equations.

\end{enumerate}

\appendix          

\chapter{Manifolds}
\setcounter{equation}{0}
\label{appendix_mdf}
\def\theequation{\thechapter{}\arabic{equation}}

This course is 
intended to give  an elementary introduction
to the subject and the student is expected only to be familiar
with basic real and complex analysis, algebra and dynamics as
covered in the undergraduate syllabus.
In particular no knowledge of differential geometry is assumed.
One obvious advantage of this approach is that the course 
is suitable for advanced undergraduate students.

The disadvantage is that the discussion
of Hamiltonian formalism  
and continuous groups of transformations in earlier chapters
used phrases like   `spaces coordinatised by $(p, q)$', 
`open sets in $\R^n$' or `groups whose elements
smoothly depend on parameters' instead calling these
object by their real name - manifolds. 
This Appendix is intended to fill this gap.
\begin{defi}
\label{manifold_defi}
 An $n$--dimensional smooth manifold is a set $M$ together
with a collection of open sets $U_\alpha$ called the coordinate charts 
such that
\begin{itemize}
\item The open sets $U_\alpha$ labeled by a   countable index $\alpha$
cover $M$.
\item There exist one-to-one maps $\phi_\alpha:U_\alpha\rightarrow 
V_\alpha$ onto open sets in $\R^n$ such that for
any pair of overlapping coordinate charts the maps 
\[
\phi_\beta\circ\phi_{\alpha}^{-1}: \phi_\alpha(U_\alpha\cap U_\beta)
\longrightarrow \phi_\beta(U_\alpha\cap U_\beta)
\]
are smooth (i.e. infinitely differentiable) functions from $\R^n$ to
$\R^n$.
\end{itemize}
\end{defi}
\begin{center}
\includegraphics[width=12cm,height=6cm,angle=0]{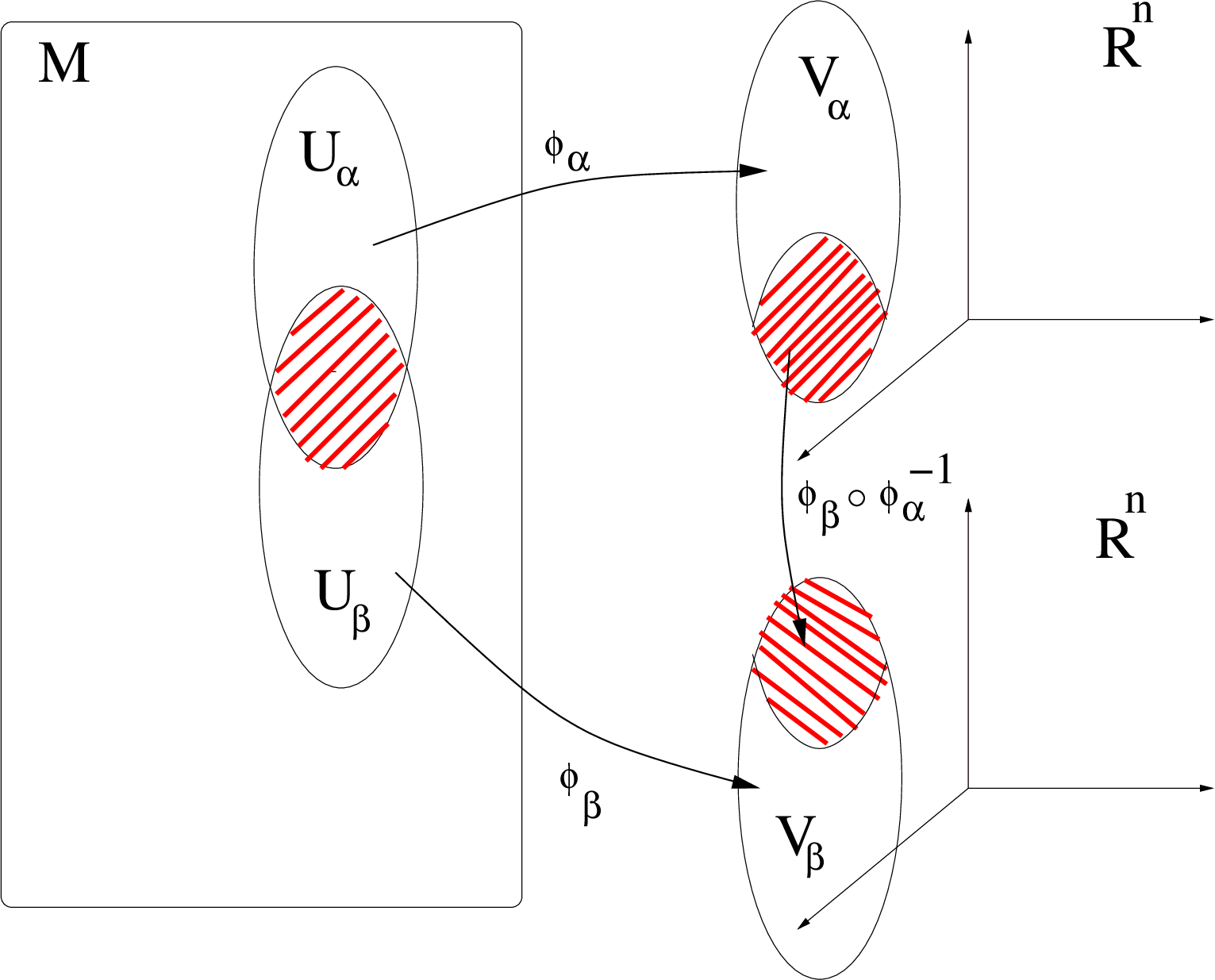}
\end{center}
Thus a manifold is a topological space together with additional structure
which makes the local differential calculus possible.
The space $\R^n$ itself is of course a manifold which can be covered by one coordinate chart. 

\begin{itemize}
\item {\bf Example.} A less trivial example is the  unit sphere 
\[
S^{n}=\{{\bf r}\in \R^{n+1}, |{\bf r}|=1\}.
\]
To verify that it is indeed a manifold, cover $S^{n}$ by two open
sets $U_1=U$ and $U_2=\widetilde{U}$
\[
U=S^n/\{0, \ldots, 0, 1\}, \quad 
\widetilde{U}=S^n/\{0, \ldots, 0, -1\}
\]
and define the local coordinates by stereographic projections
\[
\phi(r_1, r_2, \ldots, r_{n+1})=\Big(\frac{r_1}{1-r_{n+1}},
\ldots, \frac{r_n}{1-r_{n+1}}\Big)=(x_1, \ldots, x_n)\in\R^{n},
\]
\[
\tilde{\phi}(r_1, r_2, \ldots, r_{n+1})=\Big(\frac{r_1}{1+r_{n+1}},
\ldots, \frac{r_n}{1+r_{n+1}}\Big)=({x}_1, \ldots, {x}_n)\in\R^{n}.
\]
\begin{center}
\includegraphics[width=12cm,height=8cm,angle=0]{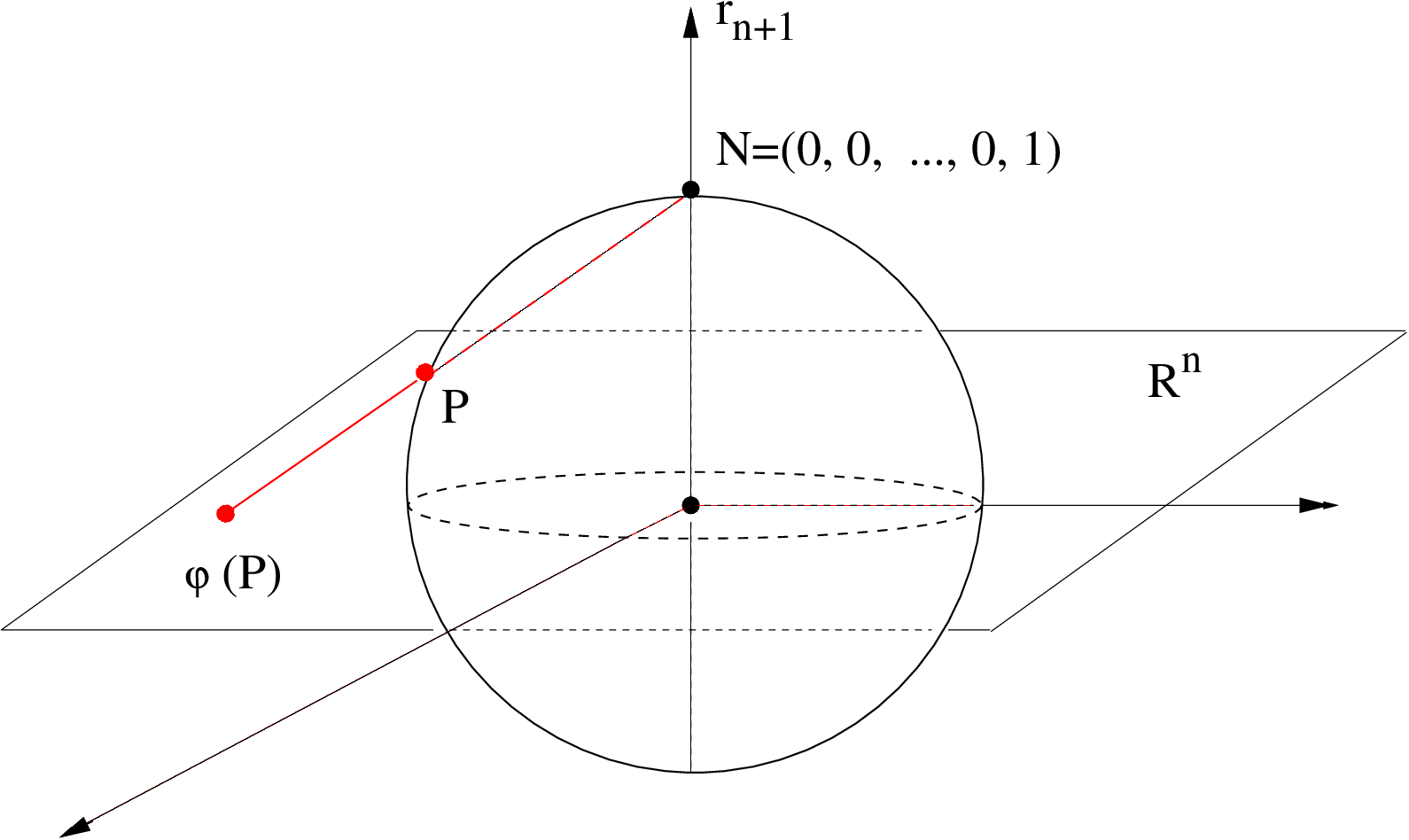}
\end{center}
Using
\[
\frac{r_k}{1+r_{n+1}}= \Big(\frac{1-r_{n+1}}{1+r_{n+1}}\Big)
\frac{r_k}{1-r_{n+1}},
\qquad k=1, \ldots, n
\]
where $r_{n+1}\neq \pm 1$ shows that
on the overlap $U \cap \widetilde{U}$ the transition functions
\[
\phi\circ\tilde{\phi}^{-1}({x}_1, \ldots, {x}_n)
=\Big(\frac{x_1}{x_1^2+\ldots+x_n^2}, \ldots,
\frac{x_n}{x_1^2+\ldots+x_n^2}\Big),
\]
are smooth.
\end{itemize}
A cartesian product of manifolds is also a manifold. For example
the $n$--torus arising in the Arnold--Liouville theorem \ref{al} is
a cartesian product on $n$ one--dimensional spheres.

Another way to obtain interesting manifolds is to define them
as surfaces in $\R^n$ by a vanishing condition for a set of 
functions.
If $f_1, \ldots, f_k:\R^n\rightarrow \R$
then the set
\be
\label{appendix_con}
M_f:=(x\in\R^n, \quad f_i(x)=0, \quad i=1, \ldots, k)
\ee
is a manifold if the rank of the $k$ by $n$ matrix of gradients
$\nabla f_i$ is constant in a neighborhood of  $M_f$ in $\R^n$.
If this rank is maximal and equal to $k$ then $\dim{M_f}=n-k$.
The manifold axioms can be verified using the implicit function theorem.
For example the sphere $S^{n-1}$ arises this way with
$k=1$ and $f_1=1-|\bf x|^2$. There is a theorem which says that
every manifold arises as some surface in $\R^n$ for sufficiently large $n$. 
If the manifold
is $m$ dimensional then $n$ is at most $2m+1$. This useful
theorem is now nearly forgotten - differential geometers like to
think of manifolds as abstract objects defined by a collections of charts as in Definition \ref{manifold_defi}.

\subsubsection{Lie groups}
We can now give a proper definition of a Lie group
\begin{defi}
\label{defi_of_lie_group_app}
A Lie group $G$ is a group and, at the same time, a smooth manifold
such that the group operations
\[
G\times G\rightarrow G, \quad (g_1, g_2)\rightarrow g_1g_2,
\qquad\mbox{and}\qquad G\rightarrow G, \quad g\rightarrow g^{-1}
\]
are smooth maps between manifolds.
\end{defi}
\begin{itemize}
\item{\bf Example.}
The general linear group $G= GL(n, \R)$ is an open set
in $\R^{n^2}$ defined by the condition $\det{g}\neq 0,
g\in G$. It is therefore a Lie group of dimension $n^2$.
The special orthogonal group $SO(n)$  is defined
by (\ref{appendix_con}), where the $n(n+1)/2$ conditions
in $\R^{n^2}$ are
\[
gg^T-{\bf 1}=0, \quad \det{g}=1.
\]
The determinant condition just selects a connected component in the set
of orthogonal matrices, so it does not count as a separate condition.
It can  be shown that the corresponding matrix of gradients has
constant rank and thus $SO(n)$ is an $n(n-1)/2$ dimensional Lie group.
\end{itemize}
In Chapter \ref{chap_lie_ref}  a Lie algebra  
$\g$  was defined as a vector space
with an anti--symmetric bilinear 
operation which satisfies the Jacobi 
identity (\ref{jacobi_bracket}). 

A Lie algebra of  a Lie group $G$ is the tangent space to $G$ at
the identity element, $\g=T_eG$ with the Lie bracket defined by
a commutator of vector fields at $e$.

\subsubsection{Proof of the first part of Arnold--Liouville's 
theorem \ref{al}.}
The gradients $\nabla f_k$ are independent, thus the set
\[
M_f:=\{(p, q)\in M; f_k(p, q)=c_k\}
\]
where $c_1, c_2, \dots, c_n$ are constant defines a manifold of dimension $n$. Let $\xi^a=(p, q)$ be local coordinates on
$M$ such that the Poisson bracket is
\[
\{f, g\}=\omega^{ab}\frac{\p f}{\p \xi^a}\frac{\p g}{\p \xi^b},
\qquad a, b=1, 2, \dots, 2n
\]
where $\omega$ is a constant anti--symmetric matrix 
\[
\left(\begin{array}{cc}
0&1_n\\
-1_n&0
\end{array}
\right).
\]
The vanishing of Poisson brackets $\{f_j, f_k\}=0$ implies
that each Hamiltonian vector field
\[
X_{f_k}=\omega^{ab}\frac{\p f_k}{\p \xi^b}\frac{\p}{\p \xi^a}
\]
is orthogonal (in the Euclidean sense) to any of the gradients $\p_a f_j,  a=1, \dots, 2n, j,k=1, \dots, n$. The gradients
are perpendicular to $M_f$, thus the Hamiltonian vector fields
are  tangent to $M_f$. They are also commuting
as
\[
[X_{f_j}, X_{f_k}]=-X_{\{f_j, f_k\}}=0,
\]
so the vectors generate an action of the abelian group $\R^n$ on
$M$. This action restricts to an $\R^n$  action on $M_f$.
Let $p_0\in M_f$, and let $\Gamma$ be a lattice consisting of
all vectors in $\R^n$ which fix $p_0$ under the group action.
Then  $\Gamma$ is a finite subgroup of $\R^n$ and
(by an 
intuitively clear modification of the orbit--stabiliser theorem)
we have
\[
M_f=\R^n/\Gamma.
\]
Assuming that $M_f$ is compact, this quotient space is
diffeomorphic to a torus $T^n$.\koniec
In fact this argument shows that we get a torus for any
choice of the constants $c_k$. Thus, varying the constants,
we find that the phase--space $M$ is foliated by $n$--dimensional tori.


\begin{thebibliography}{jafsdl}


\bibitem{ARS} Ablowitz, M. J., Ramani, A., \& 
Segur, H.  (1980) 
A connection between nonlinear evolution equations and ordinary 
differential equations of $P$-type. I, II.
J. Math. Phys.  {\bf 21} 715--721 and 1006--1015.

\bibitem{AC92} Ablowitz, M.J. \& Clarkson, P.A. (1992) {\em Solitons,
Nonlinear evolution equations and inverse scattering}, L.M.S.
Lecture note series, {\bf 149}, CUP.

\bibitem{Ablowitz_Fokas} 
Ablowitz, M. J. \& Fokas, A. S. (2003) 
Introduction and Applications of Complex Variables, Cambridge University Press, second edition.


\bibitem{arnold} Arnold, V. I.  (1989)
{\em Mathematical Methods of Classical Mechanics.}, second edition.
Graduate Texts in Mathematics, {\bf 60}, Springer.


\bibitem{DJ} Drazin, P. G., \& Johnson, R. S. (1989)
{\em Solitons: an introduction.}
Cambridge Texts in Applied Mathematics. 
Cambridge University Press, Cambridge.

\bibitem{Dunajski} Dunajski, M. (2009) {\em Solitons, Instantons and Twistors},
  Oxford Graduate Texts in Mathematics {\bf 19},  OUP, Oxford.

\bibitem{Dunajski1} Dunajski, M. (2024)
{\em Solitons, Instantons, and Twistors},
Oxford Graduate Texts in Mathematics {\bf 31},  OUP, Oxford.

\bibitem{GGKM_paper} Gardner, C.  Green, J., Kruskal, M. \& 
Miura, R. (1967) 
Method for Solving the Korteweg-deVries Equation 
Phys. Rev. Lett. {\bf 19}, 1095.


\bibitem{GLM_paper}  Gelfand, I. M  \&  Levitan, B. M. (1951) 
On the determination of a differential equation from its spectral function
Izv. Akad. Nauk SSSR, Ser. Mat. {\bf 15}, 309.


\bibitem{Hydon} Hydon P. E. (2000) {\em Symmetry Methods for Differential Equations: 
A Beginner's Guide}, CUP.
 
\bibitem{Ince} Ince, E. L. (1956) 
{\em Ordinary Differential Equations}, Dover.

\bibitem{landau} Landau, L. D., \& Lifshitz, E. M. (1995)
{\em Course of Theoretical Physics, Vol I, II}. 
Butterworth-Heinemann.

\bibitem{Lax_cite}
Lax, P. (1968), Integrals of nonlinear equations of evolution and solitary
 waves, Comm. Pure Applied Math. {\bf 21} 467-490.

\bibitem{Ma78} Magri, F. (1978) A simple model of the integrable Hamiltonian
equation, J. Math. Phys., {\bf 19},  1156-1162.

\bibitem{Marchenko_ref} Marchenko, V. A. (1955) 
Reconstruction of the potential energy from the phases of scattered 
waves Dokl. Akad. Nauk SSSR {\bf 104.}  


\bibitem{NMPZ} 
 Novikov, S., Manakov, S. V., Pitaevskii, L. P. 
\& Zakharov V. E. (1984) {\em Theory of Solitons: The Inverse
Scattering Method}, Consultants Bureau, New York.

\bibitem{olver} Olver, P. J. (1993) {\em Applications of Lie groups to differential equations.} Springer-Verlag, New--York.

\bibitem{schieff_book} Schiff, L, I (1969) 
{\em Quantum Mechanics}, 3rd ed. McGraw-Hill

\bibitem{schuster} Schuster, H.G. (1988)
{\em Deterministic Chaos: An Introduction}, second edition,
VCH Publishers, New York.



\bibitem{woodhouse} Woodhouse, N. M. J. (1987)
{\em Introduction to analytical dynamics} OUP.

\bibitem{ZM} Zaharov, V. E., \& Shabat, A. B. (1979) 
Integration of the nonlinear equations of mathematical physics by the method of the inverse scattering problem. II. Funct. Anal. Appl. {\bf 13},   13--22.

\end{thebibliography}
\end{document}